\documentclass[useAMS,usenatbib]{mn2e}
\usepackage{graphicx}

\title[Nuclear Stellar Disks in Early-type Galaxies]{A Census of Nuclear Stellar Disks in Early-type Galaxies}
\author[H. R. Ledo \textit{et al.}]{H. R. Ledo$^{1}$\thanks{E-mail:
h.ledo@herts.ac.uk}, M. Sarzi$^{1}$, M. Dotti$^{2}$, S. Khochfar$^{3}$ and L. Morelli$^{4}$\\
$^{1}$Centre for Astrophysics Research, University of Hertfordshire, College Lane, Hatfield AL10 9AB, United Kingdom\\
$^{2}$Max-Planck-Institut f\"ur Astrophysik, Karl-Schwarzschild-Str. 1, D-85748 Garching, Germany\\
$^{3}$Max-Planck-Institute f\"ur extraterrestrische Physik, Giessenbachstra$\beta$e, D-85748 Garching, Germany\\
$^{4}$Dipartimento di Astronomia, Universit\`a di Padova, vicolo dell Osservatorio 2, 35122 Padova, Italy}

\begin{document}


\pagerange{\pageref{firstpage}--\pageref{lastpage}} \pubyear{2010}

\maketitle

\label{firstpage}

\begin{abstract}
Nuclear Stellar Disks (NSDs), of a few tens to hundreds of parsec across, are a common and yet poorly studied feature of early-type
galaxies. Still, such small disks represent a powerful tool to constrain the assembling history of galaxies, since they can be used
to trace to the epoch when galaxies experienced their last major merger event. By studying the fraction and stellar age of NSDs it is
thus possible to test the predictions for the assembly history of early-type galaxies according the current hierarchical paradigm for
galaxy formation. In this paper we have produced the most comprehensive census of NSDs in nearby early-type galaxies by
searching for such disks in objects within 100 Mpc and by using archival images from the Hubble Space Telescope. We found that NSDs
are present in approximately 20\% of early-type galaxies, and that the fraction of galaxies with NSDs does not depend on their Hubble type nor on their galactic environment, whereas the incidence of NSDs appears to decline in the most massive systems. Furthermore, we have separated the light contribution of twelve such disks from that of their surrounding stellar bulge in order to extract their physical properties. This doubles the number of decomposed NSDs and although the derived values for their central surface brightness and scale-length are consistent with previous studies they also give a hint of possible different characteristics due to different formation scenario between nuclear disks and other kinds of large galactic disks.
\end{abstract}

\begin{keywords}
galaxies: elliptical and lenticular, cD -- galaxies: evolution
\end{keywords}

\section[]{Introduction}

More than a decade ago, Hubble Space Telescope (HST) observations 
revealed the presence of Nuclear Stellar Disks in several galaxies \citep{boschet94}. These disks are distinct nuclear components with a few tens to hundreds of parsecs in diameter and have since then been recognised as a relatively common feature in 
early-type galaxies \citetext{e.g. \citealp{restet01}} while being notably rare in spirals \citep{pizzellaet02}. 
Despite the relatively high frequency of NSDs only a few studies were aimed at 
deriving their fundamental properties \citep{morelliet04, pizzellaet02, 
scorzaet98}, whereas the interest in NSD revolved mainly around their 
dynamically cold character, which facilitates the mass measurement of central 
supermassive black holes (SMBHs) \citep[e.g.,][]{boschzeeuw96}.

Yet, NSDs may constitute a unique tool to constrain the assembling
history of galaxies. NSDs are indeed fragile structures that should not
survive a major merger event involving their host galaxy, which makes
them useful clocks to trace the epoch since such an event occurred.
Simple N-body simulations serve to illustrate the delicate nature of
NSDs (Fig. 1). Following \citet{dotti07} we set up a stable stellar disk that is
200 pc across, $10^8$M$_{\odot}$ in mass and which is orbiting around a
central SMBH of the same mass in the total gravitational potential
dictated also by the stellar bulge and dark-matter halo. We have then let
loose a second $10^8$M$_{\odot}$ SMBH 80 pc above the galactic plane
in a nearly circular polar orbit and follow the evolution of the disk.
Given that early-type galaxies share the same black-hole mass content
\citep{ferr_m00,geb00}, the interaction with a second SMBH of the same
mass serves to explore in conservative way (without even considering
the interaction with the stars around the second SMBH) the impact on
the disk of a merging event with second galaxy of similar mass.
Fig.~1 shows how after just 2.5 Myr the interaction with the alien
SMBH has considerably disrupted the structure of the disk, which by
becoming more vertically extended and radially concentrated would be
very hard to detect.

A systematic study of the incidence of NSD in early-type galaxies
could therefore help constraining their assembling history, which,
contrary to their star-formation history \citep[e.g.][]{thomas05}, is
poorly understood. According to the standard hierarchical 
paradigm for galaxy formation, the most massive galaxies should have been the last 
to reach their final configuration as they follow the merging paths of their host dark-matter haloes. The galactic environment should also play a role since once galaxies enter very crowded environments such as galaxy clusters it is more difficult 
for them to merge due to the high relative velocities with which they cross each other.
These dependencies are illustrated in Fig.~2, which shows the predictions of the 
semi-analytical models of \citet{ks06b,ks06a} for the epoch of the last 
major merger experienced by galaxies of different masses and living in cluster or field 
environments.
Semi-analytical models can also track whether the last merging events involved 
considerable amount of gas as well as the probability that a small gas-rich 
companion was subsequently acquired, which will then determine whether a NSD
is found today. Numerical simulations \nocite{hopkins10, mayeret07, barneshern96} $($e.g. Hopkins and Quataert, submitted; Mayer \textit{et al.} 2007; Barnes and Hernquist 1996$)$ have indeed shown that when gas is involved in a merger event, it is always driven towards the centre of the remnant where it could then form a disk, depending on its angular momentum.
NSDs could also provide constraints on the assembling history of their host galaxies in a more direct way, by dating the disk stellar population. In general, we expect the age of the stellar disk to represent a lower limit to the look-back time since the epoch of the
last major merger event experienced by their host galaxy, since the NSD could have formed also after such an event, unless the last
major merger was a gas rich event that led also to the formation of nuclear disk itself.

\begin{figure}
\begin{center}
\label{merger}
\includegraphics[width=3in]{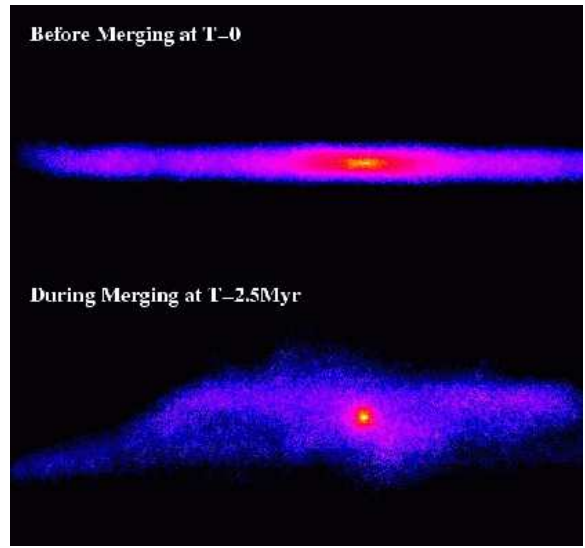}
\end{center}
\caption{N-body simulation illustrating the fragility of nuclear stellar disks during  major merger events. The top panel shows a stable nuclear disk 200 pc in diameter and 10$^8$ M$_{\odot}$ in mass that is orbiting a SMBH of the same mass in the total gravitational potential dictated also by the bulge and the dark matter particles, which are not shown for the sake of clarity. A second 10$^8$ M$_{\odot}$ SMBH is let loose in a nearly circular polar orbit 80 pc above the galactic plane, to simulate the impact with a second galaxy of similar mass. 
The lower panel illustrate the disruption suffered by the disk after just 2.5 Myr, which would be even greater if the bulge and dark matter particle from the incoming galaxy would have been included in the simulation.}
\end{figure}

\begin{figure}
\begin{center}
\label{sim}
\includegraphics[width=3in]{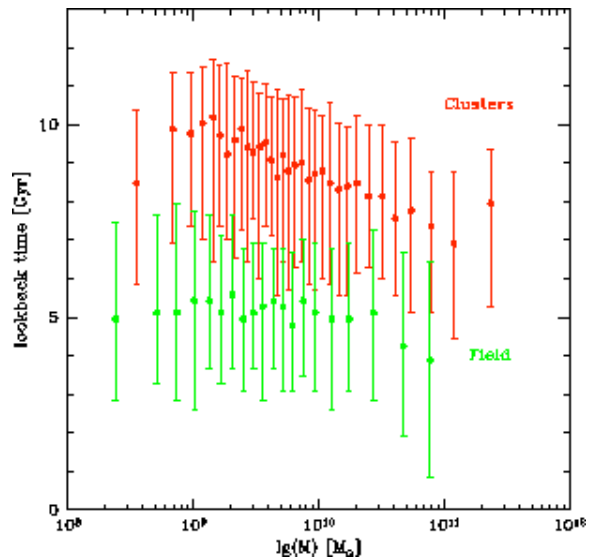}
\end{center}
\caption{Prediction of semi-analytical models from \citet{ks06b, ks06a} for the age of the last major merger event experienced by galaxies of different masses and in different environments. Field early-type galaxies appear to have assembled later than those in clusters. In clusters, the models also predict a clear dependence on mass, with larger galaxies assembling later.}
\end{figure}

As a first step to use NSDs as tools to constrain the assembling history 
of early-type galaxies, in this paper we will provide for the first time a complete census 
of the nearby NSD population, providing also estimates for basic physical properties 
such as their mass and extent. This work will provide the basis for further 
investigations which will assess the fragility of the NSD identified here using a more 
comprehensive set of numerical simulations. Such set of fragile disks will constitute a sample for spectroscopic follow-up aimed at deriving their stellar ages, while the incidence of such fragile disks could be compared with the prediction of semi-analytical 
models.

This publication is organised as follows. To compile such a census we had to define a sample and retrieve the images (\textsection 2), to look for disks in these galaxies and to separate the disks from the bulges in order to find their properties (\textsection 3). The results will be presented in \textsection 4 and the conclusions in \textsection 5.

\section{Sample Selection and Data Mining}

\subsection{Selection and Acquisition}

For this study we have selected, using the LEDA database \citep{paturelet03}, all early-type galaxies ($E$, $E$-$S0$, $S0$ and $S0a$) within 100 Mpc so as to produce a volume-limited sample. This list was then cross-correlated with the HST archive and all the galaxies with images from the WFPC2 and ACS cameras were requested, preferably the filters F555W and F606W. In some cases these filters were not available and others had to be chosen. The use of different passbands should not bias our estimates of the disk inclination and extent, however, as shown by \citet{mor10} nuclear disks do not present strong radial colour gradients.\\
After downloading the available images they were checked for their quality and those saturated, too dusty or in which the galaxy was lying at the edge of the detector were discarded since in these cases it would have been impossible to find NSDs even if they were present. In the end we collected a sample of 466 early-type galaxies with HST
images, out of a parent LEDA sample of 6801 galaxies.

\subsection{Sample Properties}

Having a sample, we retrieved the global parameters of our galaxies. Of particular interest are M$_{\mathrm{B}}$ and an estimate of the galactic environment, because they directly relate to the general questions we are trying to answer, and the galaxy inclination because it influences our ability to find the disks.The magnitude of our galaxies and an estimate for their inclination
values are available from the LEDA database whereas to obtain a value
for the galactic density we used the Nearby Galaxies Catalogue of
\citet{tully88}. 

Although the distribution of M$_{\mathrm{B}}$ and environment values for our HST sample and its parent LEDA sample are somewhat different this does not pose as a problem when we are dealing with the incidence of NSD. Indeed in this case it only matters that in each magnitude bin we have enough objects to obtain secure estimates for the NSD fraction. Also inclination differences do not matter for constraining the incidence of NSDs, as long as the distribution of inclination in each HST subsample allows a reliable correction of the fraction of NSD (that is if there are enough galaxies where the disk could have been detected, see \S4.1.1). On the other hand we need to keep in mind that any conclusion on the structural properties of the NSDs, such as their typical size or mass, will be specific to our HST sample.

\subsection{Inclination}

The more face-on a disk is, the harder it is to identify it. \citet{rw90} have studied this problem and concluded that for inclinations with $cos(i) > 0.6$ it is impossible to detect disky signatures in the galaxy isophotes. The inclination, $i$, is defined as being 0$^{\circ}$ for face-on disks and 90$^{\circ}$ if they are edge-on.

Although we do not know for sure the inclination of the nuclear disks, we will assume that the vast majority of them will have the same inclinations of their host galaxies, which is justified if we consider our sample galaxies as oblate axisymmetric spheroids. The LEDA database provides inclination estimates for the host galaxies based on the method applied by \citet{heid72} and \citet{bot83}. The inclination was calculated assuming all galaxies of a given Hubble type had an axis ratio equal to that observed in the flattest galaxy of that type. By construction, this method provides a conservative lower estimate for our galaxies' inclination, since intrinsically rounder galaxies in a given Hubble type would in reality have a higher inclination. Using such lower estimates to correct the fraction of NSD will lead to an upper limit on their incidence, whereas the uncorrected fraction provides a lower limit as if we were assuming that all galaxies where edge-on.

\section{Analysis}
\subsection{Disk Identification Process}

Having collected the HST images of 466 early-type galaxies, we
systematically searched for the signature of the presence of a NSD,
both in maps for the fine structure of the galaxy and in the shape of
the galaxy isophotes.

\noindent We have used the code of \citet{pm02} to generate structure
maps for each of our sample galaxies, taking care to use in each case
the appropriate WFPC2 or ACS point-spread function from the Tiny Tim
code \citep{kh04}.

\begin{figure}
\begin{center}
\label{stmaps}
\includegraphics[width=3.3in]{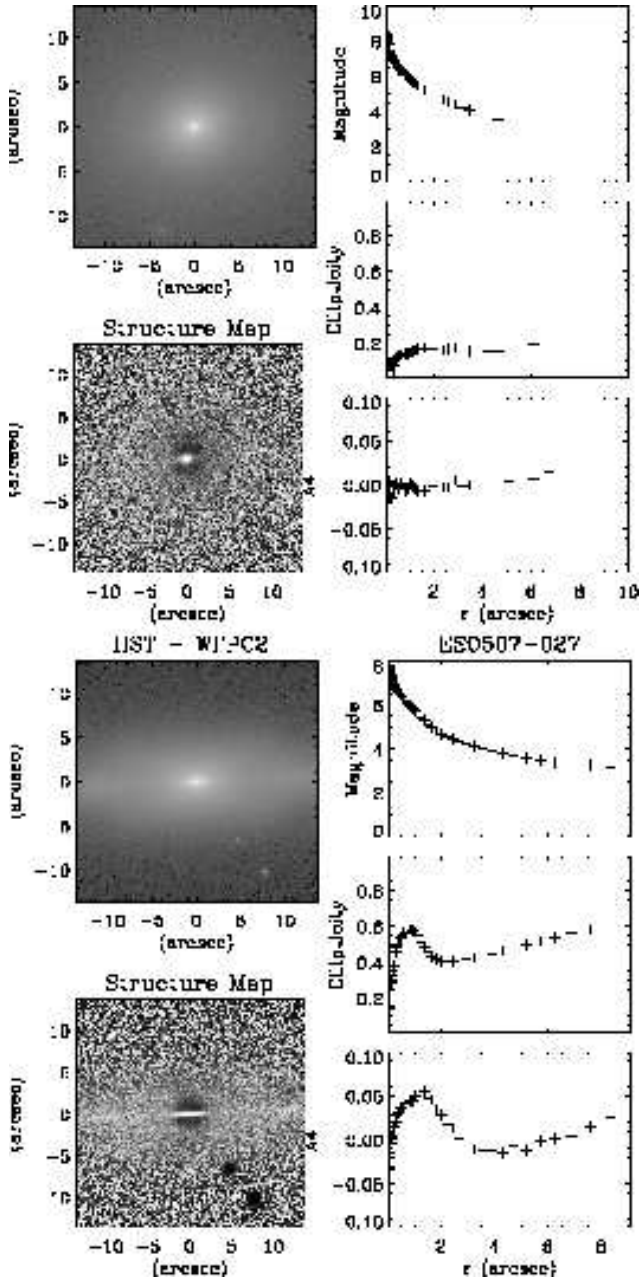}
\end{center}
\caption{On the top panel we have the original image of NGC2592, its structure map and magnitude, ellipticity and A$_4$ profiles where no Nuclear Stellar Disk is visible and, on the bottom we have the same information for ESO507-027 where its presence is detected.}
\end{figure}

We then fitted ellipses to the galaxy isophotes using the
IRAF task \textit{ellipse} and, in particular, extracted profiles for
the galaxy ellipticity, $\epsilon$, and the $A_4$ parameter which is
the $4^{th}$ coefficient of the cosine term of the Fourier expansion
of the considered isophotes \citep{carter77, jed87}. In practice, the $A_4$ parameter measures the deviation of
the isophote's shape from a perfect ellipse, and is positive in the
case of disky deviations.

The existence of a disk in our galaxies is often revealed by the structure maps, but to separate disks from otherwise simply highly
flattened central regions (e.g. a flattened bulge or a larger galactic-scale disk) we seek for a distinct peak in both the $\epsilon$ and $A_4$ profiles.

Figure 3 illustrates our disk-identification
procedure for two sample galaxies, one with and one without a
NSD. Similar figures for all the galaxies where we identified the
presence of a nuclear disk are presented in Appendix~A.

\subsection{Disk-Bulge Decomposition}

Having identified which galaxies in our sample host a NSD, we now wish
to investigate the basic properties of such disks by disentangling
their contribution to the observed surface brightness distribution of the
host galaxy. 
For early-type galaxies where the main bulge component displays simple
elliptical isophotes we can adopt the algorithm devised by
\citet{scoben95}, whereby the best disk parameters are sought by
iteratively subtracting from the galaxy image an exponential disk
model until the original signature of the disk is completely erased in
the $\epsilon$ and $A_4$ profiles that are measured in the residual
image. To perform the the Scorza \& Bender decomposition on our sample nuclear disks
we use the IDL implementation of this method of \citet{morelliet04}, where more
details about the algorithm can be found.

Fig. 4 illustrates for the ESO507-027 how the disky
deviations in the $A_4$ profile are minimised after the subtraction of
the best exponential disk model. Appendix~B shows similar figures for
all NSD that were found embedded in an elliptical bulge, whereas
Table A1 lists their basic parameters. In a few cases we report the values for the disk parameters from previous studies.

\begin{figure}
\begin{center}
\label{dbdecomp}
\includegraphics[width=3.4in]{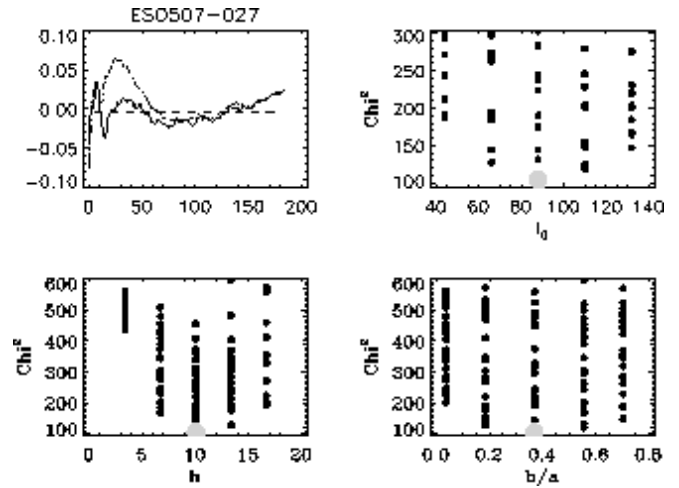}
\end{center}
\caption{On the top left panel we can see $A_4$ profiles before (dot-dashed line) and after (solid line) Scorza and Bender decomposition for ESO507-027. The other three panels show the chosen parameters with their corresponding lower $\chi^2$.}
\end{figure}

\subsection{Disk Size Estimates}

For a large fraction of the NSD that we identified we could not apply
the previously described technique efficiently, mostly owing to the presence of an intrinsically 
boxy bulge, in particular if the boxiness changes with radius, or dust. Under these circumstances we can at least estimate
the extent of the NSD by exploiting the fact that the position of the
peak of the $A_4$ deviation introduced by the disk correlates loosely
with the actual extent of the disk, as found in the objects for which
a disk-bulge decomposition has been performed (Fig. 5). This allowed us to estimate
the sizes of the disks which could not be decomposed with the Scorza and Bender method.
They are listed in Table A2.

\begin{figure}
\begin{center}
\label{h-h}
\includegraphics[width=3in]{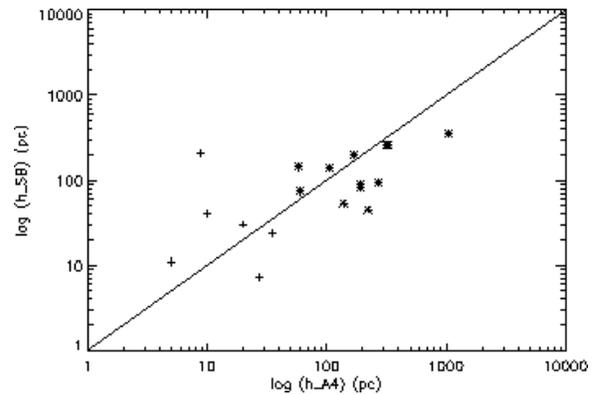}
\end{center}
\caption{Relation between the sizes derived from the peak in the $A_4$ profile and those obtained from the Scorza and Bender decomposition. Asterisks represent the disks decomposed in this work and crosses show disks decomposed in previous works \citep{sb95, scorzaet98, morelliet04, vdb94}.}
\end{figure}

\section{Results}

\subsection{The Census}

By inspecting the structure maps and the isophotal shape of the
central regions of 466 early-type galaxies we have found evidence for
a distinct disk component in 63 objects, corresponding to $13.52$\% of
our sample. Table 1 breaks down the observed
fraction of NSDs as a function of galaxy type, whereas
Tab. 2 and Tab. 3 show the incidence of NSDs as function of the B-band absolute
magnitude and galactic environment, when such quantities could be measured.

\subsubsection{Inclination correction and Final Values}

As mentioned in Sec~2.3, the inclination of the nuclear disks greatly
affects our ability to detect them. It is indeed very likely that a
considerable number of NSDs have escaped our detection because they
lie in the equatorial plane of galaxies that are only slightly inclined
from the plane of the sky, so that the NSDs appear close to face-on to
us. The fraction of NSDs deduced directly from number of observed NSDs
therefore represents only a lower limit for the true incidence of
NSDs. We can estimate such true fraction considering that the number
of NSDs that we have found should reflect the true fraction of NSD
when considering only the systems where a NSDs could have in fact been
detected, so that

\begin{equation}
f_{NSD} = \frac{f_{NSD, observed}}{f_{detectable, NSD}}
\end{equation}

where $f_{NSD, observed}$ is the observed fraction of NSDs in the entire
sample and $f_{detectable, NSD}$ is the fraction of galaxies where NSDs
can be found. Using the estimates for the inclination of our sample
galaxies that we retrieved in Sec~2.3, we can compute $f_{detectable,
  NSD}$ considering that according to \citet{rw90} the ability
to detect an embedded disk drops very quickly for $cos(i) > 0.6$,
independent of the relative disk contribution to the total light
distribution. Using the previous equation, the fraction of galaxies
with $cos(i) < 0.6$ and considering that our inclination values are
only lower estimates we can correct our face values for the incidence
of NSDs and obtain upper limits for such fractions. These are also
listed in Tabs. 1, 2 and 3.

Table 1 indicates that approximately between 13 and 23\% of nearby early-type
galaxies host nuclear stellar
disks, without significant differences depending on their Hubble
type. As regards the fraction of nuclear stellar disk as a function of
their host absolute magnitude (Tab. 2), it would
appear that the incidence of the disks peaks in the magnitude range
between ${\rm M_B}=-20$ and $-18$, decreasing sharply in particular
towards higher stellar luminosity where no NSD is found. In fact even
accounting for the smaller number of surveyed systems in the $-24$ to
$-22$ magnitude bin we should have found between 1 and 4 objects to be
consistent with the fraction estimated in the $-22$ to $-20$
bin. Finally, there appears to be no significant trend with
environment, at least as defined in \citet{tully88}, although we need
to keep in mind that unfortunately only less than half of our sample
was found in this particular catalogue.

\begin{table}
  \begin{minipage}[!h]{\columnwidth}
    \label{table:diskstypc}
    \centering
    \renewcommand{\footnoterule}{}  
    \begin{tabular}{cccccc}
      \hline \hline
       Type & HST & NSDs & ${\rm f_{NSDs,obs}}$& ${\rm f_{HST,detect}}$ & ${\rm f_{NSDs}}$ \\
      (1)        & (2) & (3)  & (4)             & (5)                    & (6)                   \\
      \hline
	E & 219 & 22 & 10.04$\pm$2.03 & 52.97 & 18.95$\pm$4\\
	E-S0 & 68 & 9 & 13.24$\pm$4.11 & 63.24 & 20.9$\pm$6.2\\
	S0 & 109 & 21 & 19.27$\pm$3.78 & 66.97 & 28.8$\pm$5.3\\
	S0-a & 70 & 11 & 15.71$\pm$4.35 & 64.29 & 24.44$\pm$6.8\\
	Total & 466 & 63 & 13.52$\pm$1.58 & 59.44 & 22.75$\pm$2.52\\
      \hline
    \end{tabular}
    \caption{As a function of galaxy Hubble type (1) the number of objects with HST images and of the NSDs found in them (2)-(3) yield a lower limit on the incidence of NSDs in early-type galaxies (4). Using the fraction of objects where disk detection is possible (5) we can correct for inclination biases and obtain an upper limit on the fraction of NSDs (6)}
  \end{minipage}
\end{table}

\begin{table}
  \begin{minipage}[!h]{\columnwidth}
    \label{table:disksmagc}
    \centering
    \renewcommand{\footnoterule}{}  
    \begin{tabular}{cccccc}
      \hline \hline
      ${\rm M_B}$& HST & NSDs & ${\rm f_{NSDs,obs}}$& ${\rm f_{HST,detect}}$ & ${\rm f_{NSDs}}$ \\
      (1)        & (2) & (3)  & (4)             & (5)                    & (6)                   \\
      \hline
	-24 to -22 & 16  & 0  & 0            & 81.3 &  0            \\
	-22 to -20 & 175 & 23 & 13.1$\pm$2.6 & 59.4 & 22.1$\pm$4.1 \\
	-20 to -18 & 174 & 34 & 19.5$\pm$3.0 & 62.6 & 31.2$\pm$4.4 \\
	-18 to -16 & 70  & 6  &  8.6$\pm$3.4 & 51.4 & 16.7$\pm$6.2 \\
	-16 to -15 & 17  & 0  & 0            & 53.0 & 0            \\
      \hline
    \end{tabular}
    \caption{Same as Tab.~1 but as a function of the galaxy absolute B-band magnitude (1), which was not available for 14 objects.}
  \end{minipage}
\end{table}

\begin{table}
  \begin{minipage}[!h]{\columnwidth}
    \label{table:disksenvc}
    \centering
    \renewcommand{\footnoterule}{}  
    \begin{tabular}{cccccc}
      \hline \hline
      Density & HST & NSDs & ${\rm f_{NSDs,obs}}$& ${\rm f_{HST,detect}}$ & ${\rm f_{NSDs}}$ \\
      (1)        & (2) & (3)  & (4)             & (5)                    & (6)                   \\
      \hline
	0-1 & 128 & 25 & 19.5$\pm$3.5 & 19.5 & 33.3$\pm$5.4\\
	1-2 & 23 & 6 & 26.1$\pm$9.2 & 26.1 & 35.3$\pm$11.6\\
	2-3 & 18 & 6 & 33.3$\pm$11.1 & 33.3 & 42.9$\pm$13.2\\
	3-4 & 22 & 5 & 22.7$\pm$8.9 & 22.7 & 41.7$\pm$14.2\\
      \hline
    \end{tabular}
    \caption{Same as Tab.~1 but as a function of galactic environment (1), from \citet{tully88}. Only 191 objects were listed in this catalogue.}
  \end{minipage}
\end{table}

\subsection{Properties of the Decomposed Disks}

We have applied the Scorza \& Bender disk-bulge decomposition to 12
disks, which doubles the number of NSD that were previously analysed
in this way. Although a few more NSDs appeared embedded in well
defined elliptical bulges, it was not possible to disentangle their
light contribution. This is mainly because the observed $A_4$ profile
does not appear to be well described by the simple model we used for the
disks. \citet{svdb98} compiled all previous measurements for
the structural properties of NSDs and plotted them together with other
kinds of galactic disks in a $\mu^0_c - h$ plot which has later been updated by \citet{morelliet04} and 
that we present in Figure 6, to which we now add our own 12 objects. Although
our values fall within the range found in previous results, confirming
that NSDs follow a similar $\mu^0_c - h$ trend as embedded
disks or main galactic disks, many of our decomposed NSDs appear to
have a lower central surface brightness or smaller scale radius. In
fact, the position of all NSDs in Fig~. 6 suggests that these may
follow a somewhat steeper and offset relation compared to bigger
disks.

\begin{figure}
\begin{center}
\label{miu-c}
\includegraphics[width=3.8in]{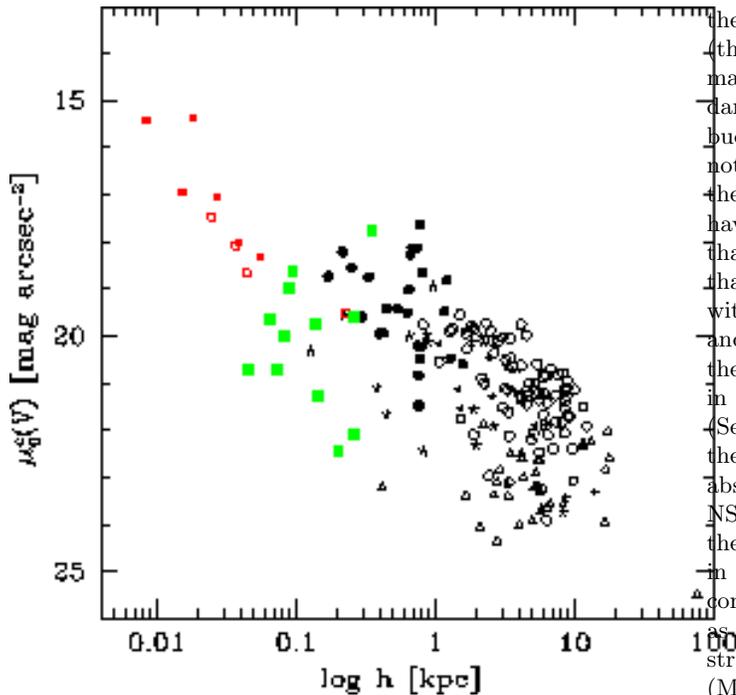}
\end{center}
\caption[Disks $\mu_0^c - h$ diagram.]{Disks $\mu_0^c - h$ diagram from \citet{morelliet04}. The open circles correspond to high surface-brightness spiral galaxies whereas low surface-brightness spirals are represented by triangles. Stars denote S0s and filled circles disky ellipticals. NSDs in elliptical and lenticular galaxies are indicated by filled squares and in spirals by open squares. The larger green squares are those added in this work and the smaller red ones are the pre-existing ones.}
\end{figure}

\section{Conclusions}

Nuclear stellar disks have long been regarded as common features in
early-type galaxies, but a quantitative assessment of such statements
was still lacking prior to this work. By performing the most extensive
volume-limited census of NSDs to date we have now shown that between
13 to 23\% of nearby early-type galaxies host such systems. The
incidence of nuclear stellar disks appears to decline in the most
massive systems, consistent with the expectations that dry mergers
dominated the most recent history of such objects \citep{ks09,kb03} and also with the
finding that NSDs are exceedingly rare in brightest cluster galaxies \citep[2\%,][]{laine03}.
On the other hand, it is less clear what may cause the decrease in the fraction of NSDs
that is found also at the faint end of our sample.  Similarly, the lack of a dependence on
galactic environment seems to contrast the first-order predictions of
semi-analytical models (Fig.~2). In fact, according to these galaxies
in clusters should have assembled much earlier and thus there should
have been more time to regrow a nuclear disk.
On the other hand, galaxies in cluster may not have ample reservoirs of cold gas at their disposal,
in particular if orbiting within the hot-gas halo of the cluster, so that the early assembly of clusters does not make the presence of nuclear disks in cluster galaxies much more likely than in their field counterparts.

Using the well established technique of \citet{scoben95} to
disentangle the light contribution of embedded disks from that of
their surrounding bulges, we have extracted the structural parameters
of 12 out of the 63 NSDs that we have identified. By doubling the
number of NSDs we have shown even more convincingly that NSDs, like other kinds of large galactic disks, follow a  correlation
between the central surface brightness $\mu^0_c$ and scale length $h$,
although it would appear that NSDs obey a somewhat different relation
with smaller scale lengths or fainter central surface brightness
values. Considering that disk galaxies obey a very precise relation
between their total luminosity and their maximum rotation velocity
(the Tully-Fisher relation, \citet{tully77}) that may ultimately derive from
the fact that they form in the dark matter halos that surround and
dominate the mass budget of galaxies \citep{steinmetz99}, it
may not be completely unexpected that NSDs, by forming in the central
and bulge-dominated regions of galaxies, may have different structural
and dynamical properties than that of their larger relatives. In fact, it is also possible that the structural properties of NSDs vary when dealing with different kinds of early-type galaxies such as disky and boxy systems \citep{kb96}. Although the
main reason for failing to isolate the disk component in the presence of a bulge of varying boxiness with radius (Sec 3.3) has
mostly to do with an unclear a priori for the intrinsic $A_4$ profile that the bulge should have in the absence of a disk, we cannot
exclude that in boxy ellipticals NSDs may have a different profile than exponential. Along the lines first suggested by \citet{bs99}, NSDs in boxy ellipticals could be akin to the rapidly rotating components that in these objects are generally identified
as kinematically decoupled cores, although in general such structure tend to extend over much larger scales than NSDs \citep{m06}, in particular for massive ellipticals with the oldest stellar populations.

Finally, although for the majority of the NSDs we could not clearly separate the disk and bulge contributions to the central light of their host galaxies, we have nonetheless estimated their sizes using as a reference radius the distance from the centre where the diskyness of the central isophotes appears to peak. Such scale length estimate, coupled with the $\mu^0_c - h$ relation that NSDs appear also to obey and the N-body simulations, will be used in future investigations to assess which of our sample NSDs are indeed fragile against major merging events, thus providing good targets for a spectroscopic follow up aimed at constraining their stellar ages and thus, the time of their last major merger.

\bibliography{refs.bib}

\appendix

\section{Nuclear Stellar Disks}

\begin{table*}
  \begin{minipage}[H!]{\columnwidth}
    \caption[Properties of the decomposed disks]{Properties of the decomposed disks where $h$ is the scale-length, the inclination is as defined in \textsection 2.3 and $\mu_{0,V}^c$ is the central surface brightness.}
    \label{table:disks}
    \centering
    \renewcommand{\footnoterule}{}  
    \begin{tabular}{cccc}
      \hline \hline
      \textbf{Galaxy name} & \textbf{h} & \textbf{inclination} & \boldmath$\mu_{0,V}^c$\\
       & (pc) & ($^\circ$) & (mag arcsec$^{-2}$)\\
      \hline
	ESO352-057 & 44.74$\pm^{23.16}_{16.05}$ & 76.26$\pm^{8.58}_{22.42}$ & 20.74$\pm^{1.05}_{1.04}$\\
	ESO378-020 & 53.39$\pm^{33.51}_{18.54}$ & 65.98$\pm^{16.78}_{19.69}$ & 19.1$\pm^{1.27}_{0.58}$\\
	ESO507-027 & 94.87$\pm^{23.22}_{16.8}$ & 68.28$\pm^{7.83}_{9.61}$ & 18.61$\pm^{0.47}_{0.37}$\\
	IC0875 & 259.67$\pm^{59.41}_{77.71}$ & 84.84$\pm^{2.29}_{2.31}$ & 22.08$\pm^{0.63}_{0.4}$\\
        NGC0584 & 90.18$\pm^{24.82}_{18.37}$ & 73.86$\pm^{6.93}_{4.96}$ & 18.97$\pm^{0.6}_{0.28}$\\
	NGC3385 & 351.19$\pm^{52.33}_{55.07}$ & 61.71$\pm^{1.55}_{5.08}$ & 17.78$\pm^{0.06}_{0.16}$\\
	NGC3610 & 260.38$\pm^{44.81}_{79.84}$ & 81.81$\pm^{4.19}_{1.6}$ & 19.60$\pm^{0.77}_{0.2}$\\
	NGC4128 & 138.01$\pm^{43.84}_{38.68}$ & 85.76$\pm^{1.95}_{0.92}$ & 19.77$\pm^{0.67}_{0.21}$\\
	NGC4474 & 143.5$\pm^{59.28}_{59.21}$ & 86.49$\pm^{2.36}_{2.23}$ & 21.29$\pm^{1.21}_{0.54}$\\
	NGC4621 & 73.87$\pm^{13.71}_{24.26}$ & 83.97$\pm^{2.59}_{2.02}$ & 20.72$\pm^{0.6}_{0.32}$\\
	NGC4660 & 81.19$\pm^{12.32}_{27.19}$ & 78.92$\pm^{4.19}_{3.99}$ & 20.02$\pm^{0.51}_{0.33}$\\
	NGC5308\footnote{We were unable to derive the errors for the inclination.} & 201.59$\pm^{30.21}_{45.56}$ & 89.89$\pm^{-}_{-}$ & 22.46$\pm^{-}_{-}$\\
      \hline
    \end{tabular}
  \end{minipage}
\end{table*}

\subsection[]{Properties of the Galaxies With Disks}

\begin{table*}
\small
    \caption[Galaxies where disks were found and some of their properties.]{Galaxies where disks were found and some of their properties. (1) Galaxy's name; (2) Hubble type; (3) log$_{10}$ of the axis ratio; (4) log$_{10}$ of the apparent major axis diameter d25 at 25 mag.arcsec$^{-2}$; (5) disk inclination; (6) absolute B-magnitude; (7) radial velocity corrected for Local Group infall towards Virgo (208 kms$^{-1}$); (8) Distance derived from (7) and assuming H$_0$=72 kms$^{-1}$ Mpc$^{-1}$, except for Virgo cluster members for which we adopted a common distance of 18 Mpc \citep{fouque01}; (9) Density from \citet{tully88}; (10) Disk size with the indication in (11) of the source of the values, where \textit{M} are those measured with the decomposition procedure, \textit{E} have been estimated using the peak of the $A_4$ and \textit{L} come from the literature.}
    \label{table:gal_disks}
    \centering
    \renewcommand{\footnoterule}{}  
    \begin{tabular}{ccccccccccc}
      \hline \hline
      \textbf{Name} & \textbf{Type} & \textbf{logr$_{25}$} & \textbf{logd$_{25}$} & \textbf{inclination} & \textbf{mabs} & \textbf{V\_vir} & \textbf{distance} & \textbf{Density} & \textbf{Disk size} & \textbf{Source}\\
	  &  &  &  & \textbf{($^o$)} &  & \textbf{(kms$^{-1}$)} & \textbf{(Mpc)} & \textbf{(Mpc$^{-3}$)} & \textbf{(pc)} & \\
	(1) & (2) & (3) & (4) & (5) & (6) & (7) & (8) & (9) & (10) & (11)\\
      \hline
	ESO352-057 & S0 & 0.327 & 1.014 & 82.98 & -19.842 & 5736.2 & 76.11 & - & 44.74 & M\\      
	ESO378-020 & S0 & 0.243 & 1.186 & 68.17 & -19.899 & 5479.7 & 39.87 & - & 53.39 & M\\      
	ESO507-027 & S0 & 0.682 & 1.251 & 90.0 & -20.079 & 2870.7 & 43.27 & - & 94.87 & M\\      
	IC0875 & S0 & 0.234 & 1.16 & 67.23 & -19.442 & 3115.4 & 42.32 & - & 259.67 & M\\     
	NGC0584 & E & 0.179 & 1.578 & 79.78 & -20.891 & 3047.0 & 24.49 & 0.42 & 90.18 & M\\
	NGC1023 & E-S0 & 0.377 & 1.868 & 76.7 & -20.914 & 769.9 & 10.69 & 0.57 & 57 & E\\      
	NGC1129 & E & 0.42 & 1.444 & 90.0 & -21.604 & 5404.8 & 75.07 & - & 146 & E\\
     	NGC1351 & E-S0 & 0.189 & 1.532 & 65.35 & -19.004 & 1301.0 & 18.07 & 1.57 & 131 & E\\
	NGC1381 & S0 & 0.421 & 1.405 & 90.0 & -19.338 & 1520.6 & 21.12 & 1.54 & 131 & E\\      
	NGC1426 & E & 0.175 & 1.457 & 78.27 & -19.092 & 1263.2 & 17.54 & 0.66 & 43 & E\\      
	NGC1427 & E & 0.179 & 1.636 & 71.35 & -19.399 & 1159.2 & 16.10 & 1.59 & 16 & E\\      
	NGC1439 & E & 0.02 & 1.473 & 23.03 & -19.507 & 1486.8 & 20.65 & 0.45 & 130 & E\\
    	NGC2549 & S0 & 0.594 & 1.563 & 90.0 & -19.445 & 1248.6 & 17.34 & 0.13 & 42 & E\\      
	NGC2685 & S0-a & 0.272 & 1.636 & 68.79 & -19.092 & 1094.2 & 15.2 & 0.13 & 66 & E\\      
	NGC2787 & S0-a & 0.249 & 1.510 & 65.86 & -19.645 & 949.0 & 13.18 & 0.06 & 115 & E\\      
	NGC2865 & E &  0.082 & 1.382 & 45.2 & -20.792 & 2578.8 & 35.82 & 0.11 & 191 & E\\
     	NGC3115 & E-S0 & 0.383 & 1.916 & 81.6 & -20.042 & 642.6 & 8.93 & 0.08 & 30.0 & L\footnotemark[1]\\
	NGC3156 & S0 & 0.28 & 1.282 & 79.15 & -18.512 & 1340.3 & 18.62 & 0.2 & 63 & E\\
      	NGC3377 & E & 0.327 & 1.588 & 90.0 & -19.163 & 744.5 & 10.34 & 0.49 & 35 & E\\
      	NGC3384 & E-S0 & 0.349 & 1.717 & 90.0 & -19.826 & 917.1 & 12.74 & 0.54 & 49 & E\\
	NGC3385 & S0 & 0.267 & 1.182 & 72.56 & -21.927 & 7824.4 & 108.67 & - & 351.19 & M\\      
	NGC3610 & E & 0.027 & 1.375 & 25.95 & -20.664 & 1943.4 & 26.99 & 0.3 & 260.38 & M\\      
	NGC3613 & E & 0.315 & 1.544 & 90.0 & -20.864 & 2232.9 & 31.01 & - & 15 & E\\
     	NGC3706 & E-S0 & 0.167 & 1.477 & 61.64 & -21.117 & 2817.9 & 39.14 & 0.27 & 114 & E\\
	NGC3818 & E & 0.199 & 1.387 & 90.0 & -19.389 & 1678.7 & 23.32 & 0.2 & 23 & E\\      
	NGC3900 & S0-a & 0.311 & 1.413 & 70.82 & -20.168 & 1942.8 & 26.98 & 0.13 & - & E\\      
	NGC3945 & S0-a & 0.227 & 1.744 & 63.17 & -20.085 & 1496.5 & 20.78 & 0.5 & 50 & E\\
  	NGC4026 & S0 & 0.702 & 1.644 & 90.0 & -19.592 & 1206.5 & 16.76 & 1.71 & 41 & E\\
      	NGC4128 & S0 & 0.519 & 1.344 & 90.0 & -20.041 & 2591.1 & 36.0 & 0.27 & 138.01 & M\\         
	NGC4270 & S0 & 0.392 & 1.269 & 72.8 & -19.702 & 2414.3 & 18.0 & 0.83 & 43.63 & E\\      
	NGC4318 & E & 0.167 & 0.869 & 79.12 & -17.285 & 1297.6 & 18.02 & 1.47 & 70 & E\\
	NGC4342 & E-S0 & 0.297 & 1.101 & 90.0 & -16.954 & 810.7 & 11.26 & 2.64 & 7.3 & L\footnotemark[2]\\
	NGC4352 & S0 & 0.331 & 1.231 & 90.0 & -19.056 & 2164.4 & 18.0 & - & 17.45 & E\\
	NGC4458 & E & 0.03 & 1.204 & 28.77 & -17.437 & 770.0 & 10.69 & 3.21 & 11.0 & L\footnotemark[3]\\      
	NGC4473 & E & 0.235 & 1.630 & 90.0 & -21.707 & 2338.1 & 18.0 & 2.17 & 183.26 & E\\       
	NGC4474 & S0 & 0.174 & 1.368 & 57.3 & -19.652 & 1707.6 & 18.0 & 3.8 & 143.5 & M\\   
	NGC4478 & E & 0.089 & 1.243 & 50.85 & -19.602 & 1479.7 & 20.55 & 3.92 & 40.5 & L\footnotemark[3]\\      
	NGC4483 & S0-a & 0.282 & 1.265 & 71.34 & -17.632 & 958.3 & 13.31 & 3.83 & 13 & E\\
	NGC4515 & E-S0 & 0.099 & 1.123 & 45.77 & -17.775 & 1059.8 & 14.72 & - & 36 & E\\
	NGC4528 & S0 & 0.271 & 1.208 & 73.43 & -18.874 & 1453.4 & 20.19 & - & 10 & E\\
	NGC4546 & E-S0 & 0.254 & 1.510 & 75.96 & -19.726 & 1064.1 & 14.78 & 0.27 & 43 & E\\
	NGC4570 & S0 & 0.618 & 1.597 & 90.0 & -20.444 & 1813.0 & 18.0 & 2.66 & 23.5 & L\footnotemark[2]\\      
	NGC4621 & E &  0.152 & 1.658 & 71.93 & -20.789 & 527.1 & 18.0 & 2.6 & 73.87 & M\\       
	NGC4623 & S0-a &  0.477 & 1.352 & 90.0 & -19.005 & 1863.7 & 18.0 & 2.36 & 210 & L\footnotemark[4]\\      
	NGC4660 & E & 0.124 & 1.323 & 61.24 & -19.241 & 1185.1 & 18.0 & 3.37 & 81.19 & M\\      
	NGC4742 & E & 0.191 & 1.358 & 90.0 & -19.376 & 1257.0 & 17.46 & 0.73 & 51 & E\\      
	NGC4762 & S0 & 0.37 & 1.917 & 90.0 & -19.921 & 1065.4 & 18.0 & 2.65 & 226.89 & E\\      
	NGC4866 & S0-a & 0.768 & 1.761 & 90.0 & -20.727 & 2095.8 & 18.0 & 1.08 & 26.18 & E\\      
	NGC5076 & S0-a & 0.138 & 1.126 & 48.83 & -19.491 & 2975.3 & 41.32 & - & 621 & E\\      
	NGC5252 & S0 & 0.233 & 1.111 & 67.18 & -21.039 & 6967.0 & 96.76 & - & - & E\\      
	NGC5308 & S0 & 0.877 & 1.642 & 90.0 & -20.439 & 2279.5 & 31.66 & 0.45 & 201.59 & M\\     
    	\\
	\multicolumn{10}{l}{$^1$ \citet{sb95}}\\
     	\multicolumn{10}{l}{$^2$ \citet{scorzaet98}}\\
     	\multicolumn{10}{l}{$^3$ \citet{morelliet04}}\\
     	\multicolumn{10}{l}{$^4$ \citet{vdb94}}\\
	\end{tabular}
\end{table*}

\begin{table*}
\small
    \caption[]{Continuation of properties table.}
    \centering
    \renewcommand{\footnoterule}{}  
    \begin{tabular}{ccccccccccc}
      \hline \hline
      \textbf{Name} & \textbf{Type} & \textbf{logr$_{25}$} & \textbf{logd$_{25}$} & \textbf{inclination} & \textbf{mabs} & \textbf{V\_vir} & \textbf{distance} & \textbf{Density} & \textbf{Disk size} & \textbf{Source}\\
	  &  &  &  & \textbf{($^o$)} &  & \textbf{(kms$^{-1}$)} & \textbf{(Mpc)} & \textbf{(Mpc$^{-3}$)} & \textbf{(pc)} & \\
	(1) & (2) & (3) & (4) & (5) & (6) & (7) & (8) & (9) & (10) & (11)\\
      \hline
	NGC5854 & S0-a & 0.58 & 1.483 & 90.0 & -19.698 & 1833.0 & 25.46 & 0.74 & 148 & E\\
	NGC7173 & E & 0.088 & 1.275 & 47.16 & -19.897 & 2368.8 & 32.90 & 0.35 & 48 & E\\
	NGC7176 & E & 0.024 & 1.569 & 25.29 & -20.423 & 2387.5 & 33.16 & 0.39 & 64 & E\\
	NGC7562 & E & 0.137 & 1.338 & 67.29 & -21.427 & 3571.8 & 49.61 & - & - & E\\      
	NGC7585 & S0-a & 0.123 & 1.406 & 46.52 & -21.291 & 3432.0 & 47.67 & - & 69 & E\\      
	NGC7619 & E & 0.106 & 1.407 & 56.04 & -21.987 & 3798.5 & 52.76 & - & 256 & E\\      
	NGC7785 & E & 0.283 & 1.447 & 90.0 & -21.411 & 3875.7 & 53.83 & - & 497 & E\\
	PGC013343 & E-S0 & 0.039 & 1.030 & 28.55 & -16.909 & 1486.7 & 20.65 & - & 100 & E\\
	PGC036465 & S0-a & 0.307 & 0.825 & 67.59 & -19.656 & 5736.2 & 79.67 & - & 232 & E\\
	PGC044815 & S0 & 0.35 & 0.7 & 90.0 & -18.58 & 6784 & 94.22 & - & 1005 & E\\
	UGC01003 & S0 & 0.346 & 0.922 & 90.0 & -19.488 & 5192.1 & 72.11 & - & 105 & E\\
	UGC03426 & S0 & 0.096 & 1.224 & 42.78 & -20.869 & 4292.5 & 59.62 & - & 58 & E\\
    \end{tabular}
\end{table*}

\clearpage

\subsection[]{NSD Structure Maps, Ellipticity and $A_4$ profiles}

\begin{figure*}
\begin{center}
\label{fig:ESO}
\includegraphics[width=2.2in]{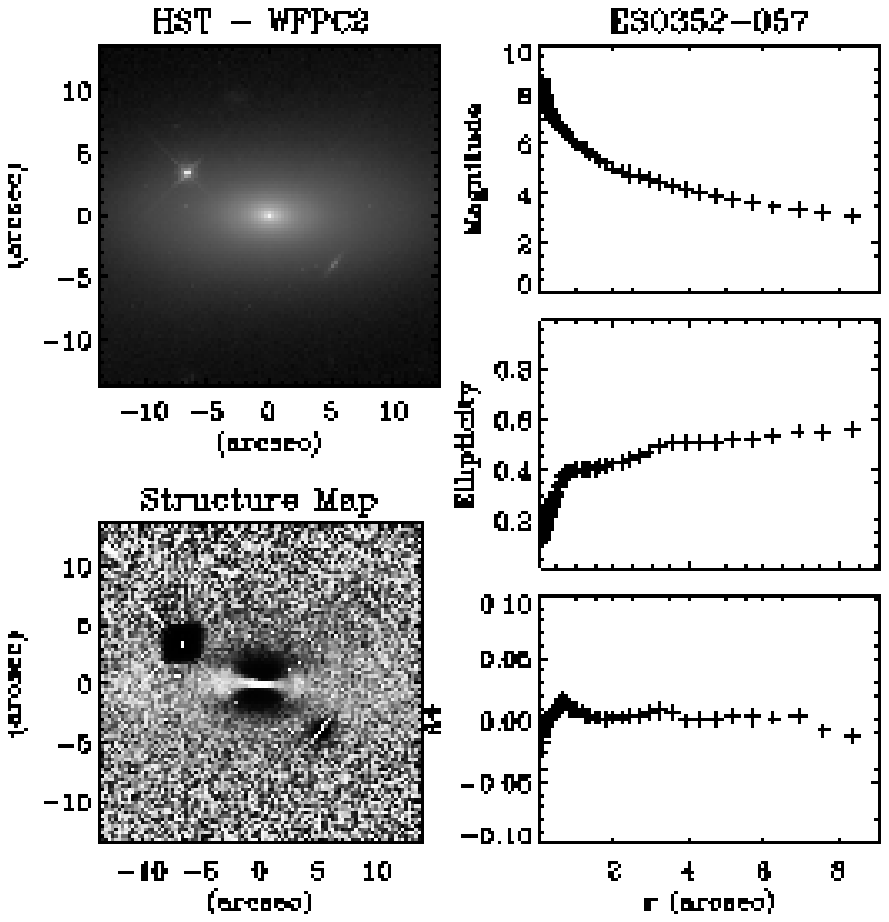}
\includegraphics[width=2.2in]{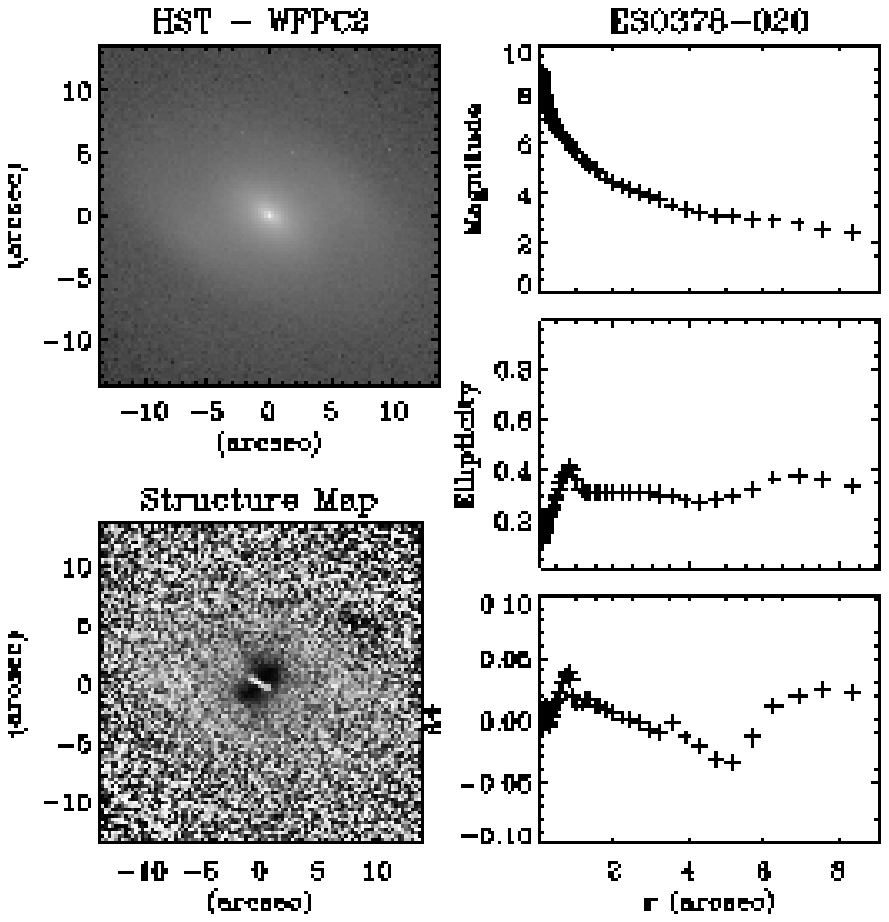}
\includegraphics[width=2.2in]{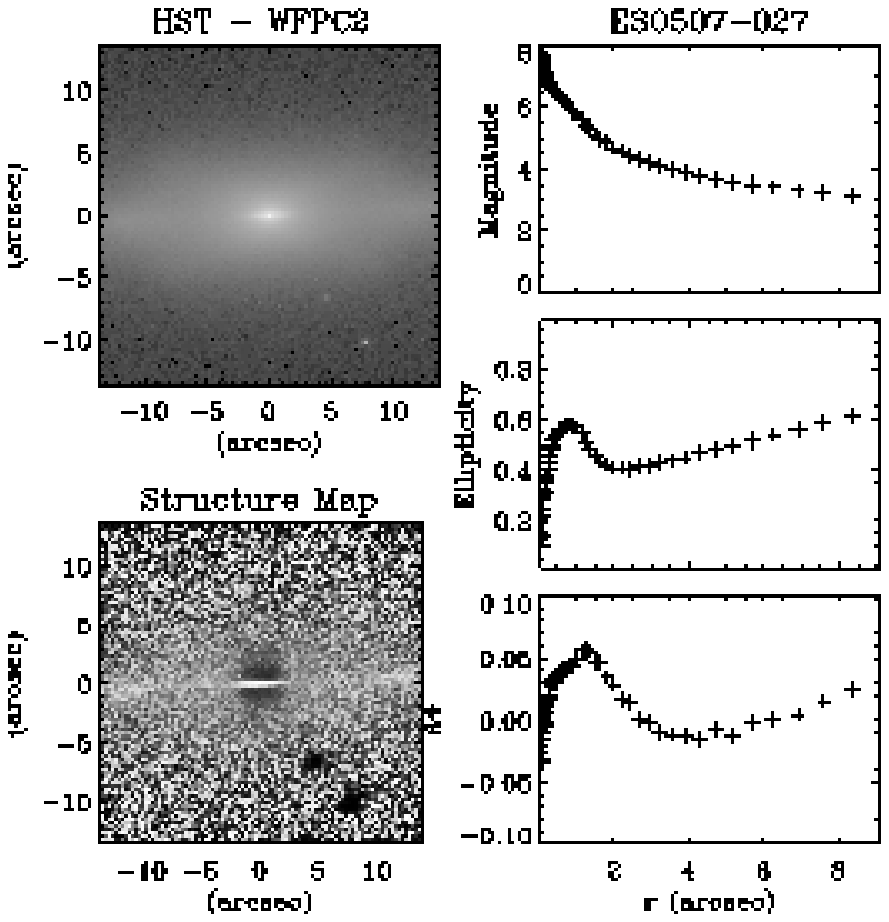}
\caption{ESO352-057 on the left panel, in the middle ESO378-020 and ESO507-027 at the right. The original images and the structure maps are shown on the left of each panel and on the right we can see the surface brightness, ellipticity and A$_4$ profiles.}
\end{center}
\end{figure*}

\begin{figure*}
\begin{center}
\includegraphics[width=2.2in]{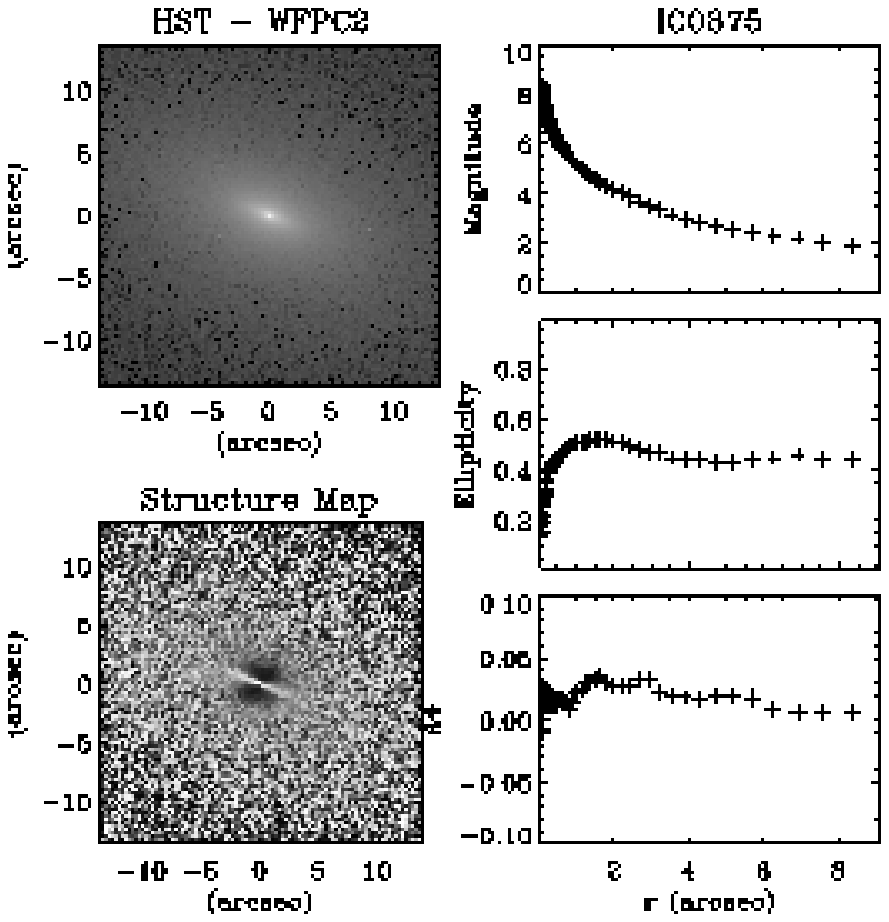}
\includegraphics[width=2.2in]{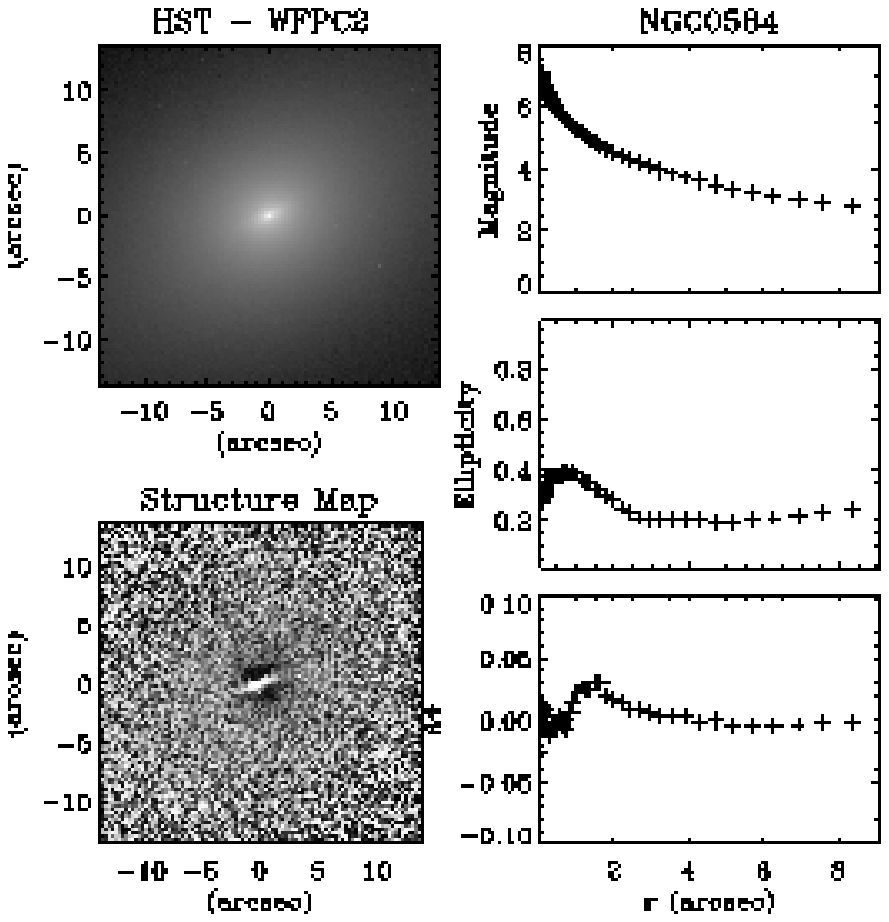}
\includegraphics[width=2.2in]{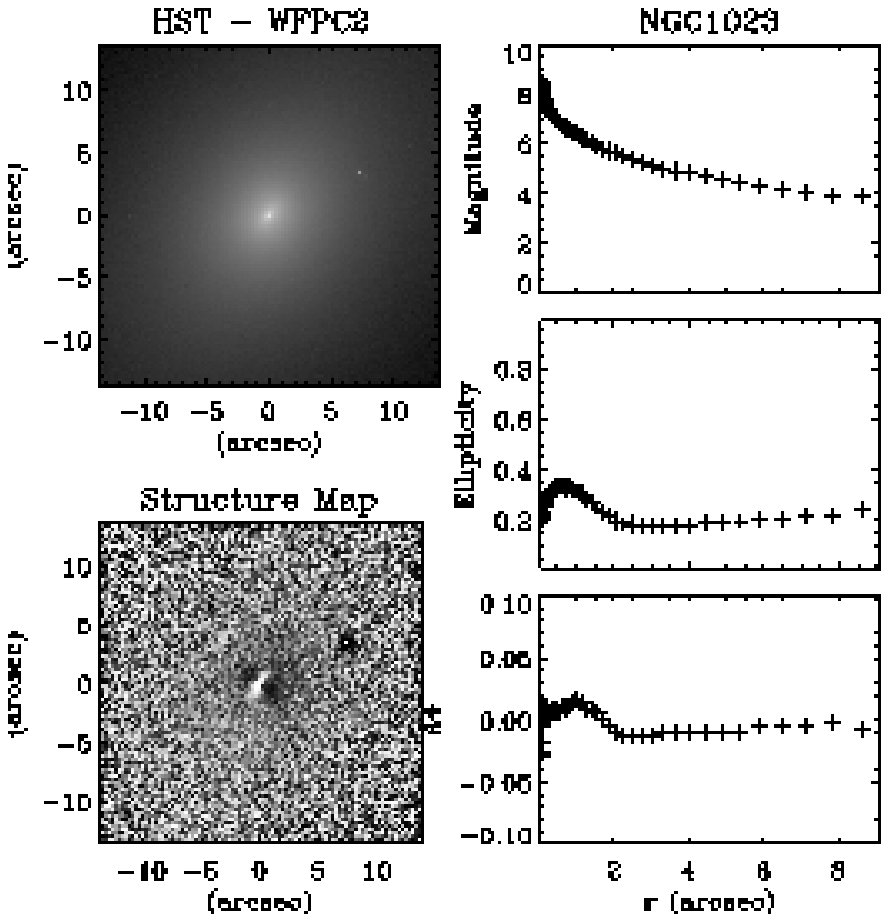}
\caption{IC0875 on the left, NGC0584 in the middle and NGC1023 on the right.}
\end{center}
\end{figure*}

\begin{figure*}
\begin{center}
\includegraphics[width=2.2in]{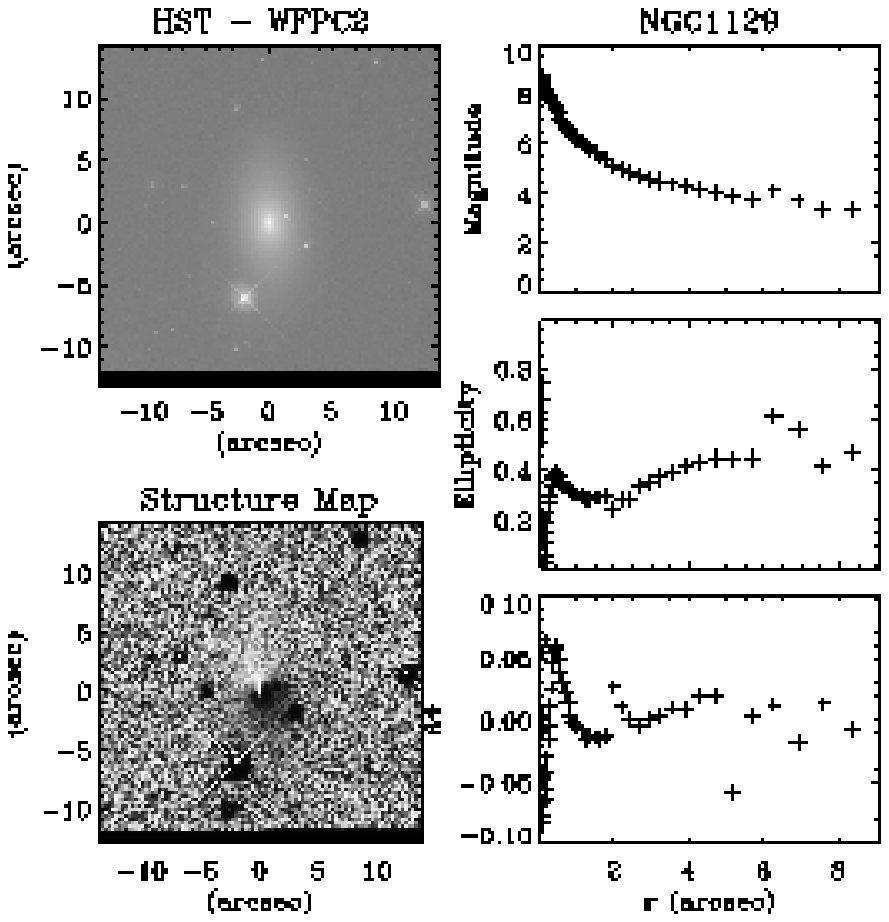}
\includegraphics[width=2.2in]{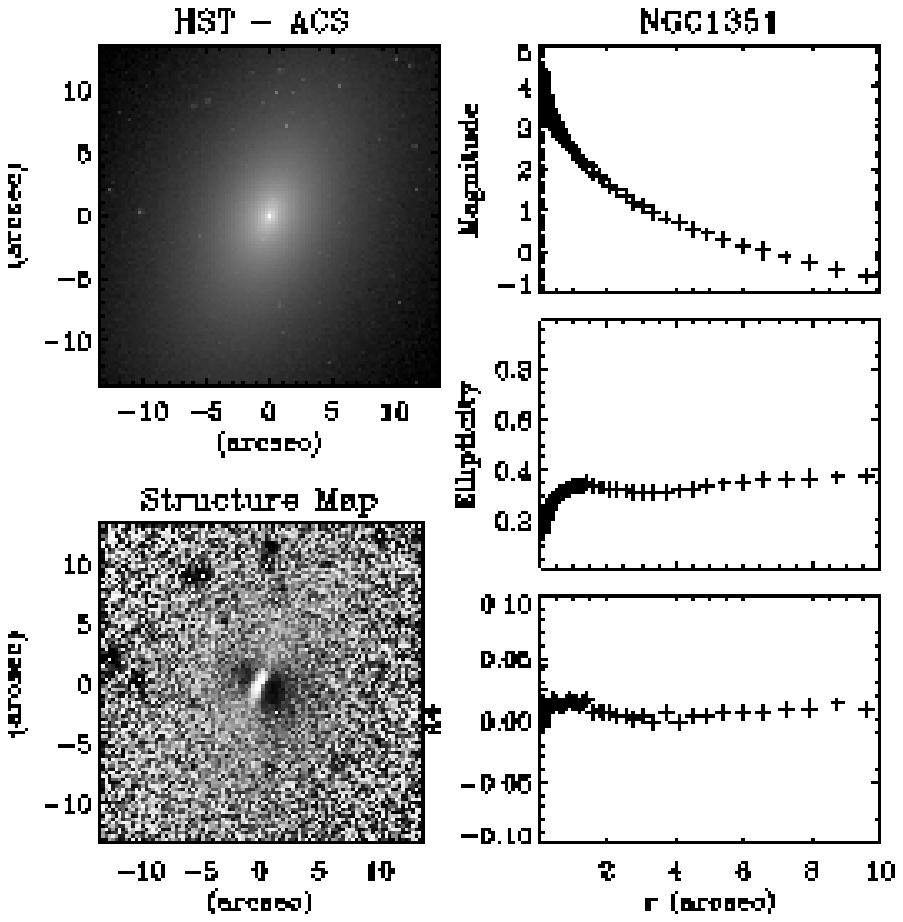}
\includegraphics[width=2.2in]{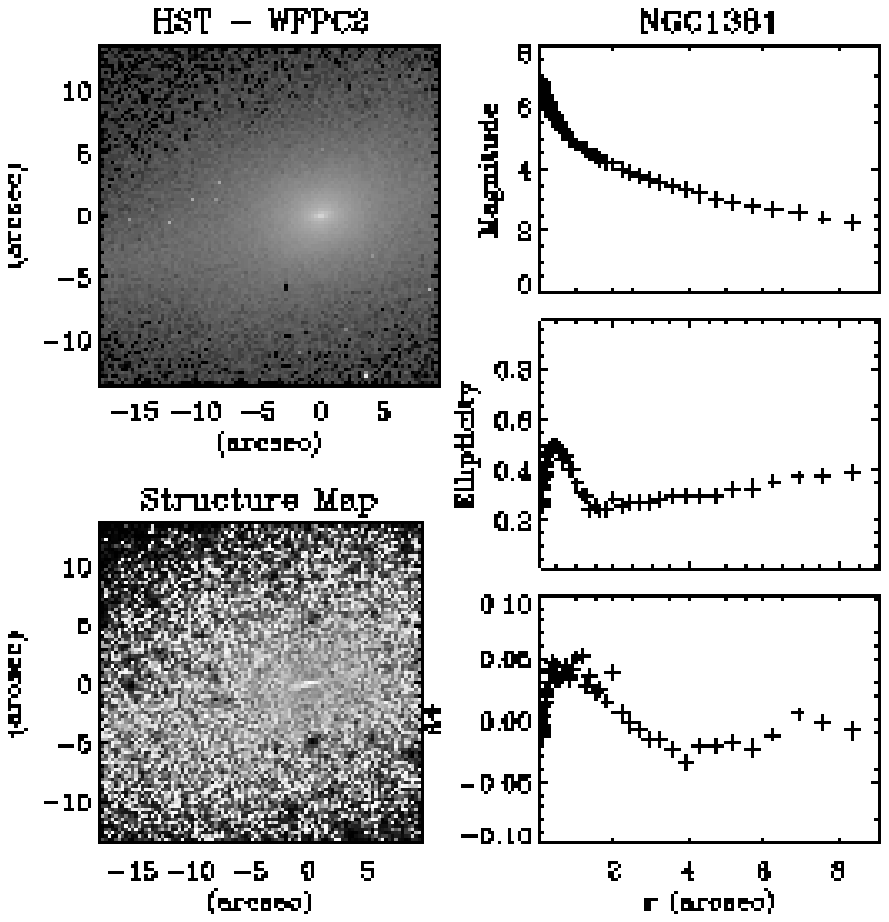}
\caption{NGC1129 on the left, NGC1351 in the middle and NGC1381 on the right.}
\end{center}
\end{figure*}

\clearpage

\begin{figure*}
\begin{center}
\includegraphics[width=2.2in]{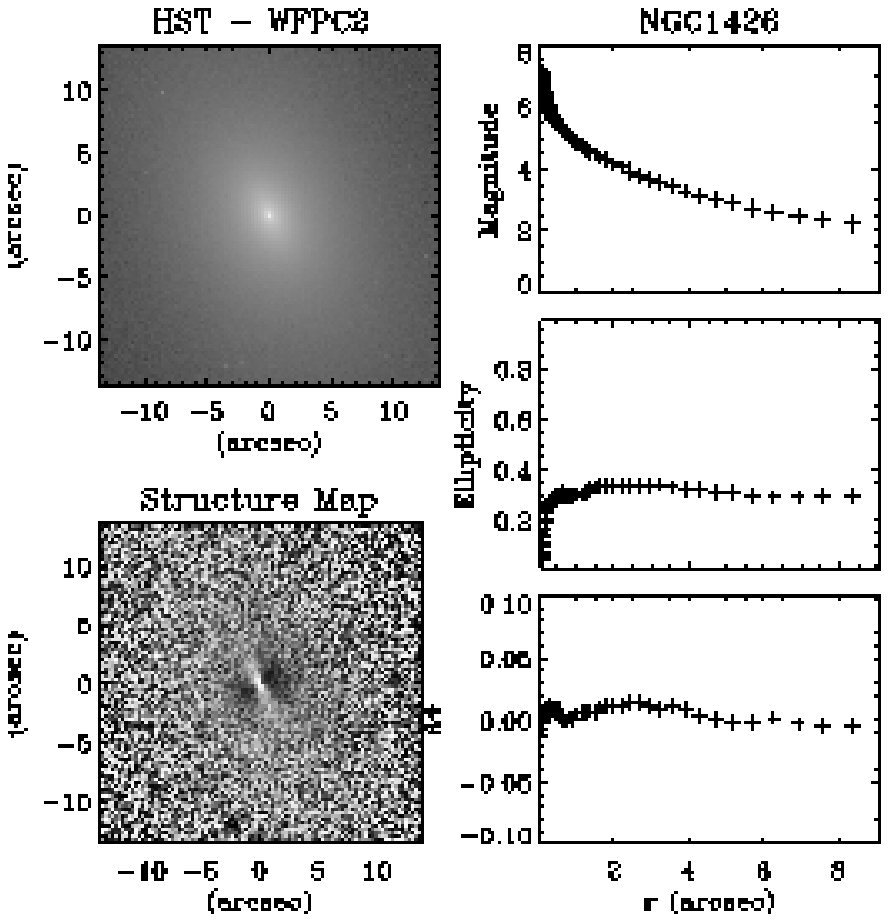}
\includegraphics[width=2.2in]{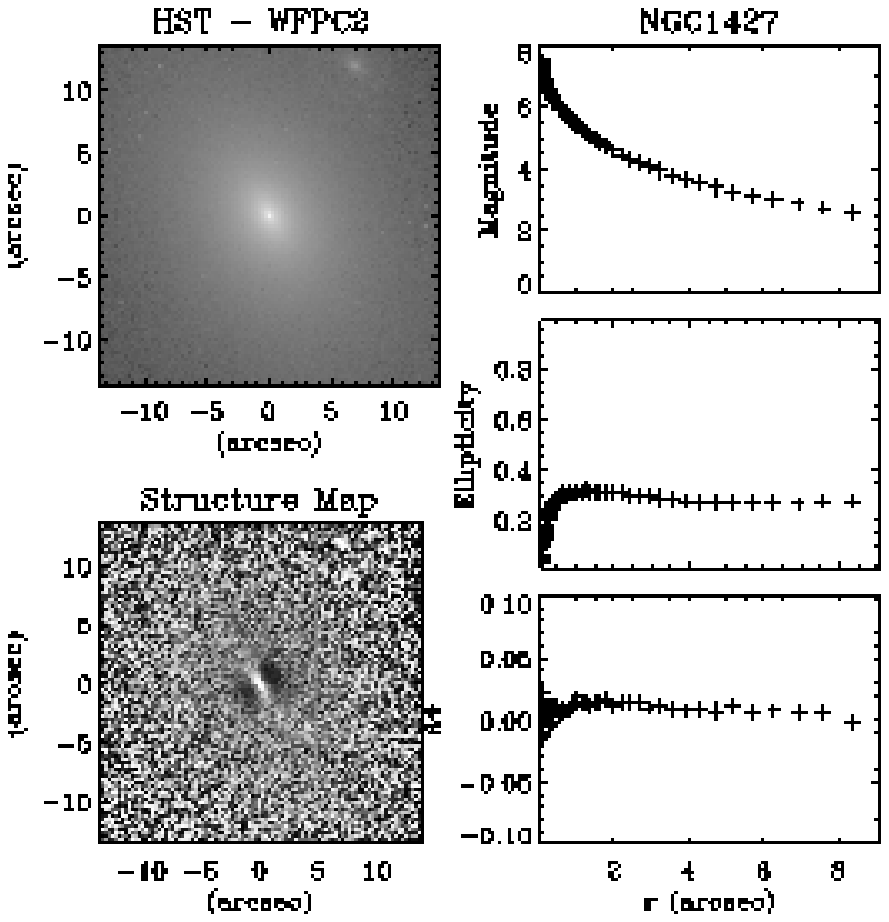}
\includegraphics[width=2.2in]{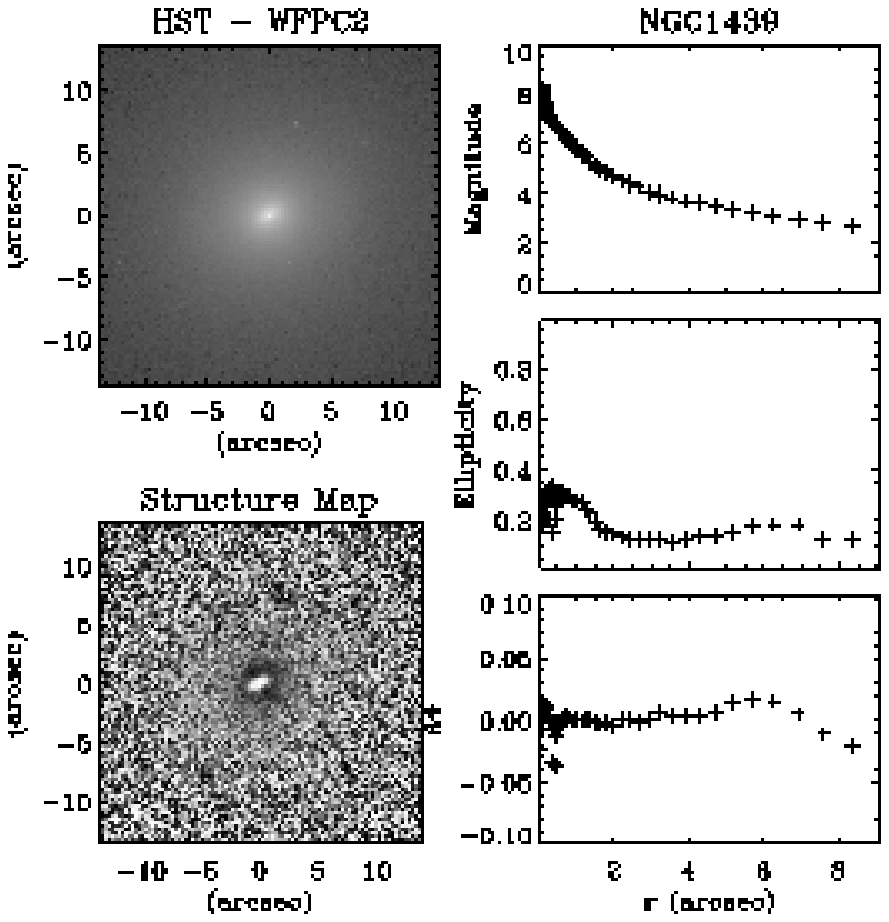}
\caption{NGC1426 on the left, NGC1427 in the middle and NGC1439 on the right.}
\end{center}
\end{figure*}

\begin{figure*}
\begin{center}
\includegraphics[width=2.2in]{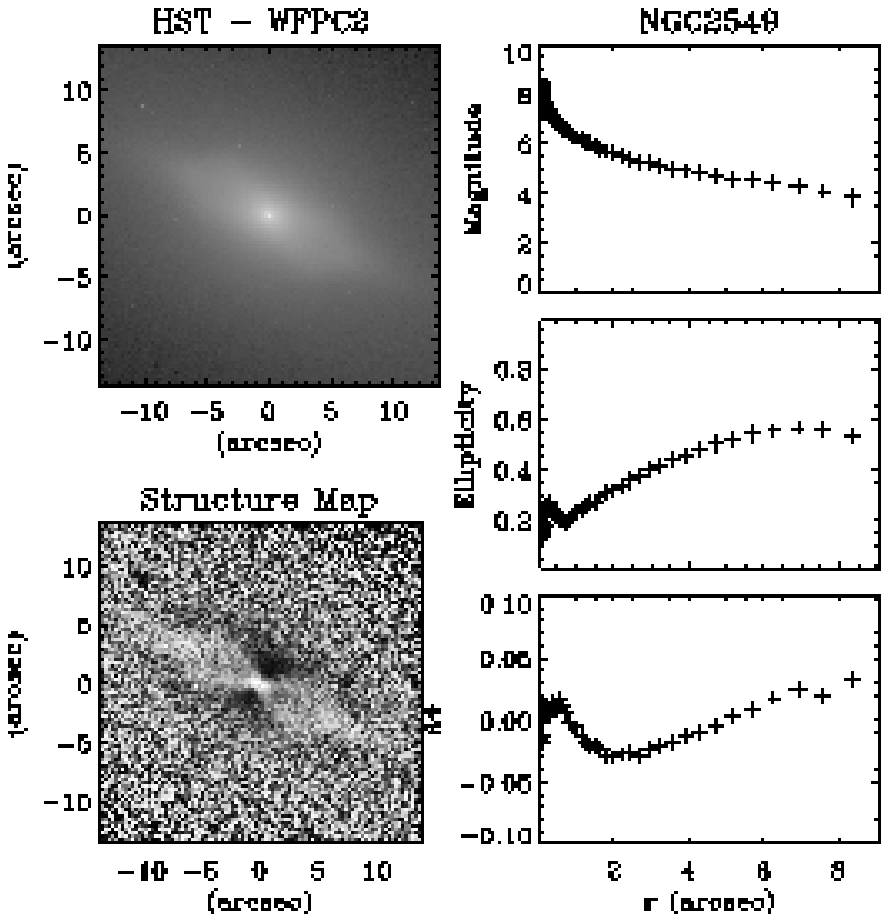}
\includegraphics[width=2.2in]{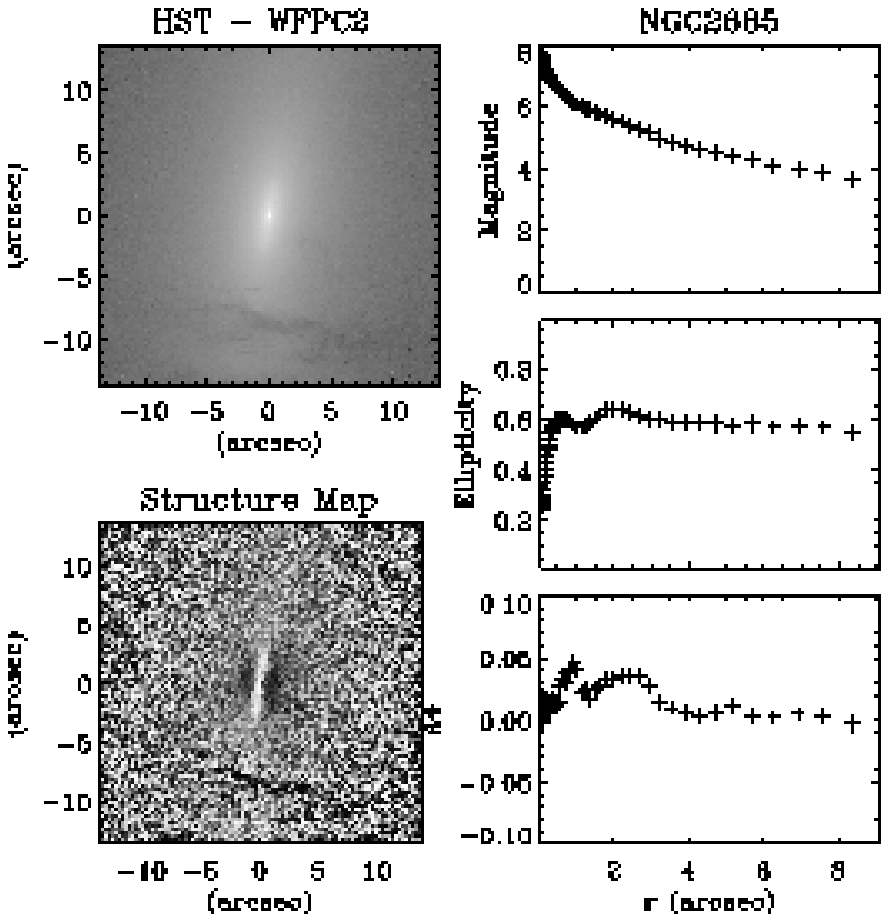}
\includegraphics[width=2.2in]{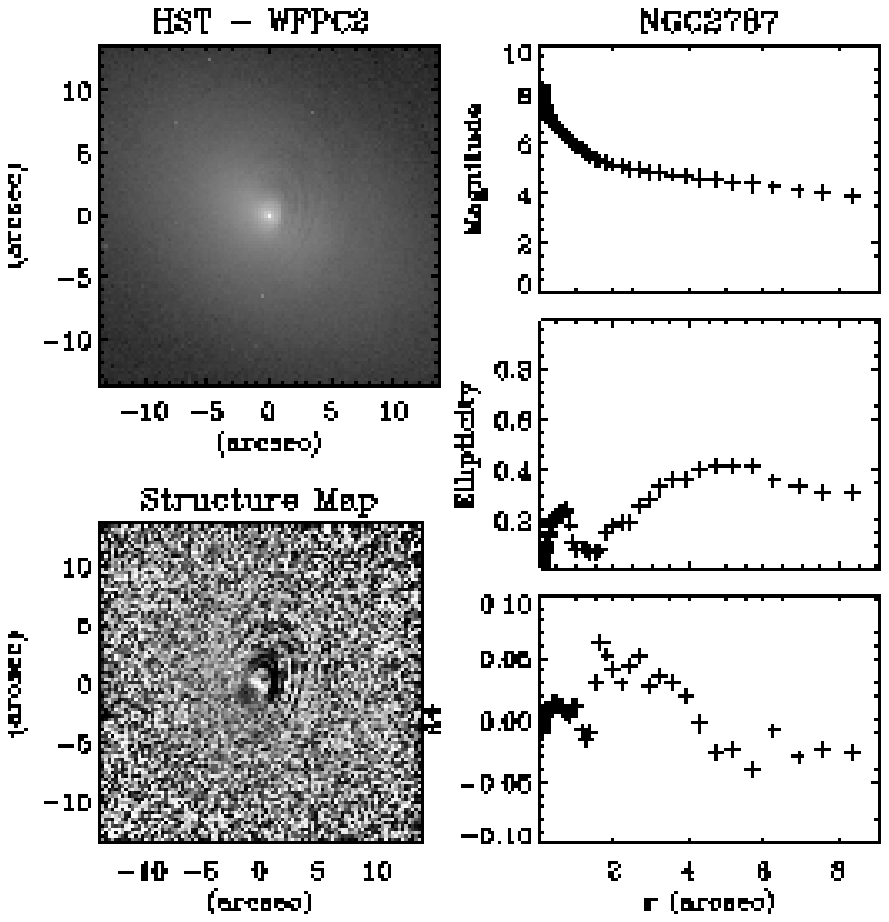}
\caption{NGC2549 on the left, NGC2685 in the middle and NGC2787 on the right.}
\end{center}
\end{figure*}

\begin{figure*}
\begin{center}
\includegraphics[width=2.2in]{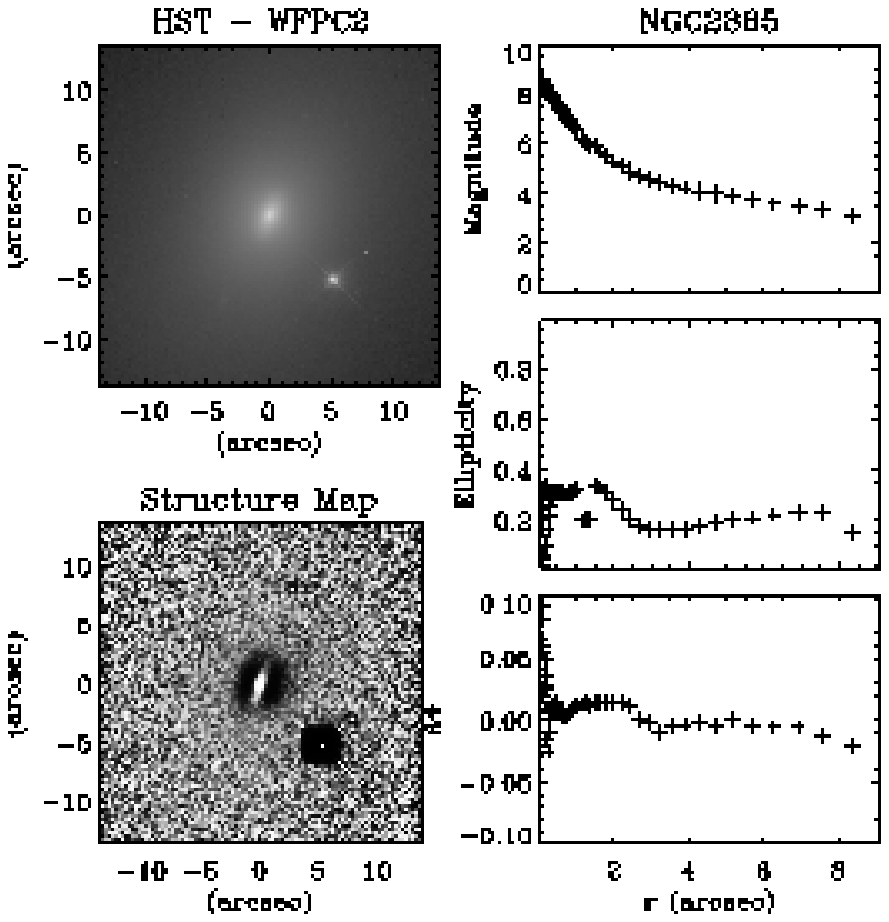}
\includegraphics[width=2.2in]{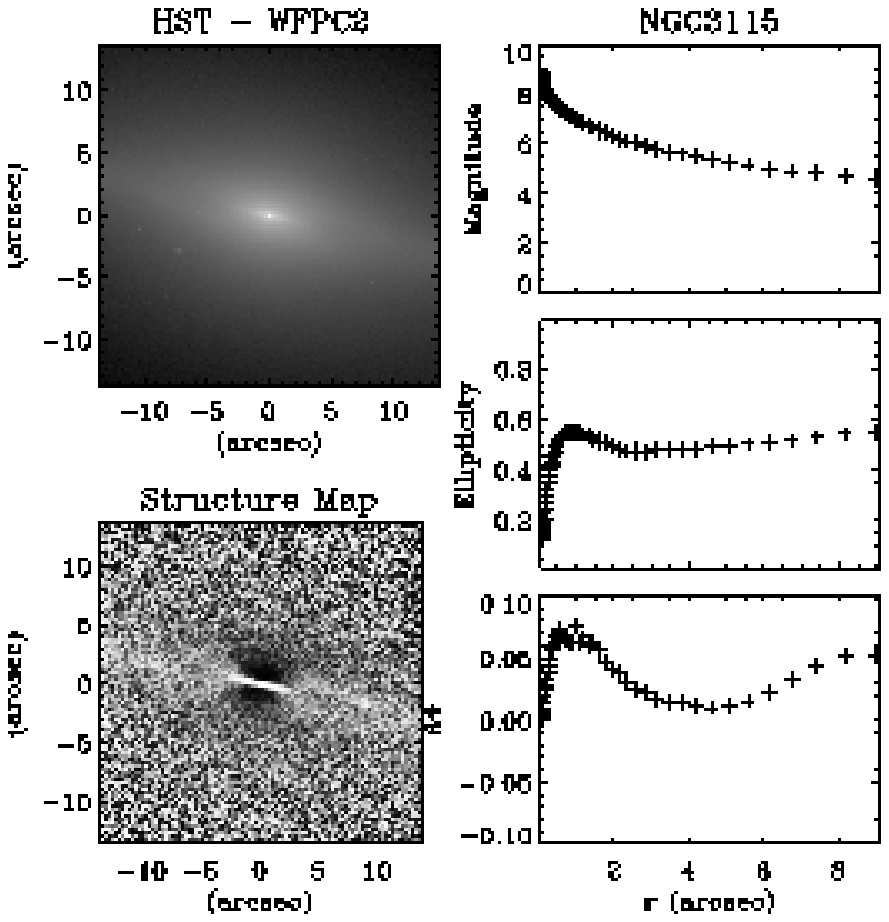}
\includegraphics[width=2.2in]{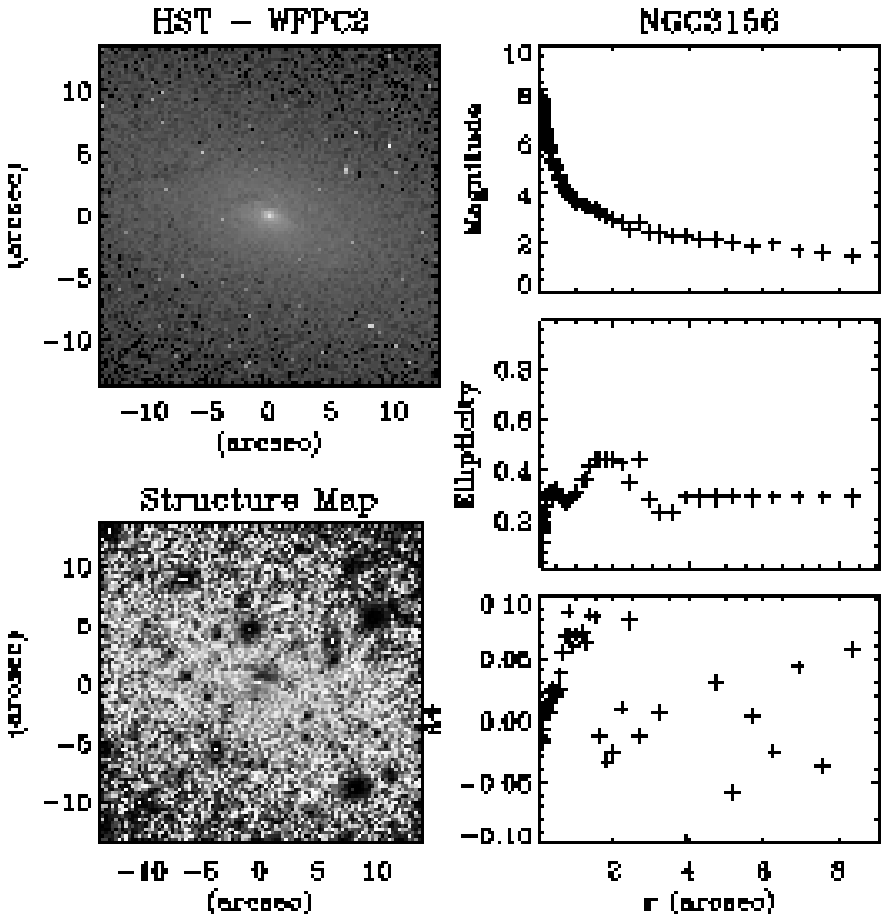}
\caption{NGC2865 on the left, NGC3115 in the middle and NGC3156 on the right.}
\end{center}
\end{figure*}

\clearpage

\begin{figure*}
\begin{center}
\includegraphics[width=2.2in]{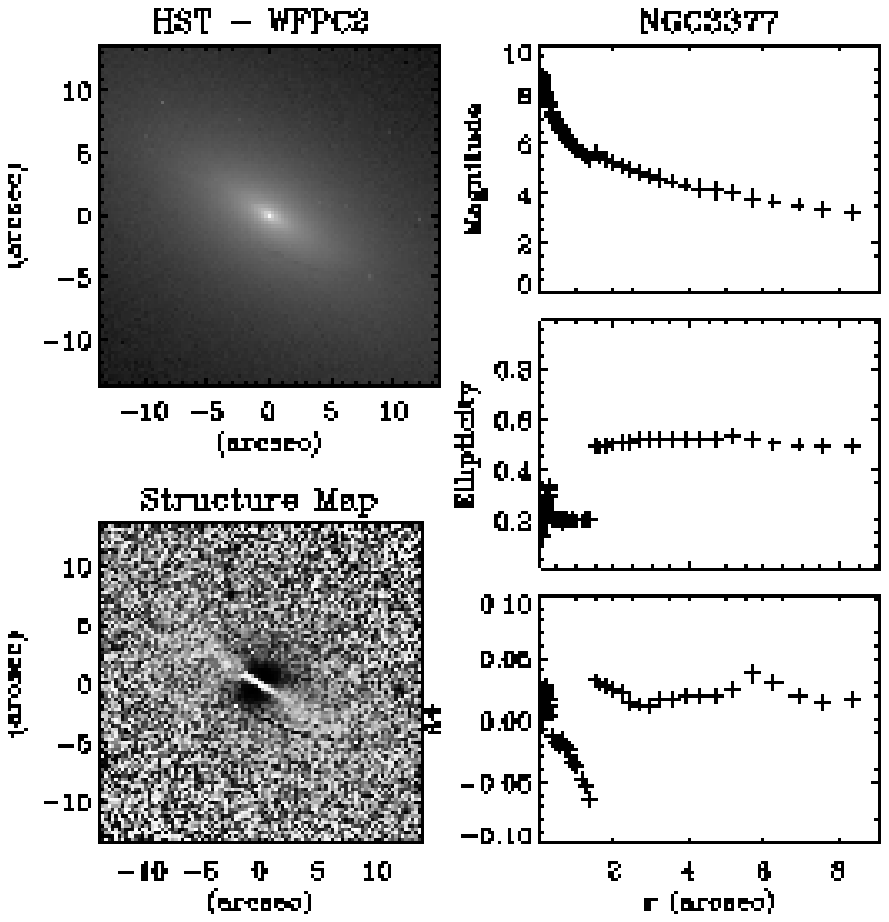}
\includegraphics[width=2.2in]{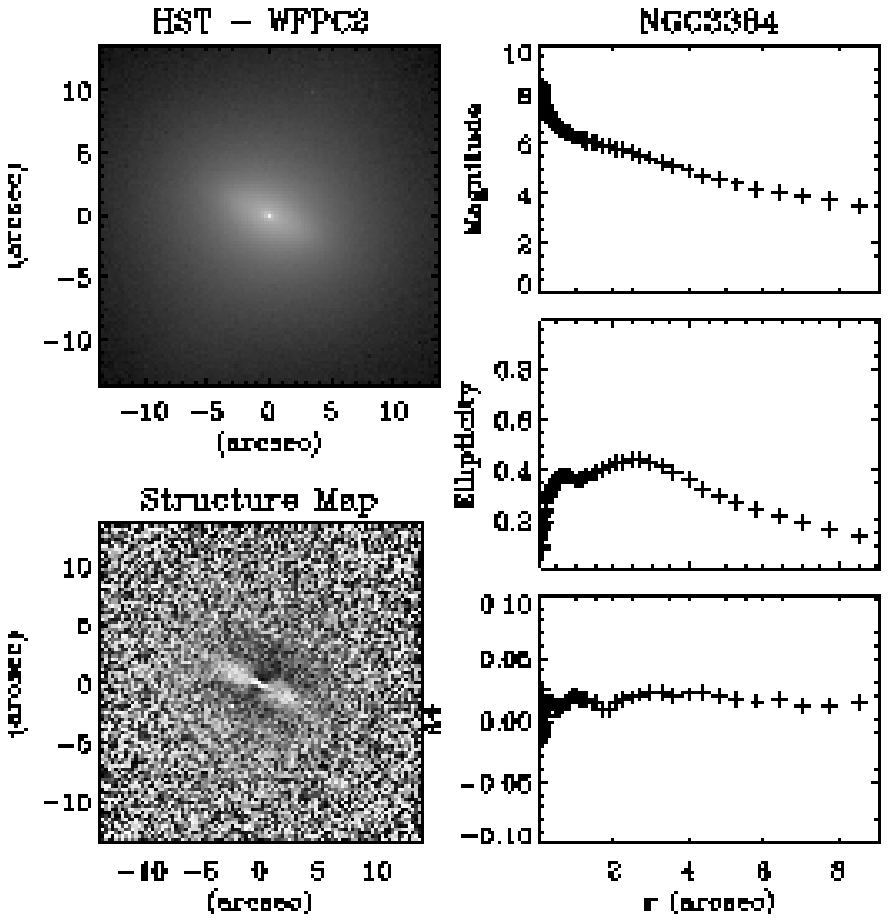}
\includegraphics[width=2.2in]{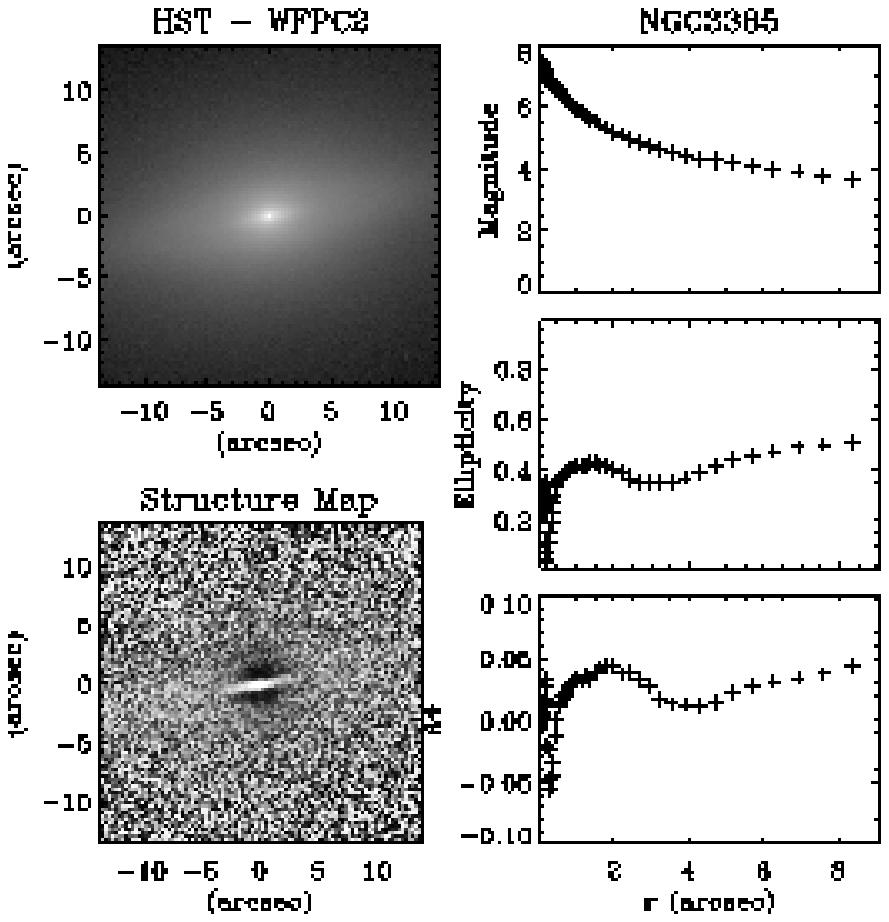}
\caption{NGC3377 on the left, NGC3384 in the middle and NGC3385 on the right.}
\end{center}
\end{figure*}

\begin{figure*}
\begin{center}
\includegraphics[width=2.2in]{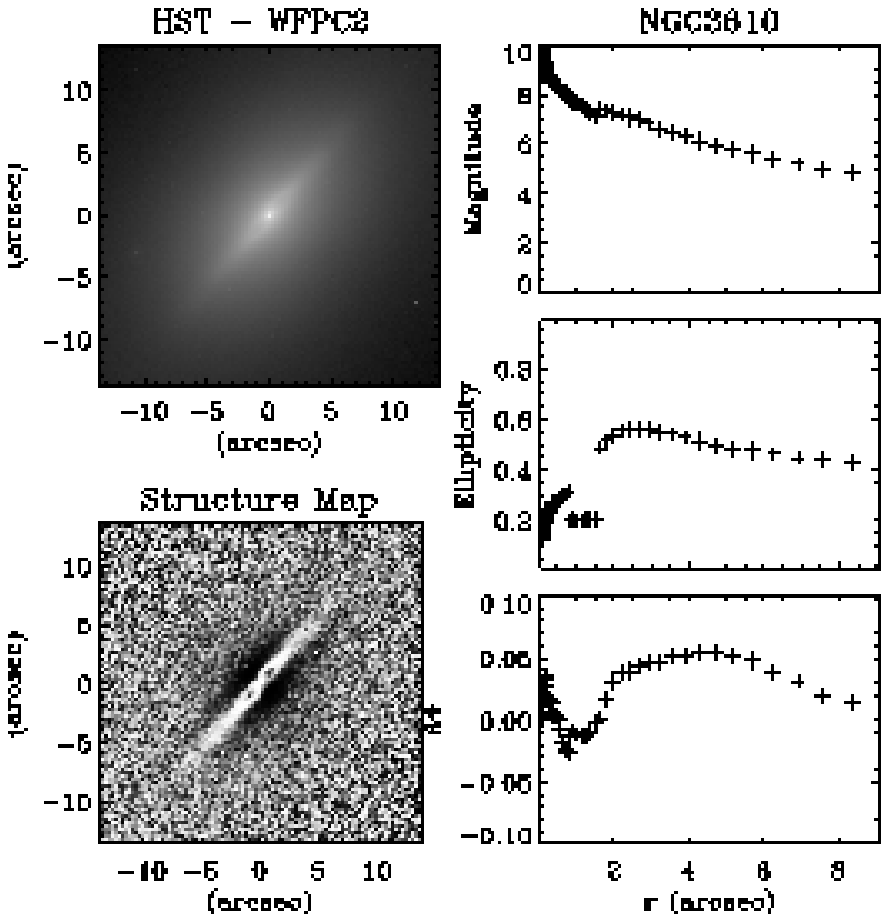}
\includegraphics[width=2.2in]{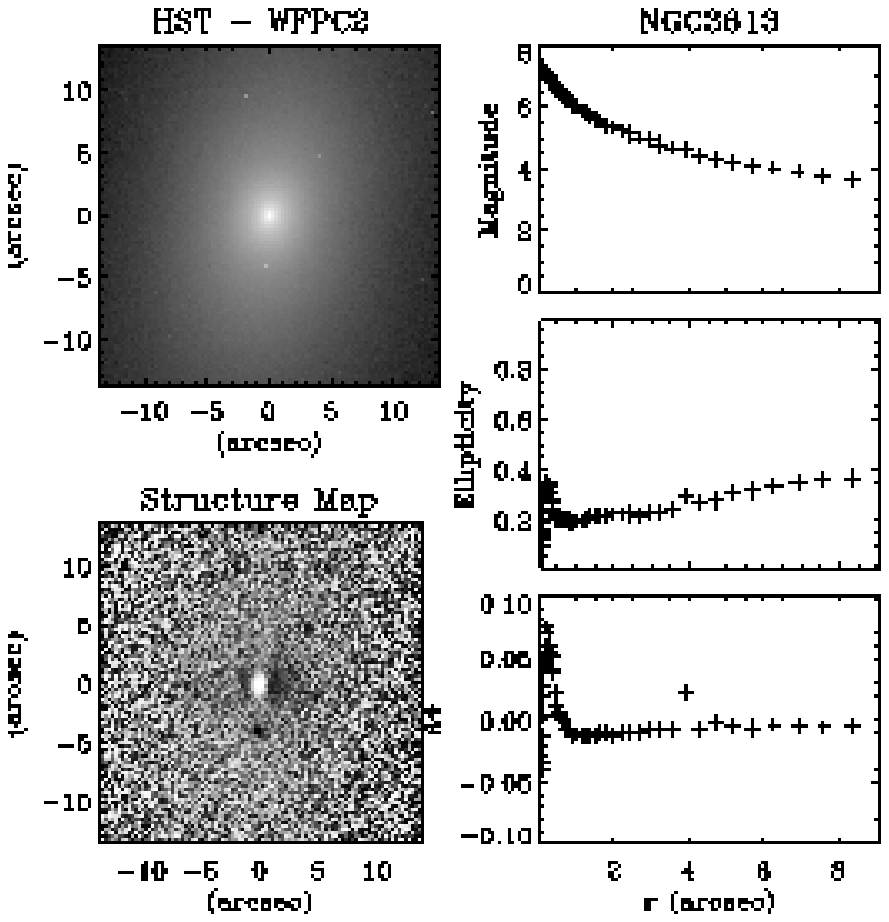}
\includegraphics[width=2.2in]{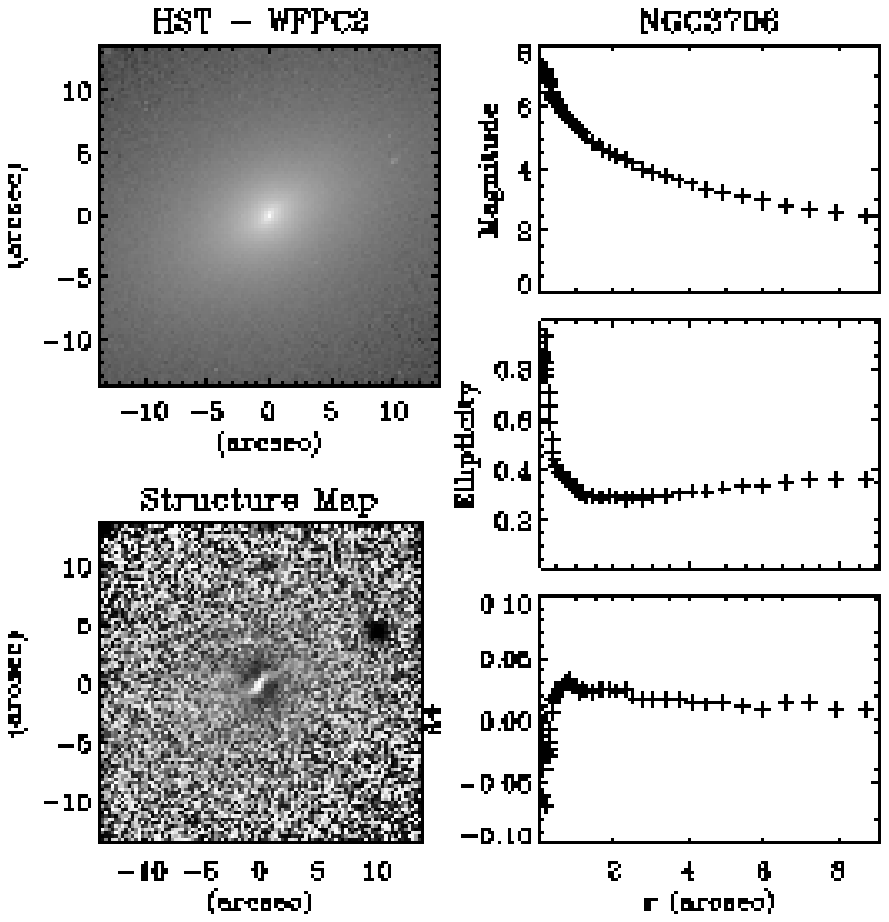}
\caption{NGC3610 on the left, NGC3613 in the middle and NGC3706 on the right.}
\end{center}
\end{figure*}

\begin{figure*}
\begin{center}
\includegraphics[width=2.2in]{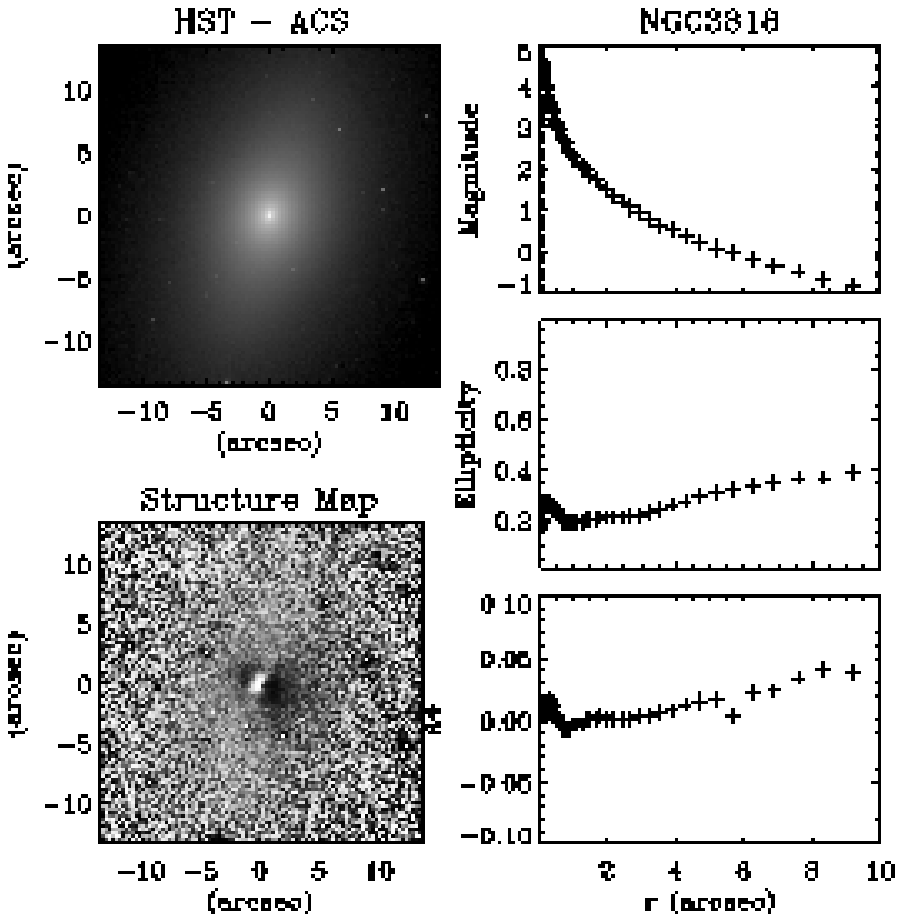}
\includegraphics[width=2.2in]{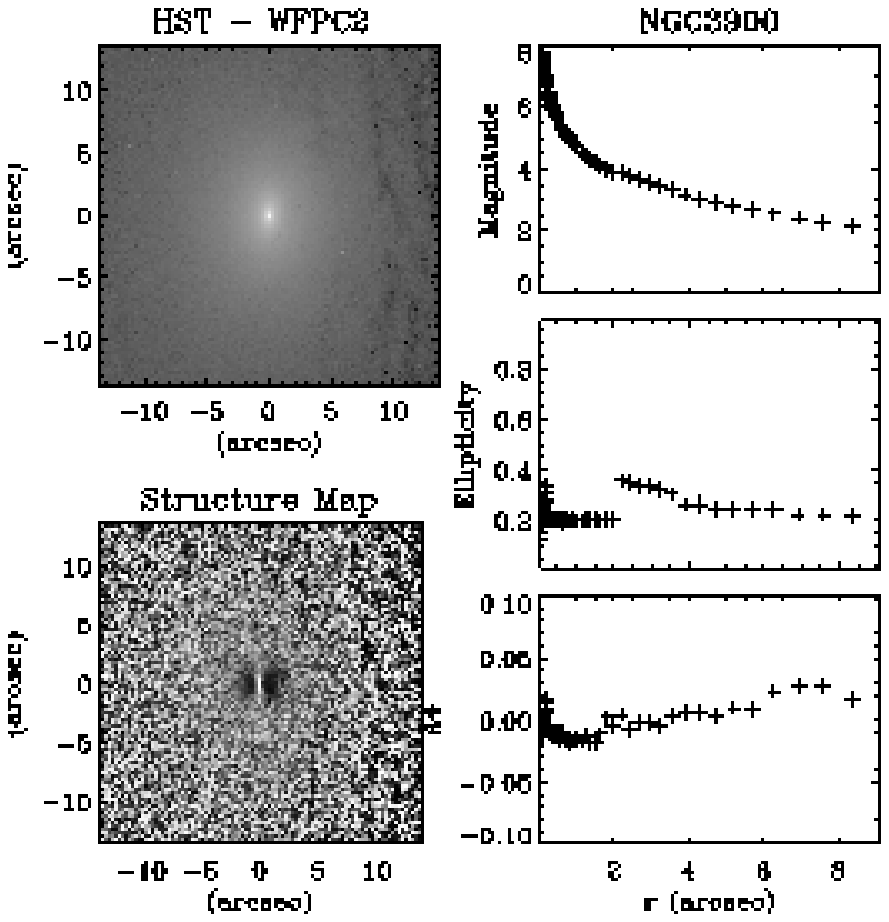}
\includegraphics[width=2.2in]{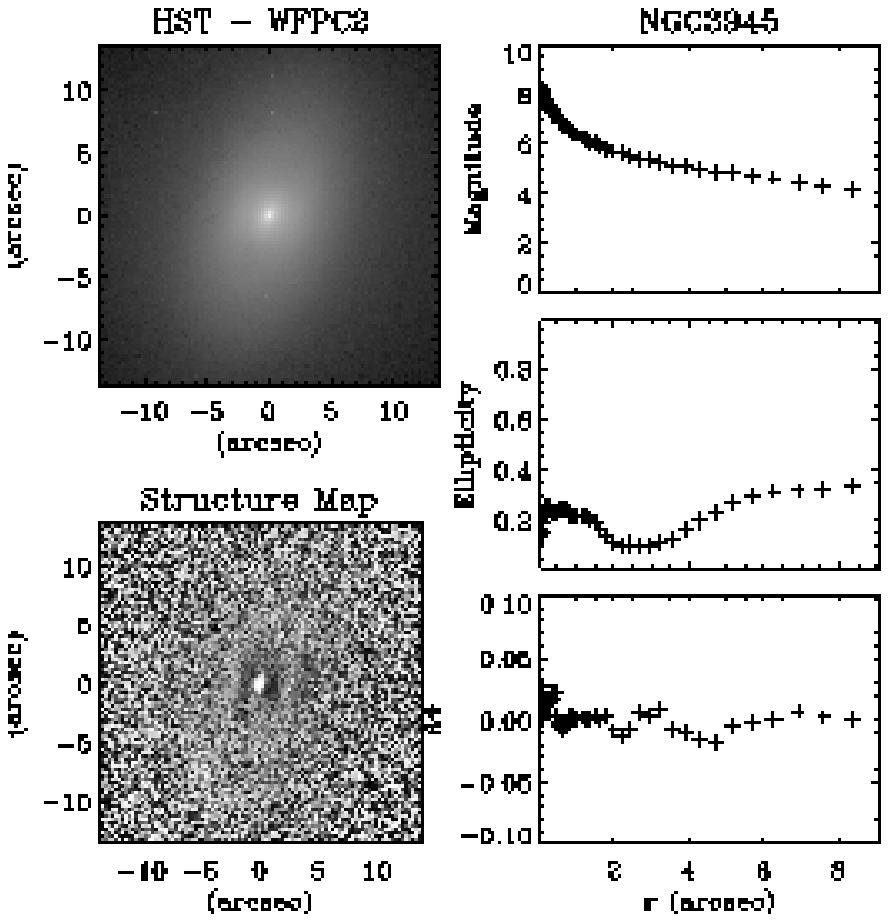}
\caption{NGC3818 on the left, NGC3900 in the middle and NGC3945 on the right.}
\end{center}
\end{figure*}

\begin{figure*}
\begin{center}
\includegraphics[width=2.2in]{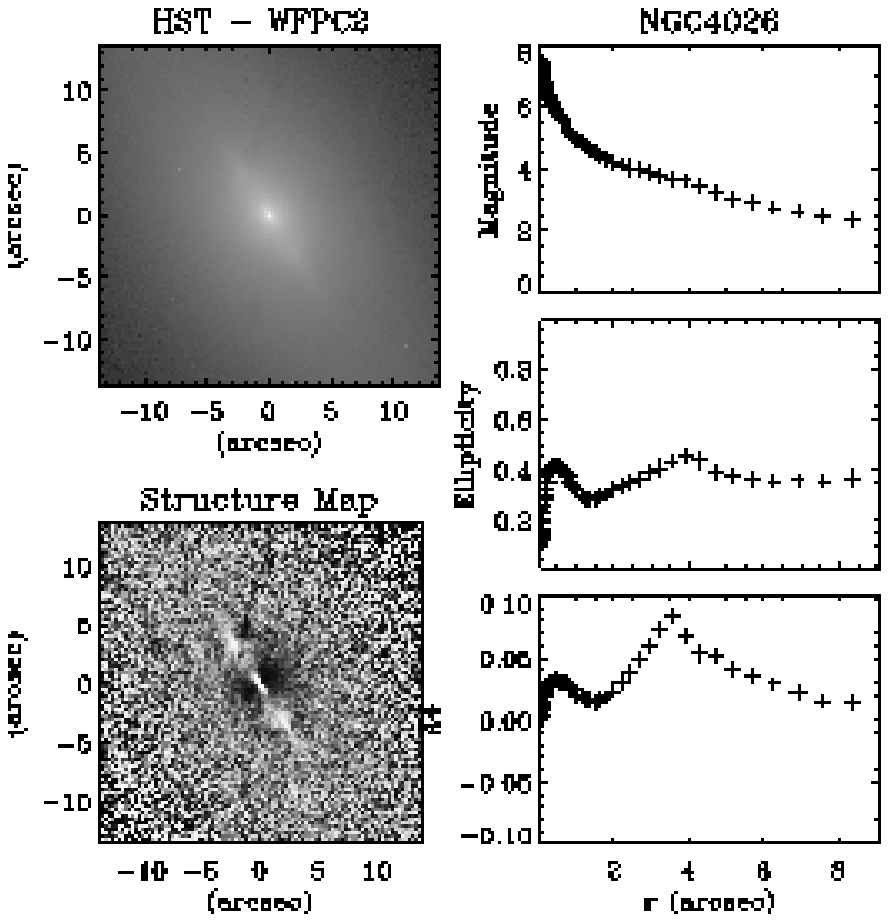}
\includegraphics[width=2.2in]{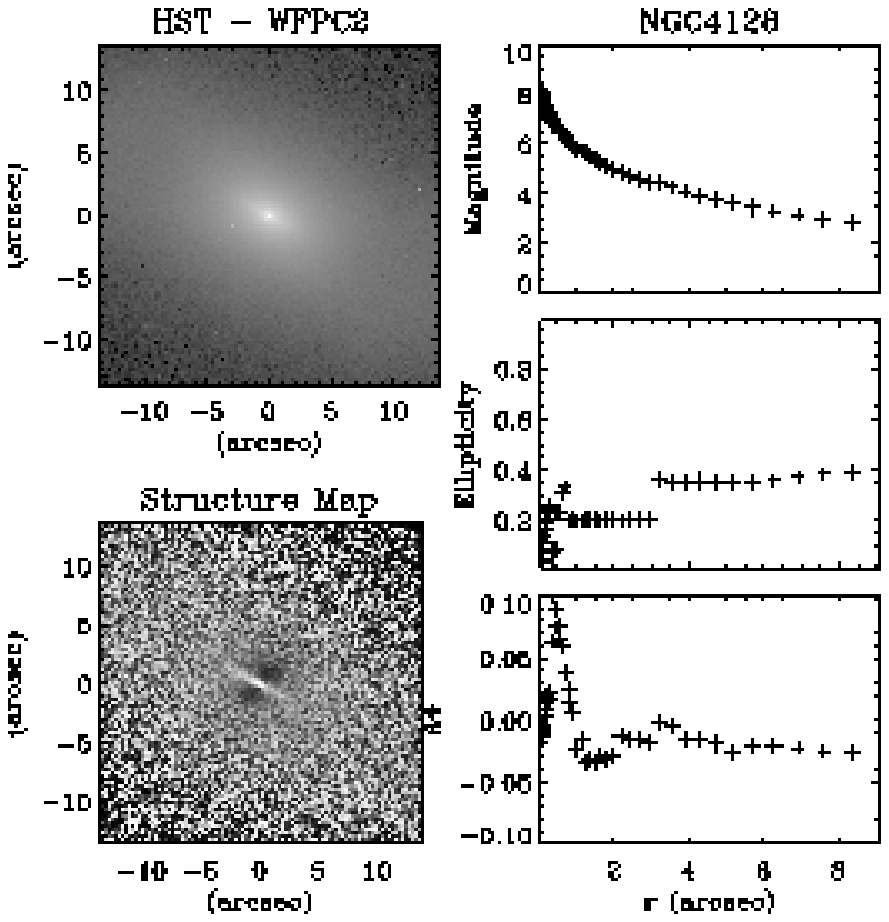}
\includegraphics[width=2.2in]{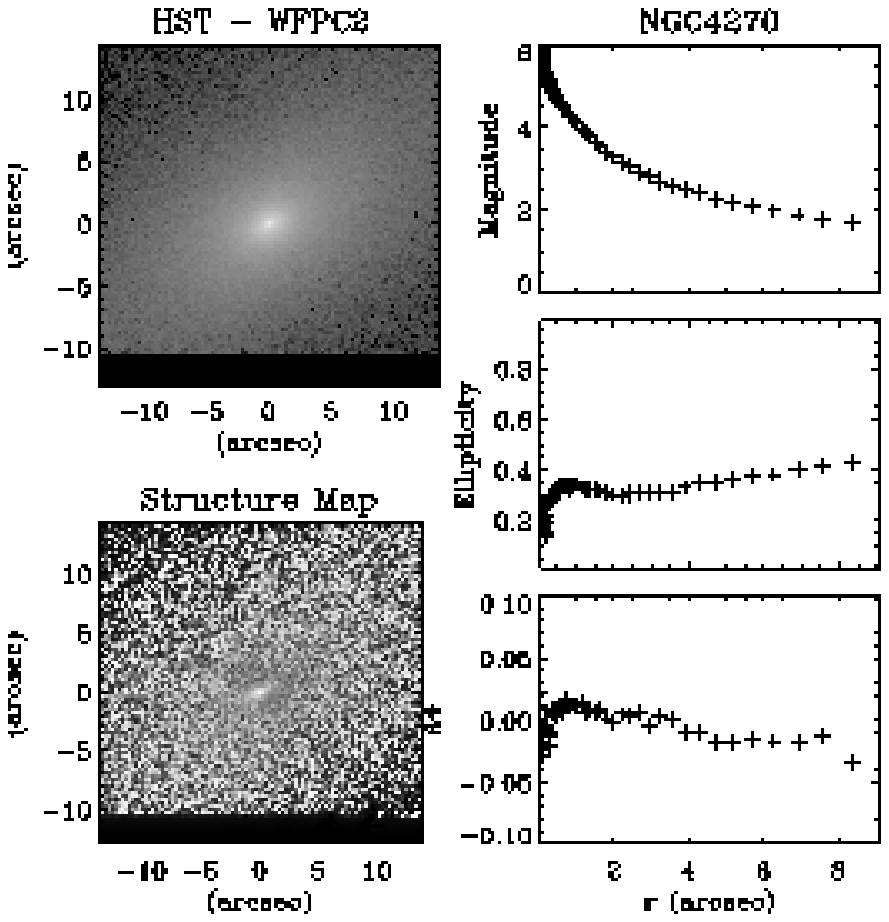}
\caption{NGC4026 on the left, NGC4128 in the middle and NGC4270 on the right.}
\end{center}
\end{figure*}

\begin{figure*}
\begin{center}
\includegraphics[width=2.2in]{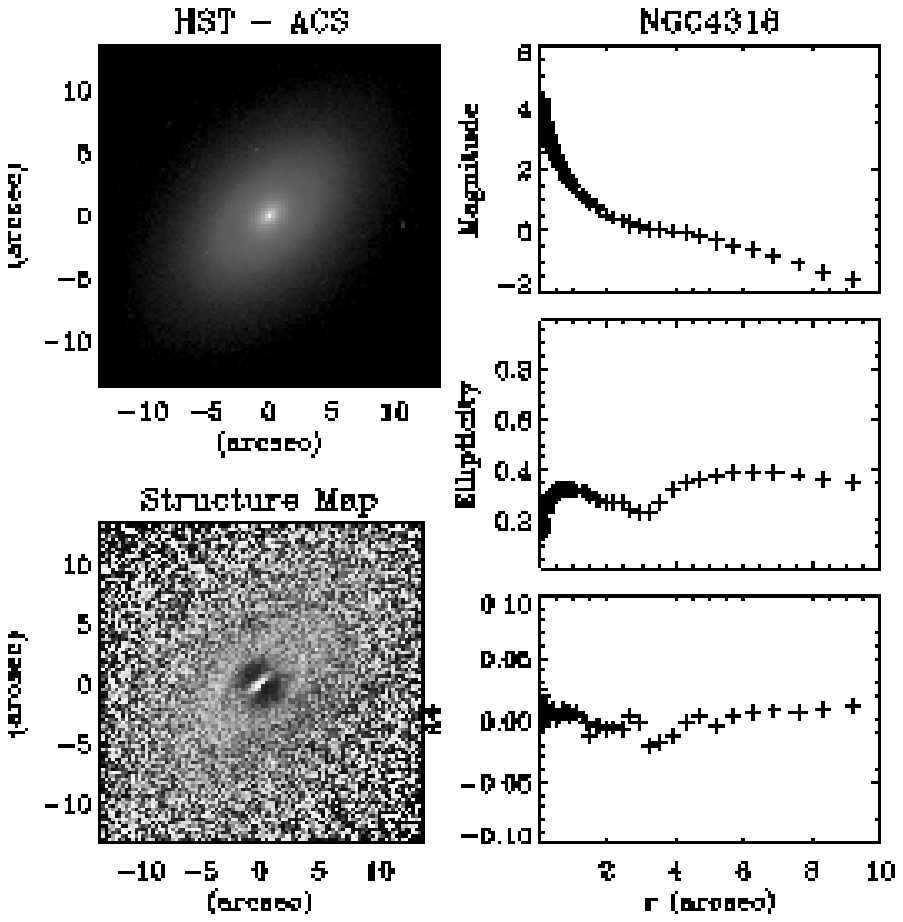}
\includegraphics[width=2.2in]{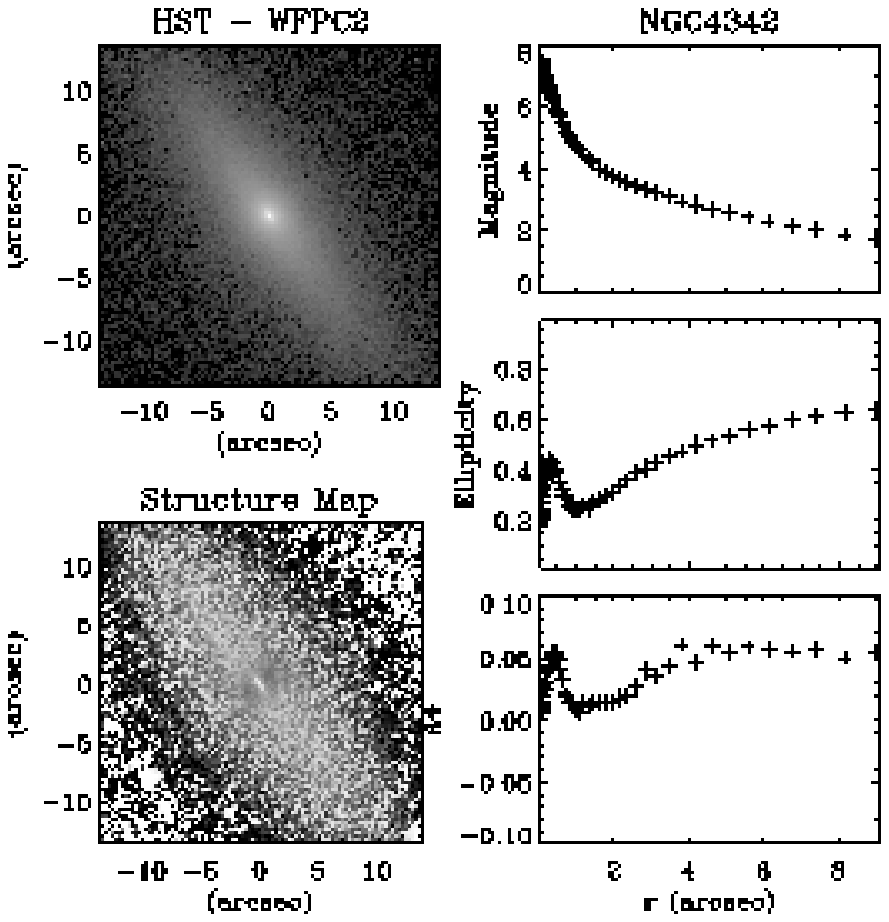}
\includegraphics[width=2.2in]{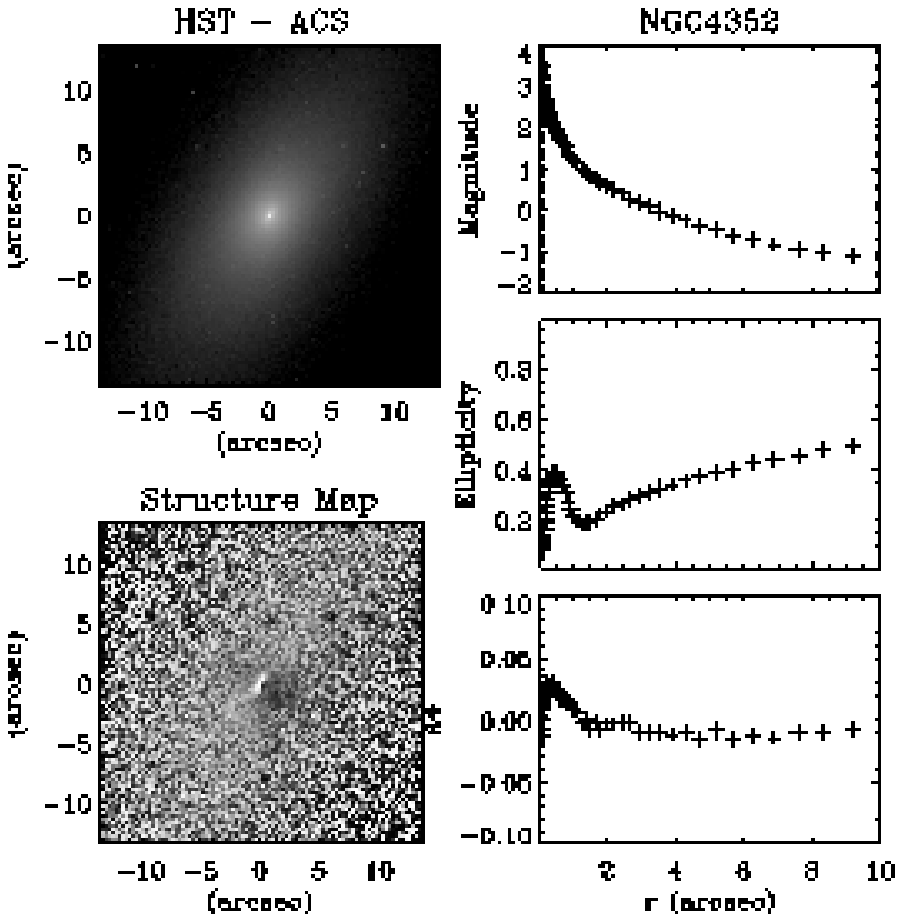}
\caption{NGC4318 on the left, NGC4342 in the middle and NGC4352 on the right.}
\end{center}
\end{figure*}

\begin{figure*}
\begin{center}
\includegraphics[width=2.2in]{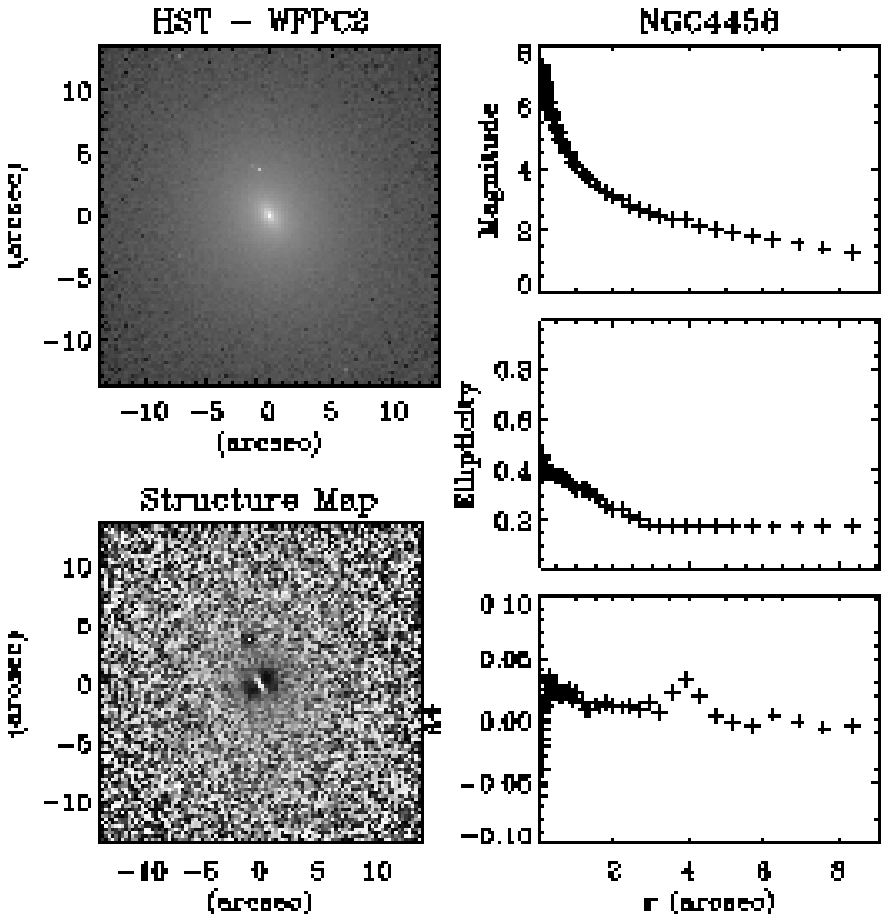}
\includegraphics[width=2.2in]{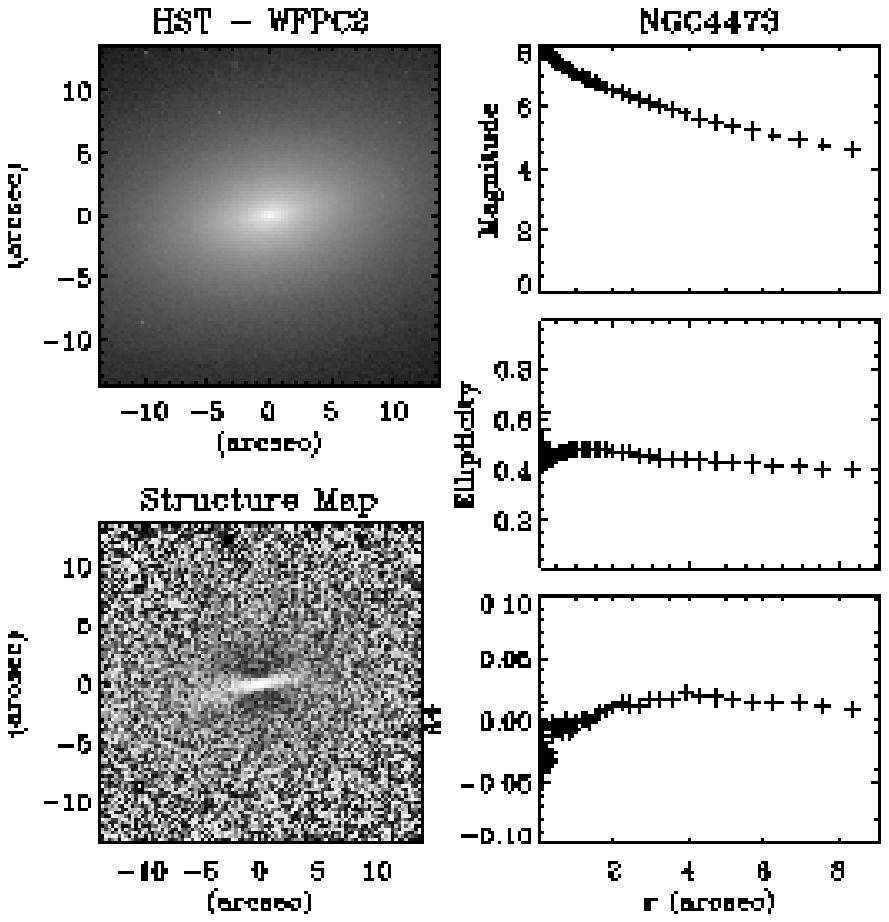}
\includegraphics[width=2.2in]{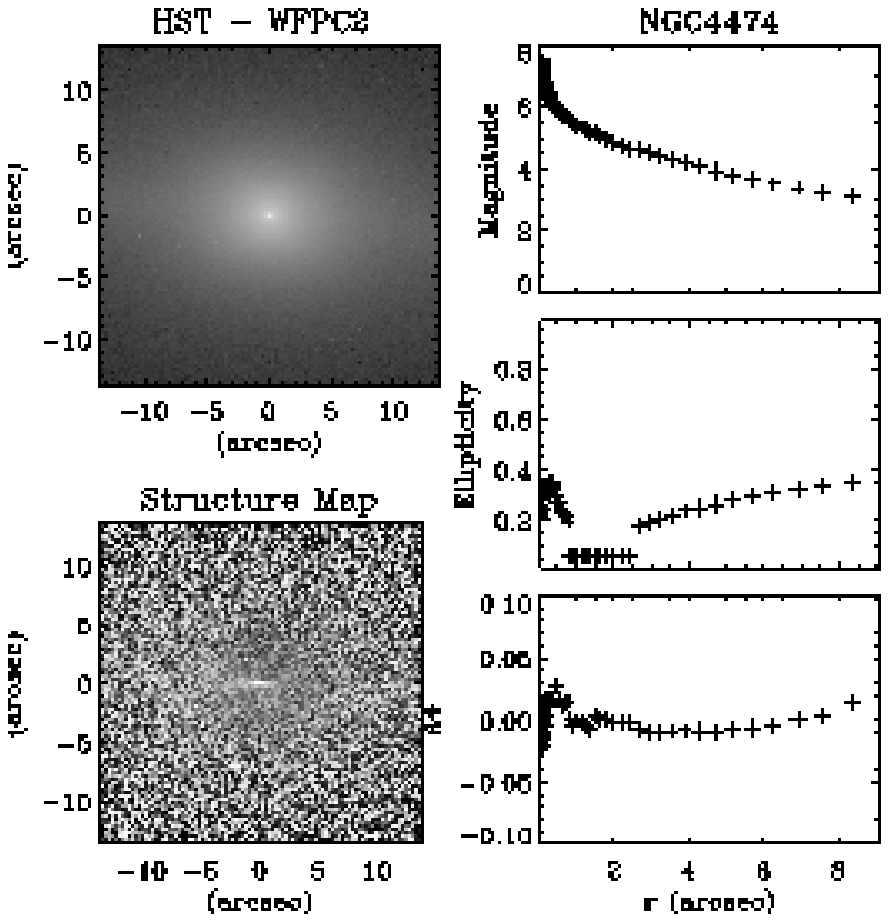}
\caption{NGC4458 on the left, NGC4473 in the middle and NGC4474 on the right.}
\end{center}
\end{figure*}

\clearpage

\begin{figure*}
\begin{center}
\includegraphics[width=2.2in]{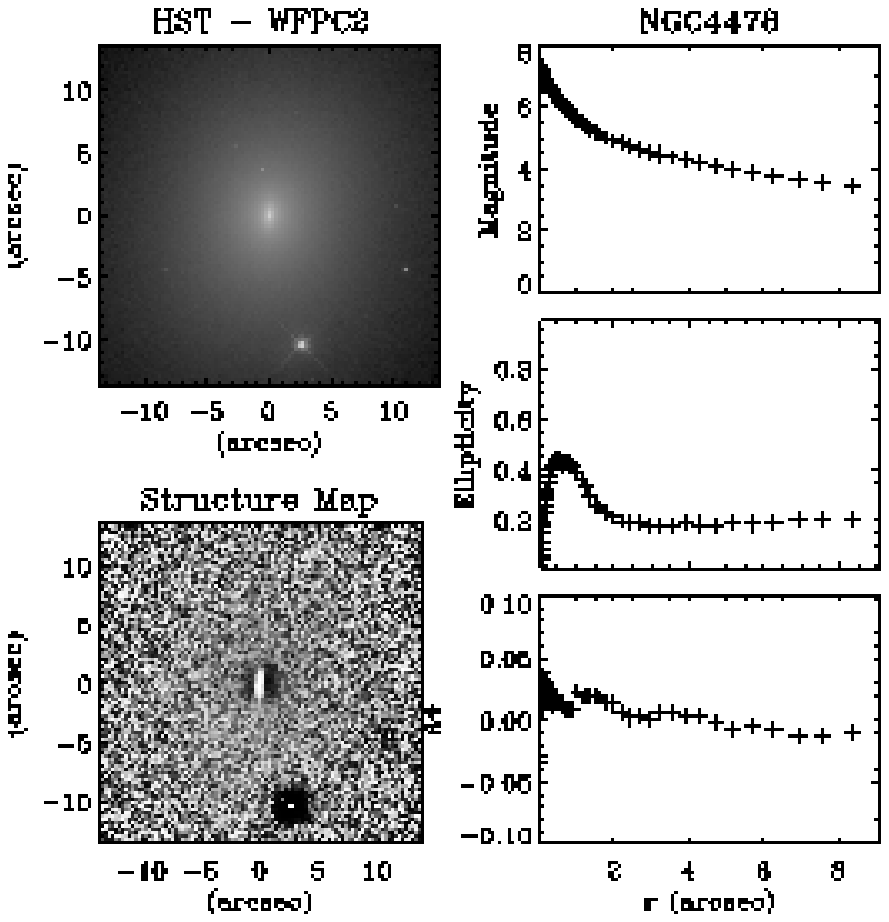}
\includegraphics[width=2.2in]{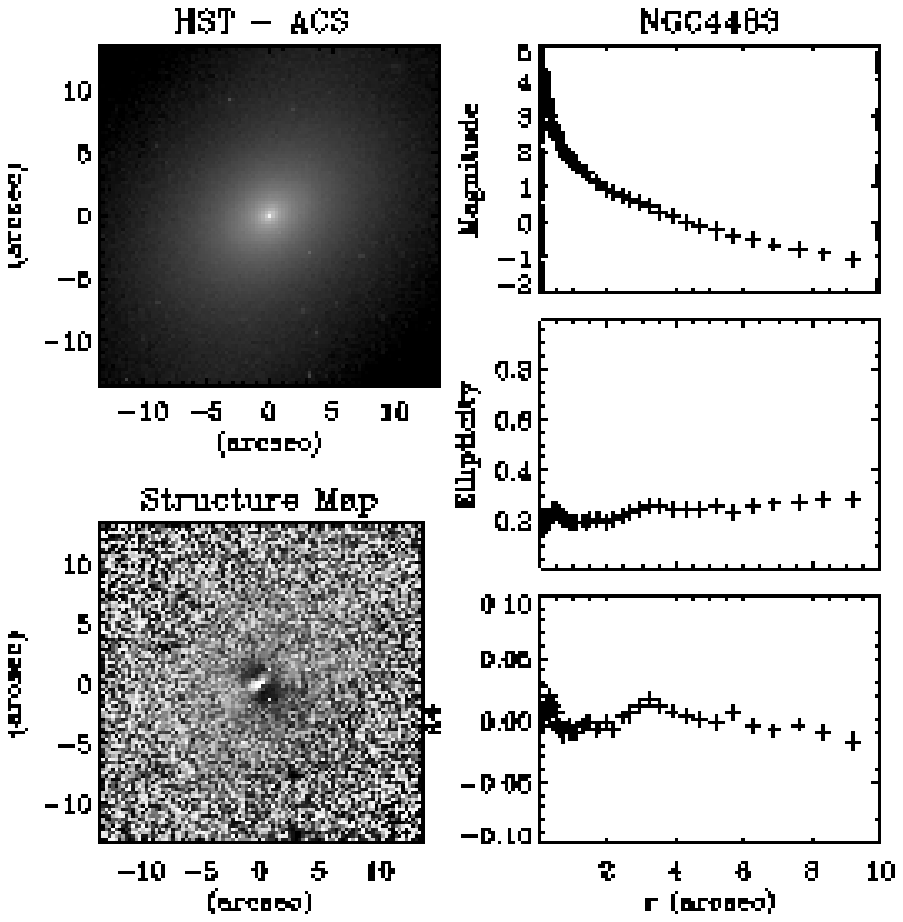}
\includegraphics[width=2.2in]{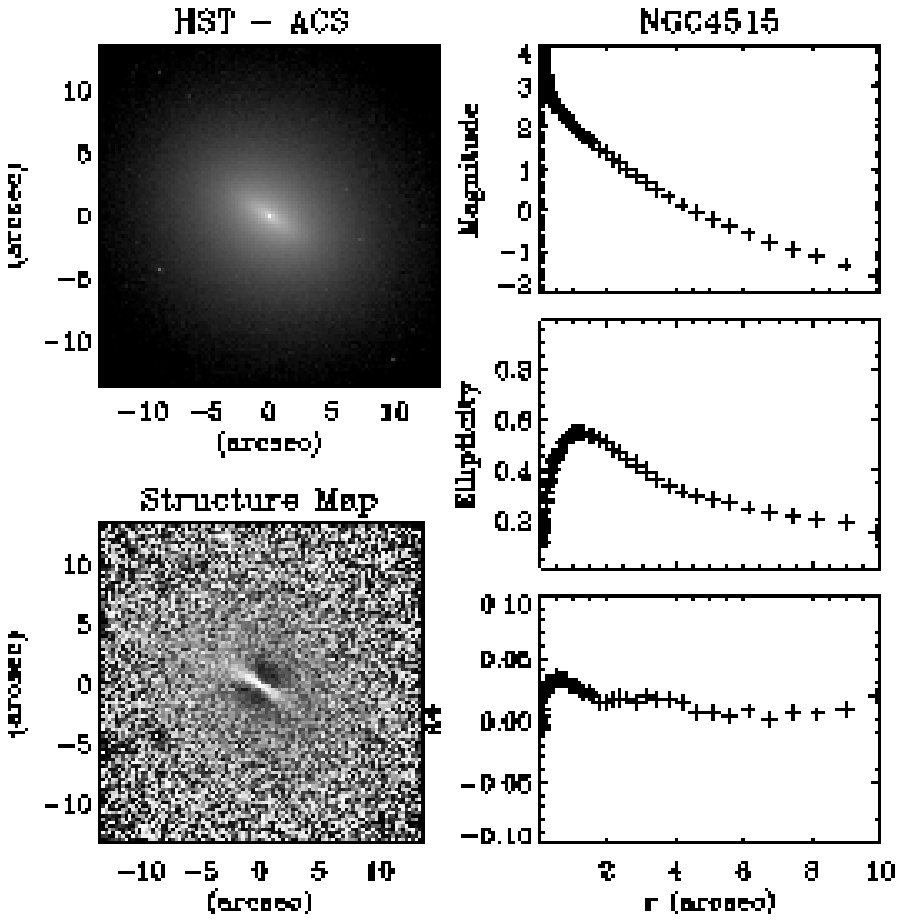}
\caption{NGC4478 on the left, NGC4483 in the middle and NGC4515 on the right.}
\end{center}
\end{figure*}

\begin{figure*}
\begin{center}
\includegraphics[width=2.2in]{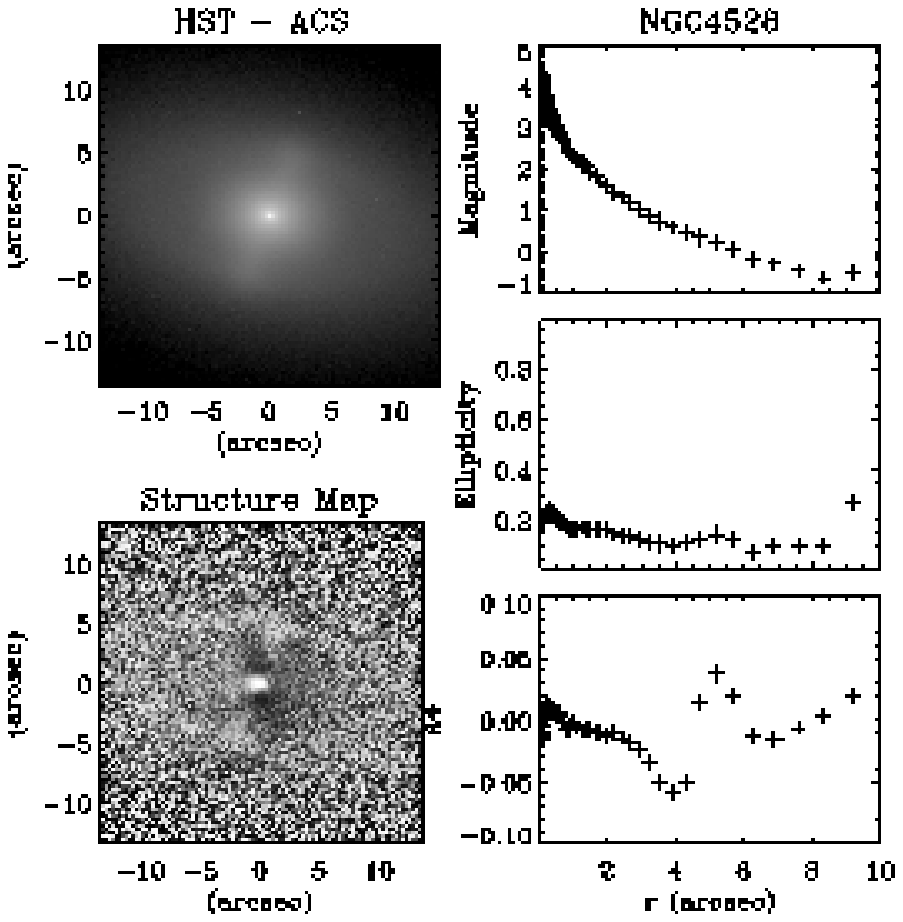}
\includegraphics[width=2.2in]{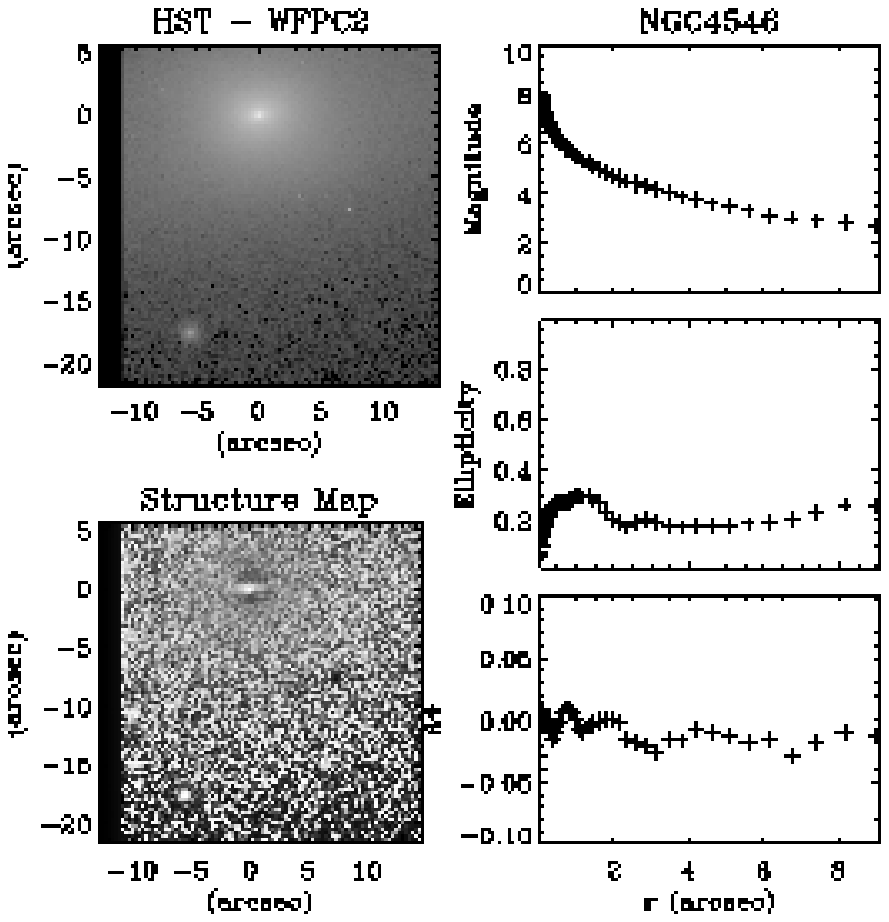}
\includegraphics[width=2.2in]{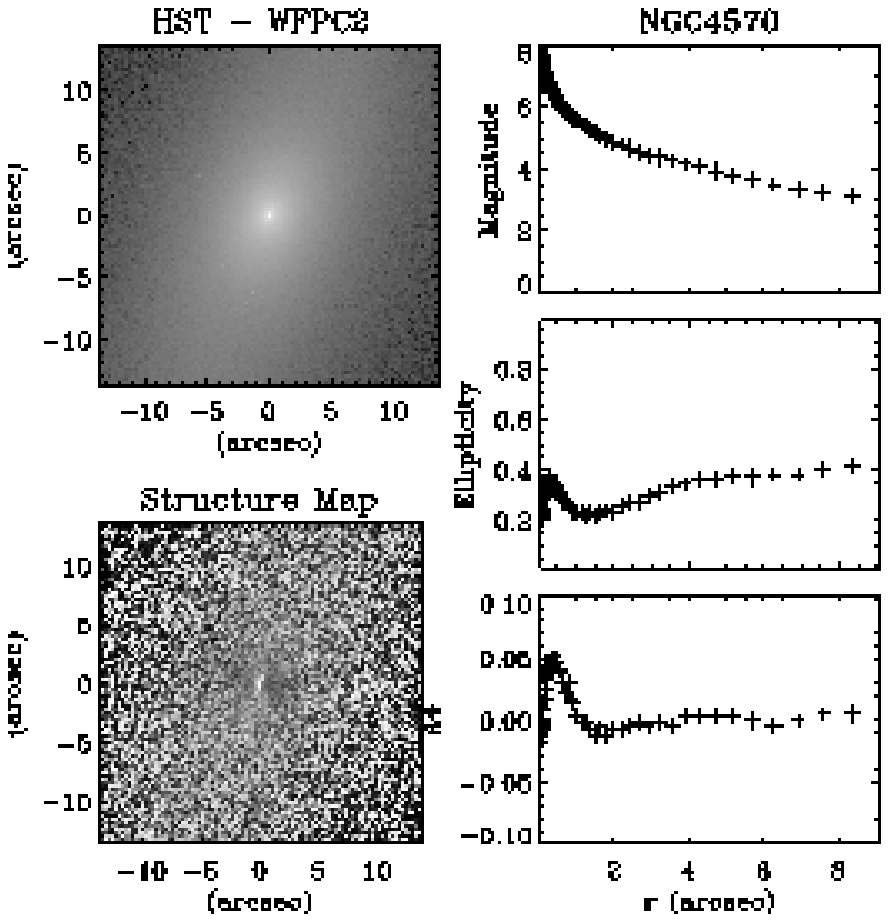}
\caption{NGC4528 on the left, NGC4546 in the middle and NGC4570 on the right.}
\end{center}
\end{figure*}

\begin{figure*}
\begin{center}
\includegraphics[width=2.2in]{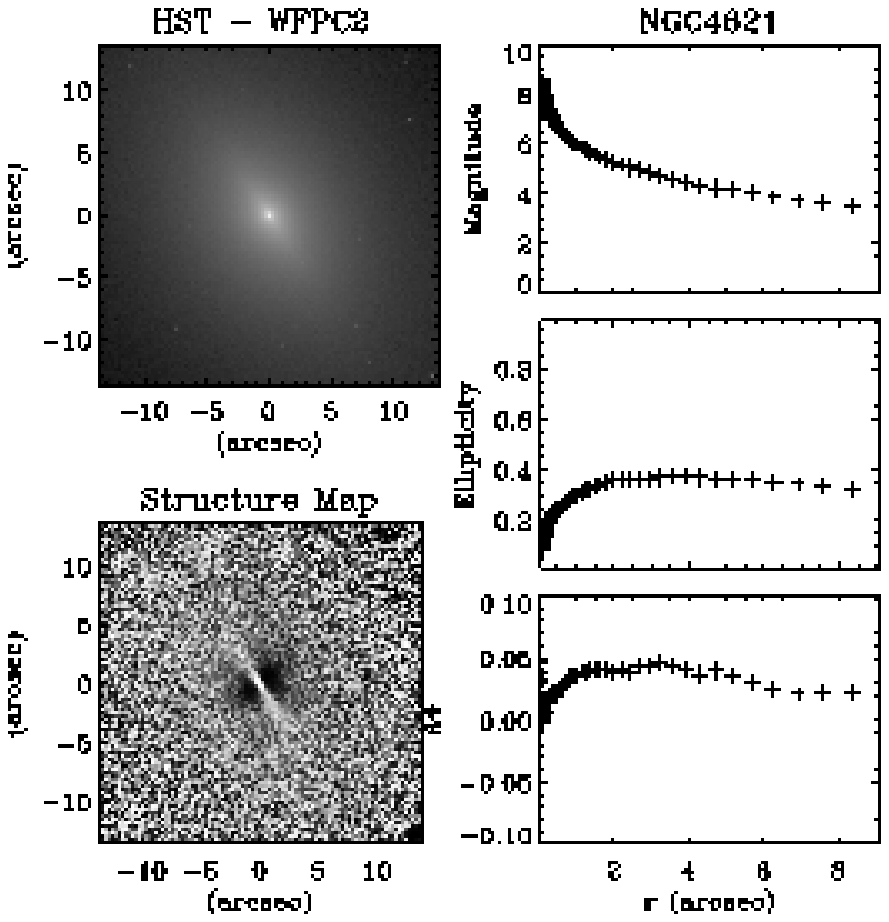}
\includegraphics[width=2.2in]{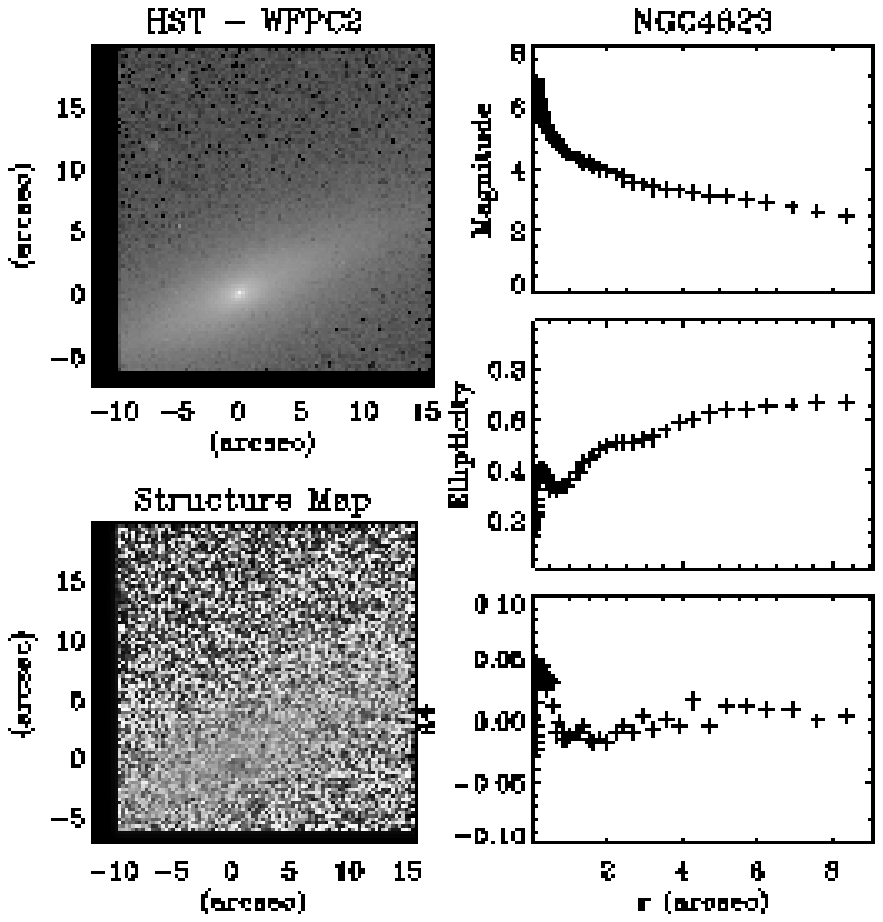}
\includegraphics[width=2.2in]{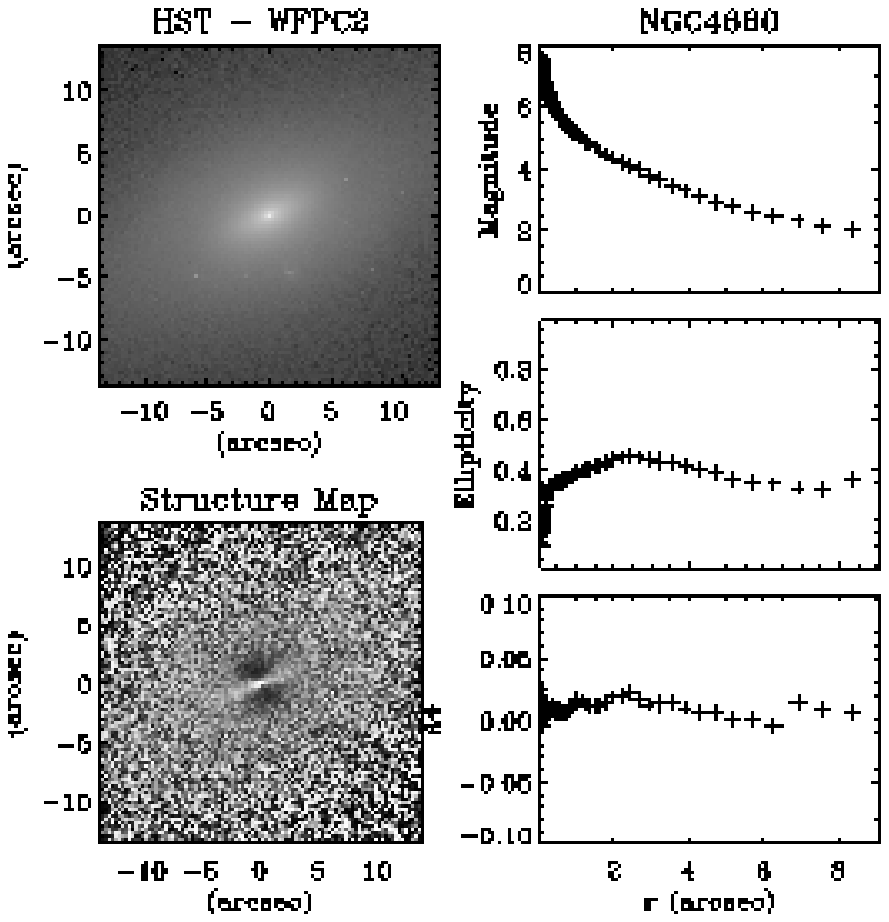}
\caption{NGC4621 on the left, NGC4623 in the middle and NGC4660 on the right.}
\end{center}
\end{figure*}

\begin{figure*}
\begin{center}
\includegraphics[width=2.2in]{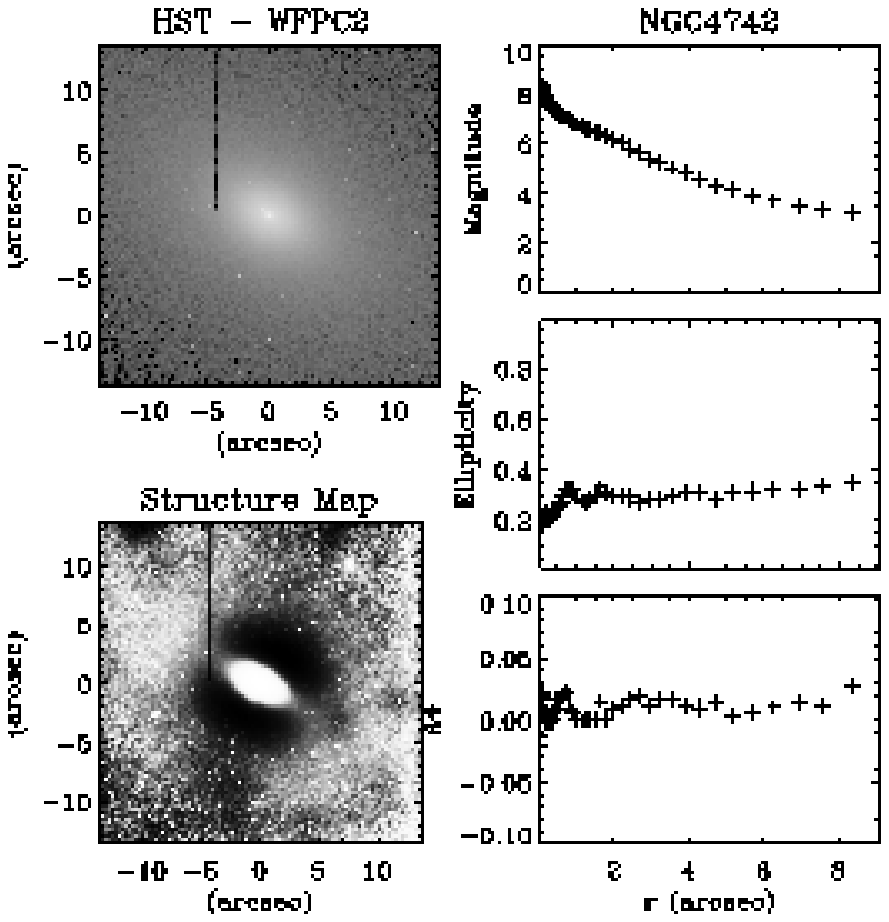}
\includegraphics[width=2.2in]{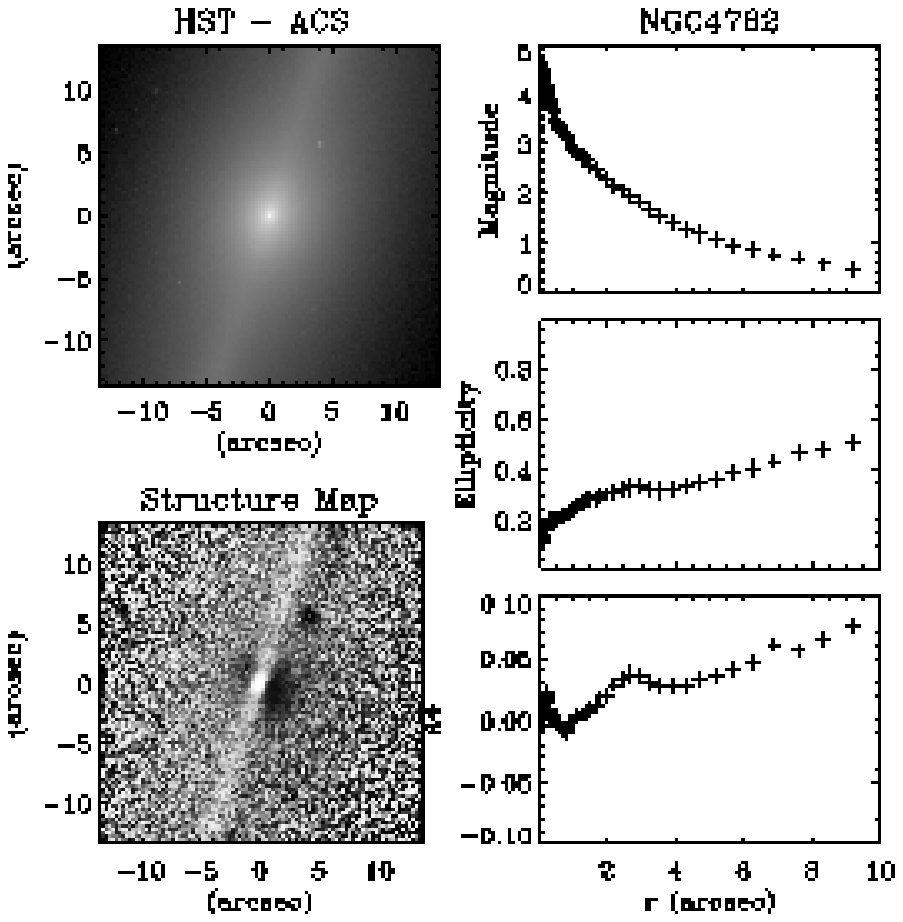}
\includegraphics[width=2.2in]{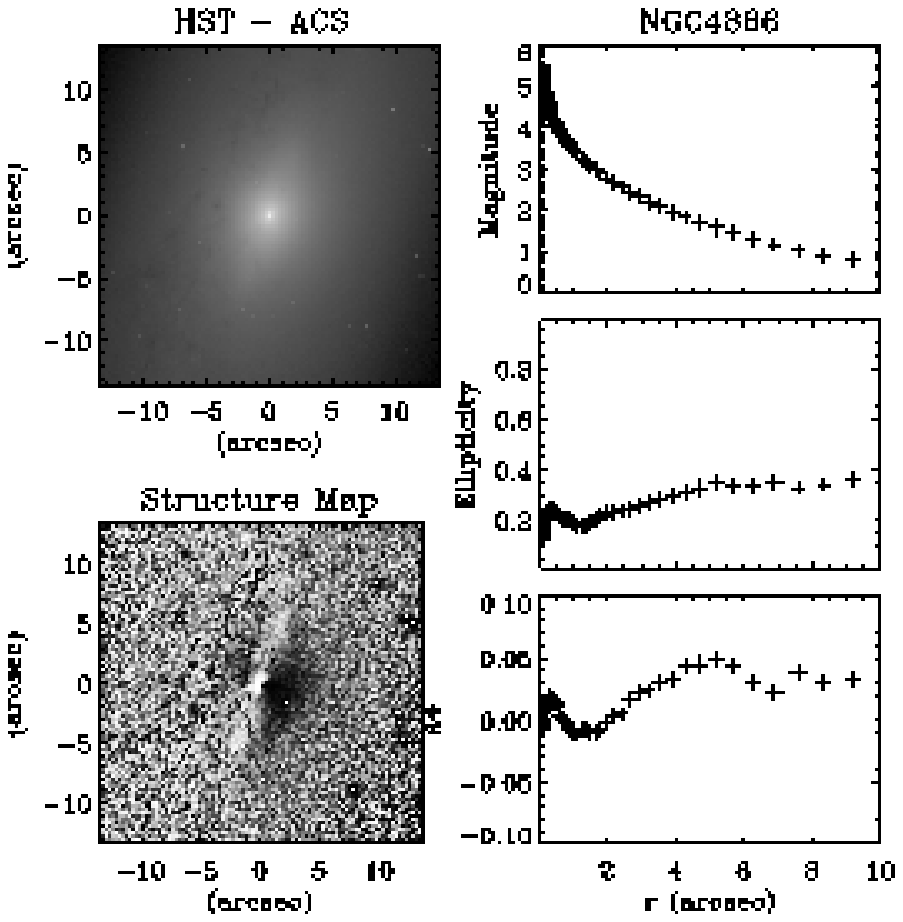}
\caption{NGC4742 on the left, NGC4762 in the middle and NGC4866 on the right.}
\end{center}
\end{figure*}

\begin{figure*}
\begin{center}
\includegraphics[width=2.2in]{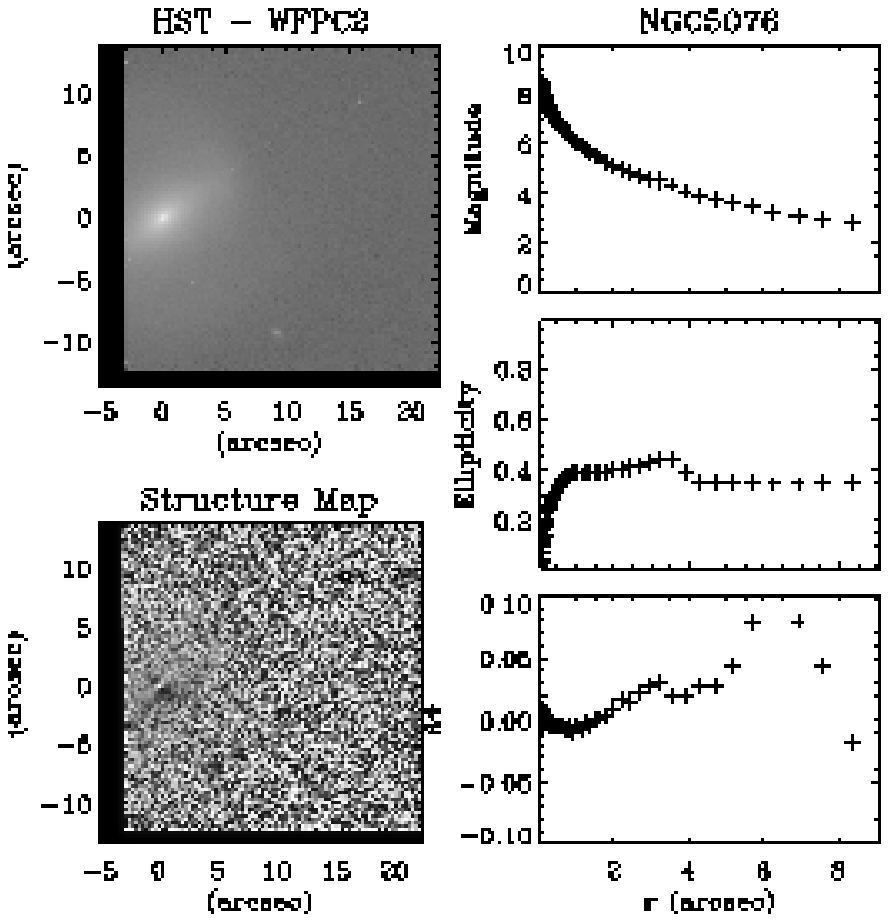}
\includegraphics[width=2.2in]{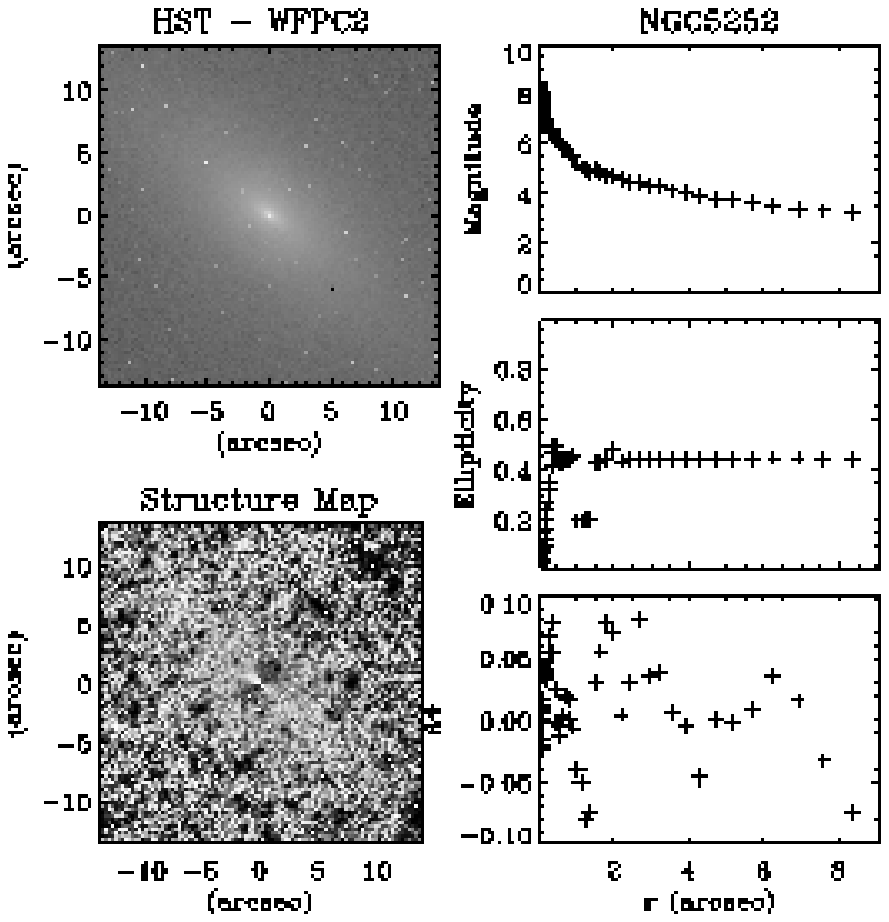}
\includegraphics[width=2.2in]{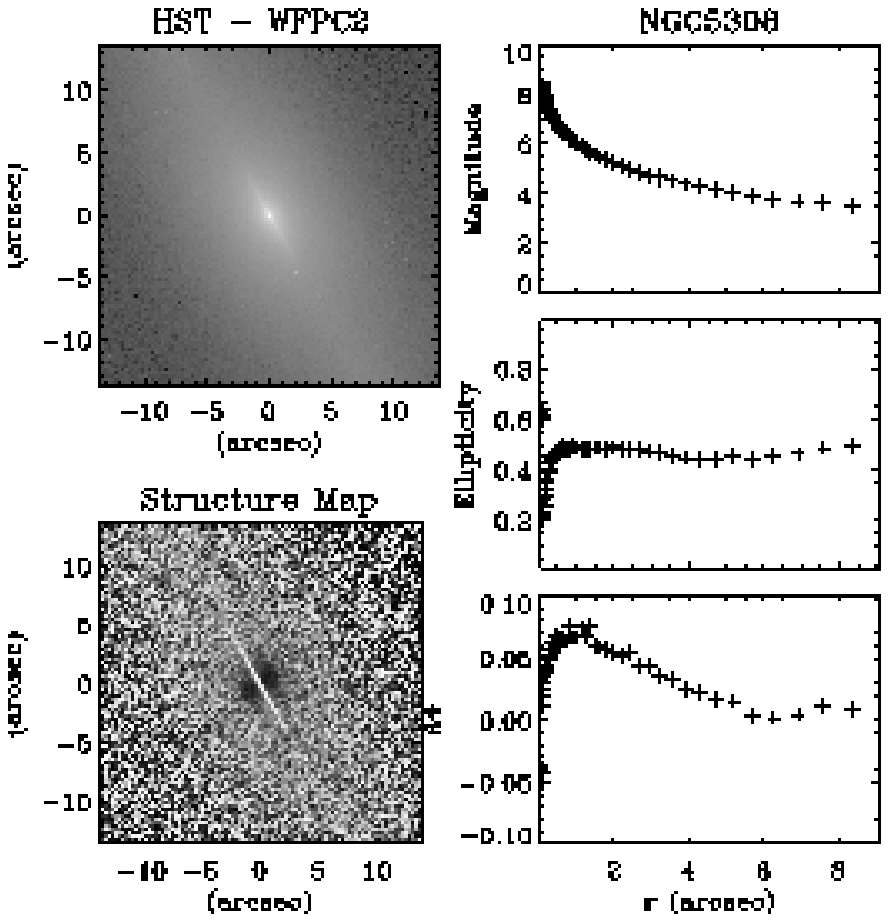}
\caption{NGC5076 on the left, NGC5252 in the middle and NGC5308 on the right.}
\end{center}
\end{figure*}

\begin{figure*}
\begin{center}
\includegraphics[width=2.2in]{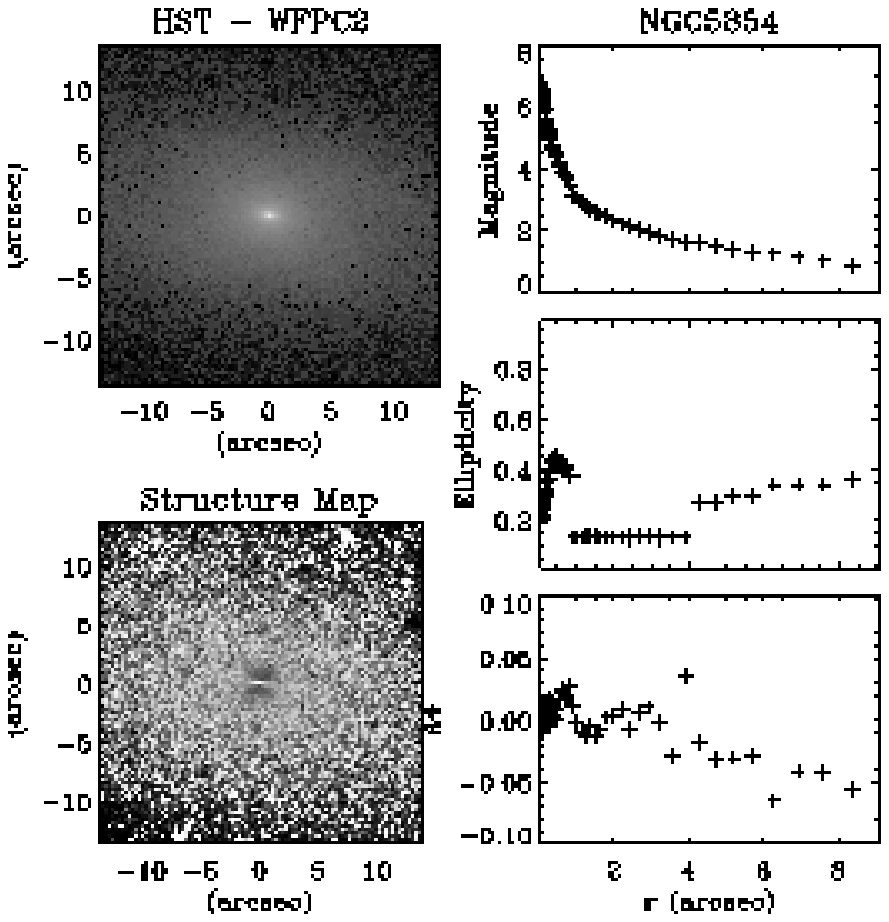}
\includegraphics[width=2.2in]{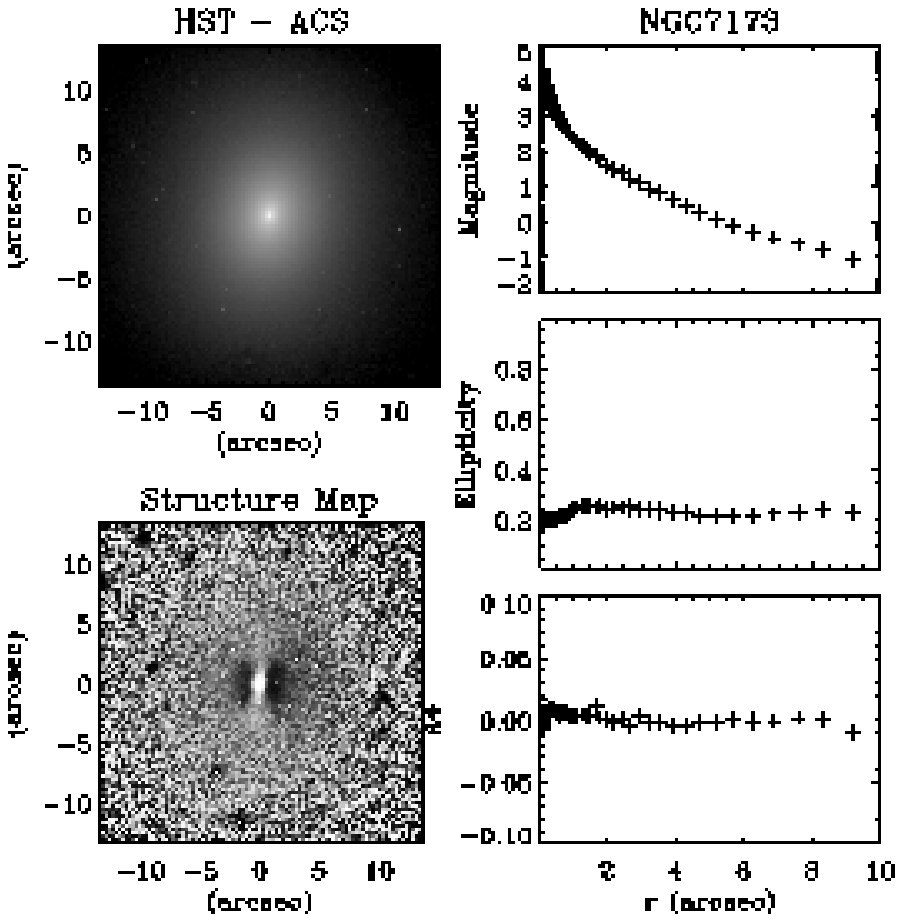}
\includegraphics[width=2.2in]{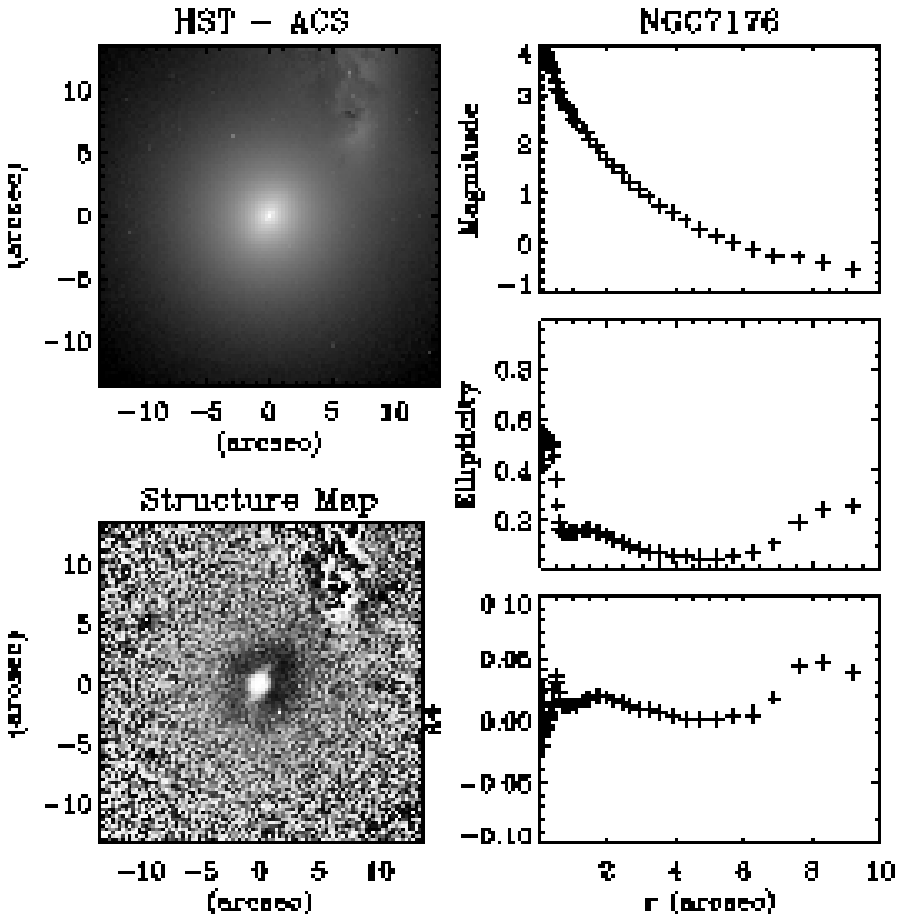}
\caption{NGC5854 on the left, NGC7173 in the middle and NGC7176 on the right.}
\end{center}
\end{figure*}

\begin{figure*}
\begin{center}
\includegraphics[width=2.2in]{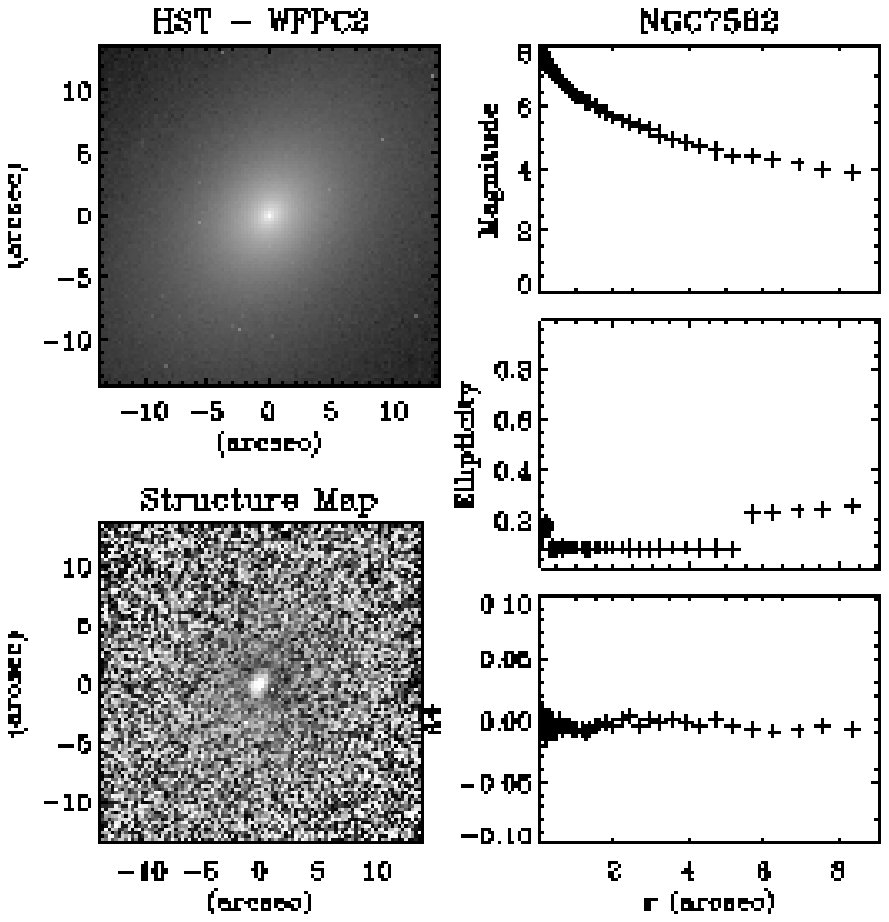}
\includegraphics[width=2.2in]{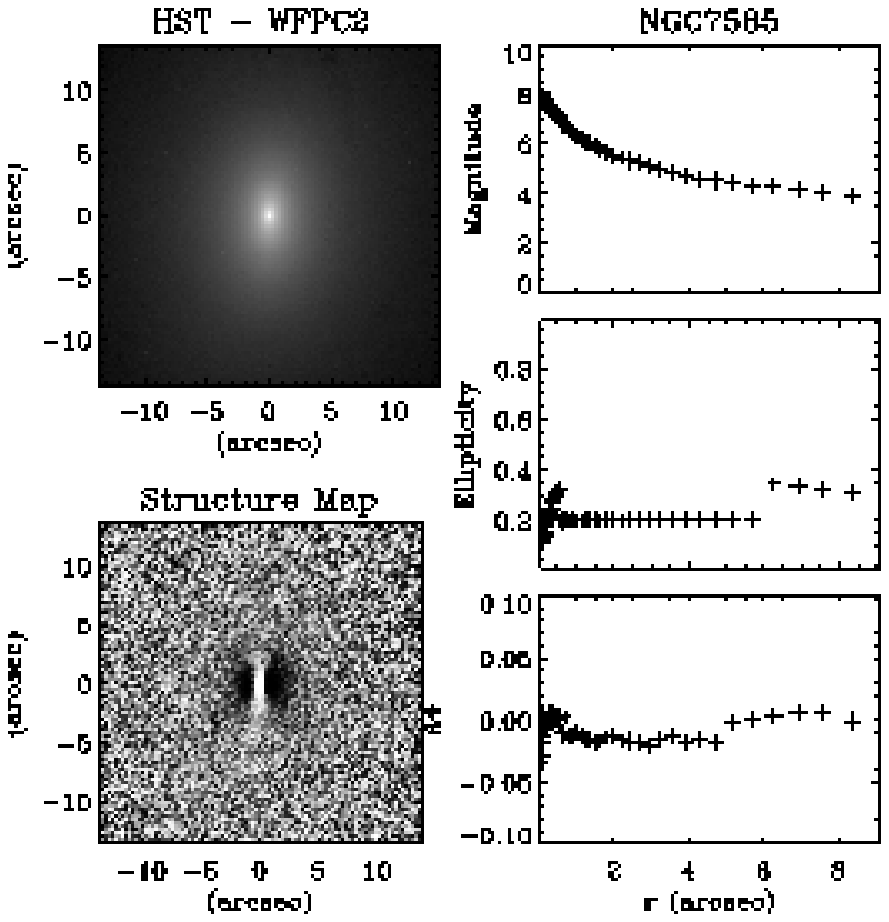}
\includegraphics[width=2.2in]{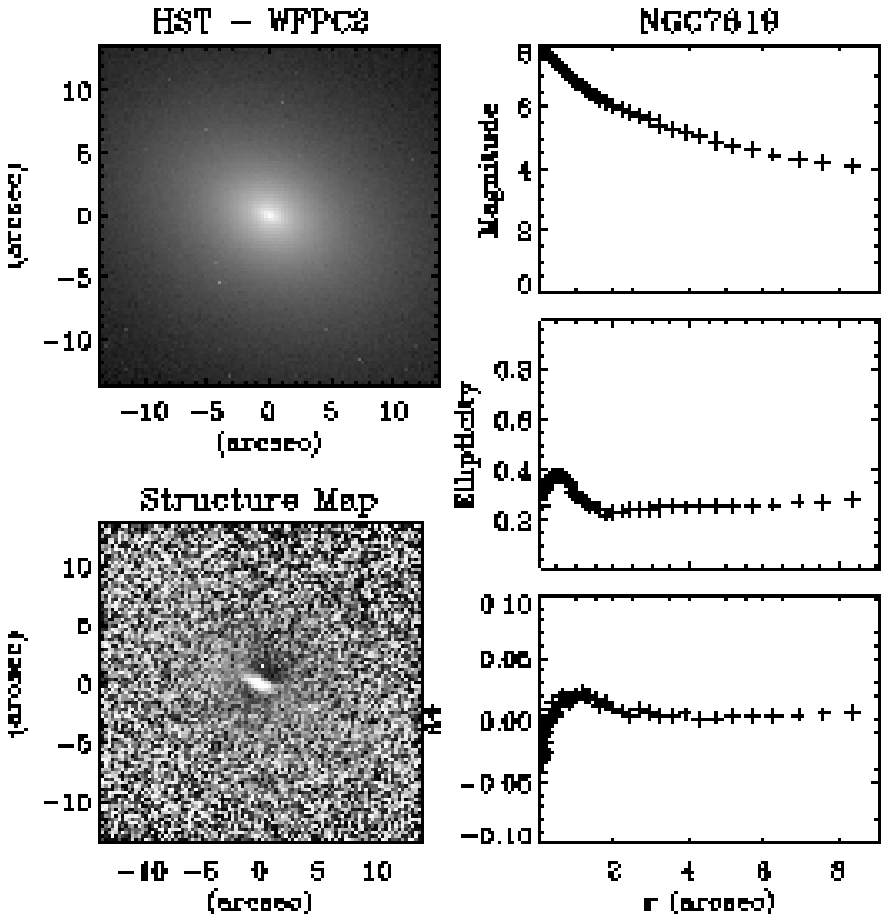}
\caption{NGC7562 on the left, NGC7585 in the middle and NGC7619 on the right.}
\end{center}
\end{figure*}

\begin{figure*}
\begin{center}
\includegraphics[width=2.2in]{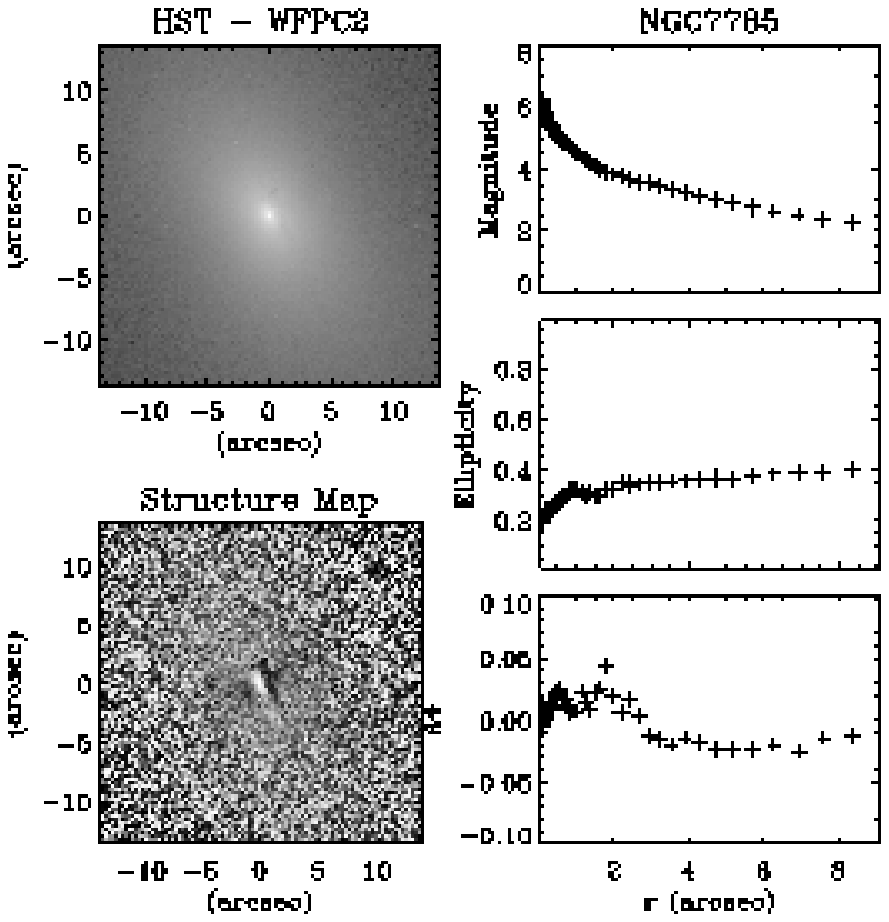}
\includegraphics[width=2.2in]{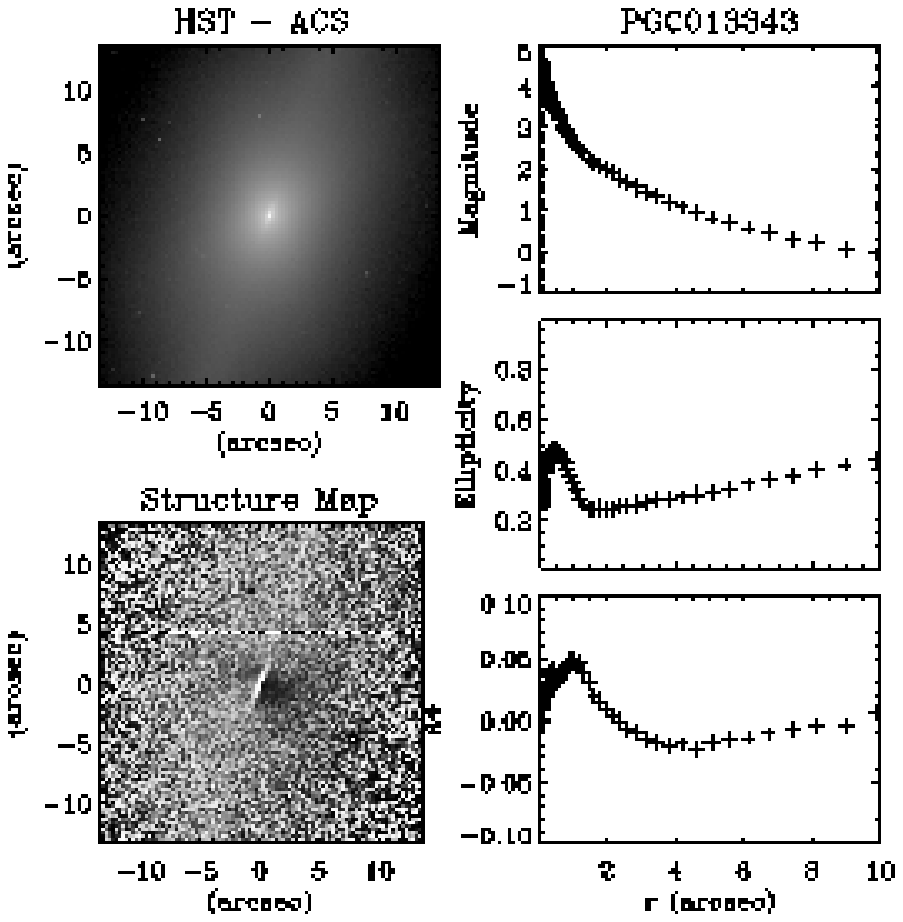}
\includegraphics[width=2.2in]{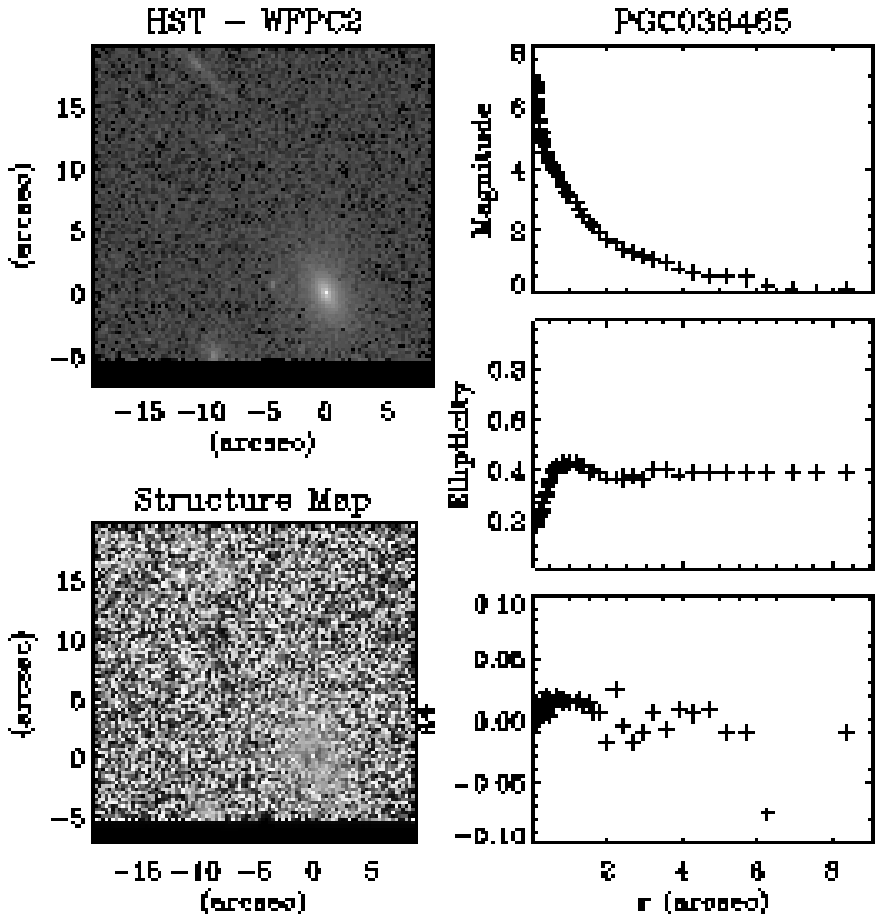}
\caption{NGC7785 on the left, PGC013343 in the middle and PGC036465 on the right.}
\end{center}
\end{figure*}

\begin{figure*}
\begin{center}
\includegraphics[width=2.2in]{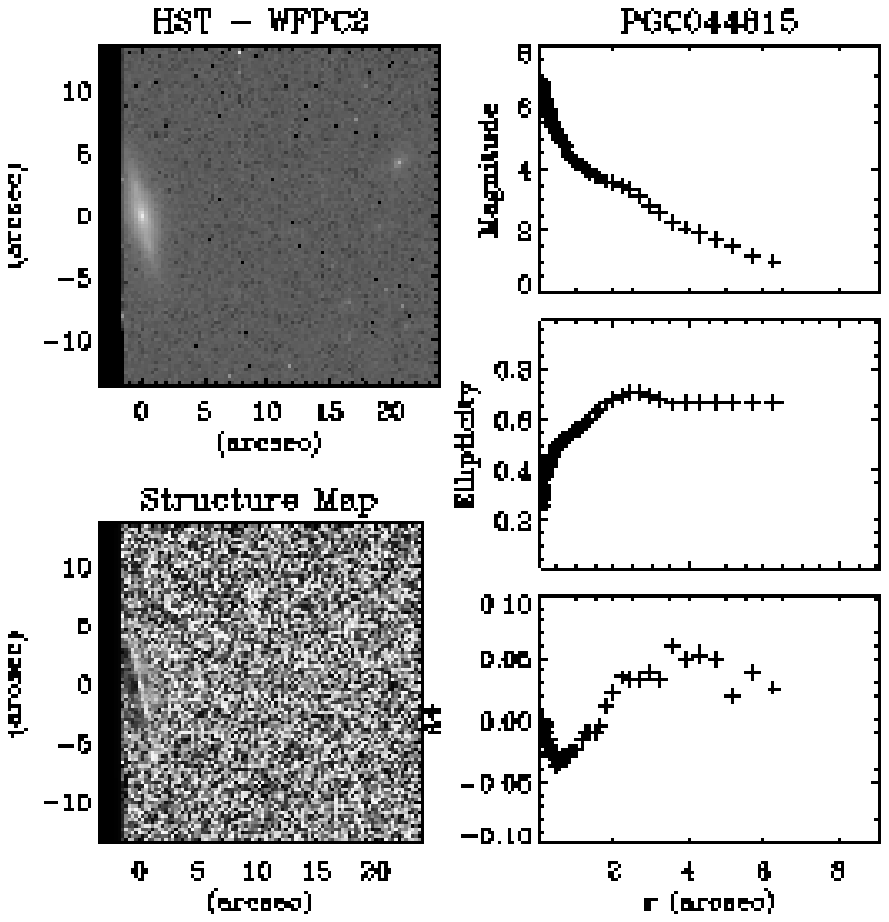}
\includegraphics[width=2.2in]{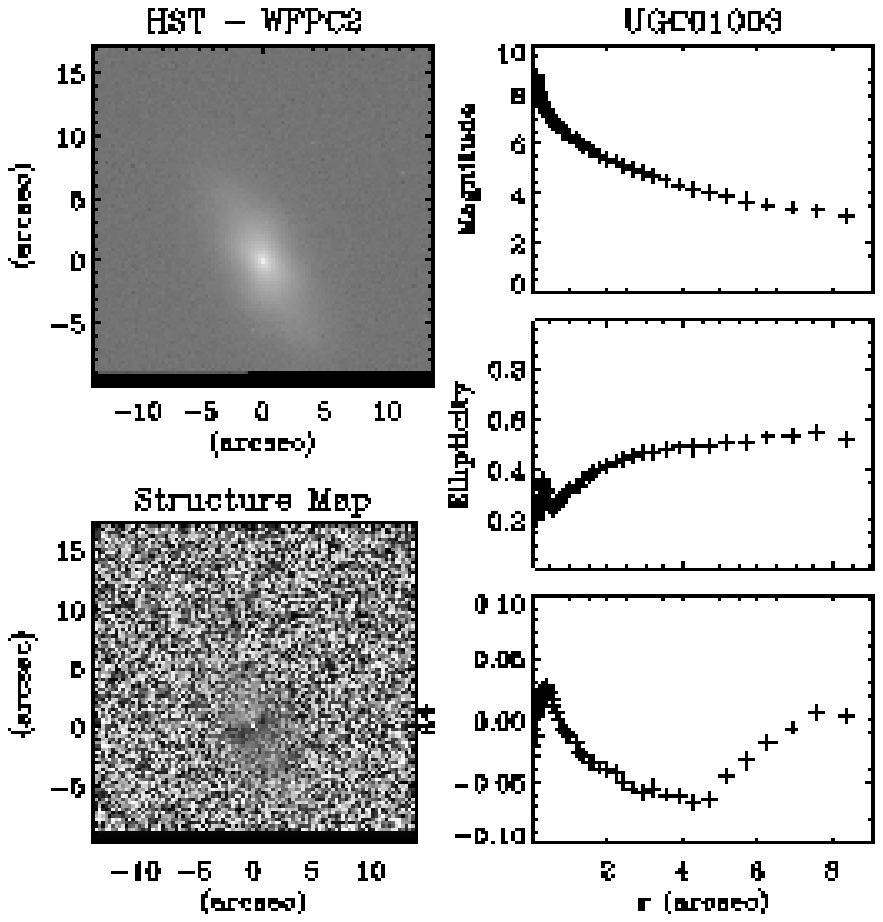}
\includegraphics[width=2.2in]{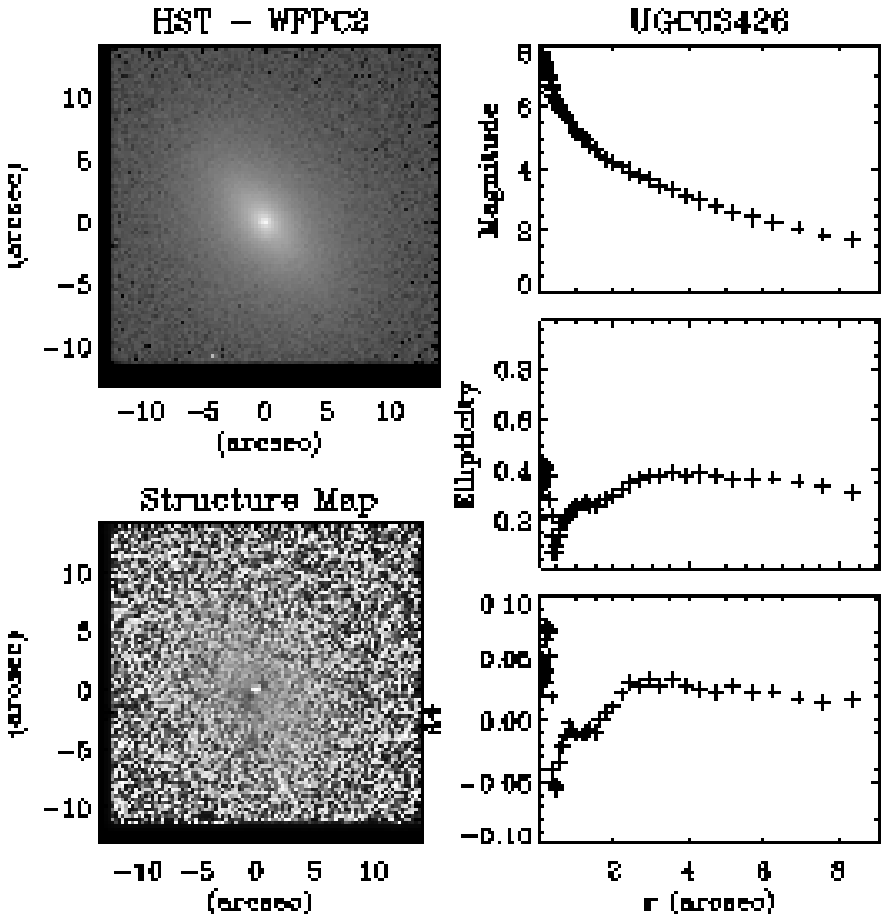}
\caption{PGC044815 on the left, UGC01003 in the middle and UGC03426 on the right.}
\end{center}
\end{figure*}

\clearpage

\subsection{Disk Decomposition}

\begin{figure*}
\begin{center}
\includegraphics[width=3in]{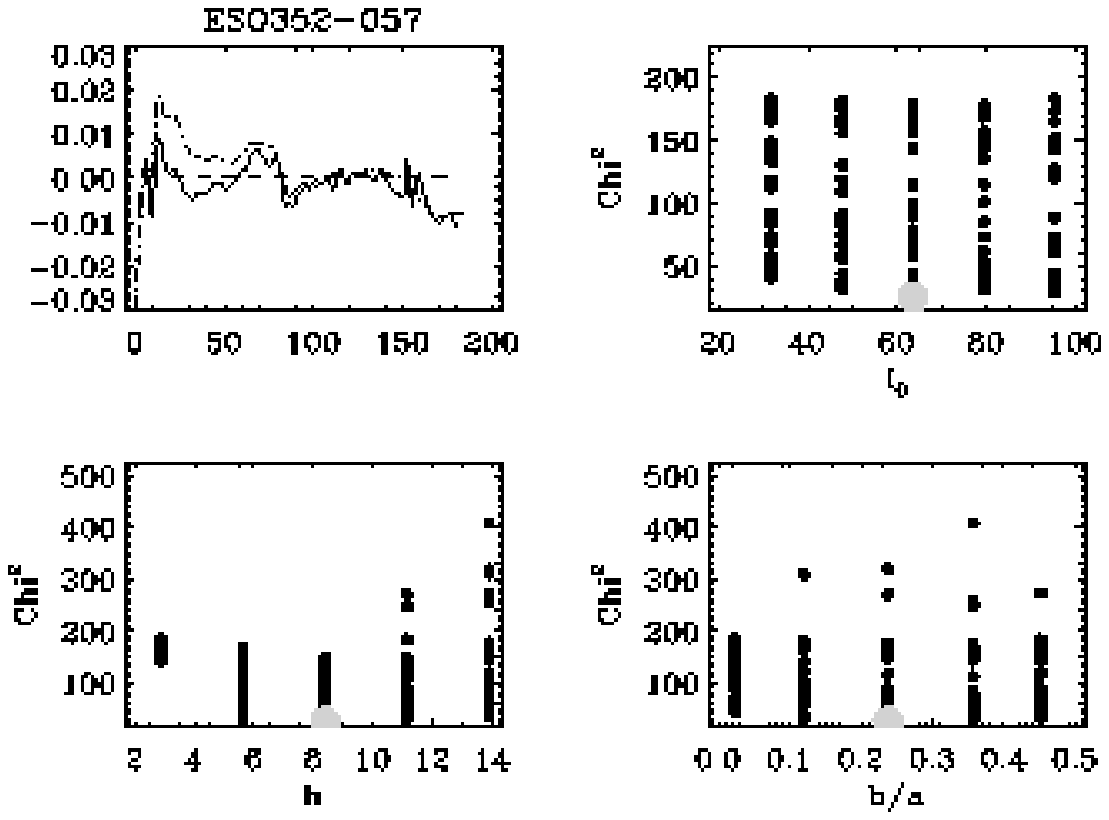}
\includegraphics[width=3in]{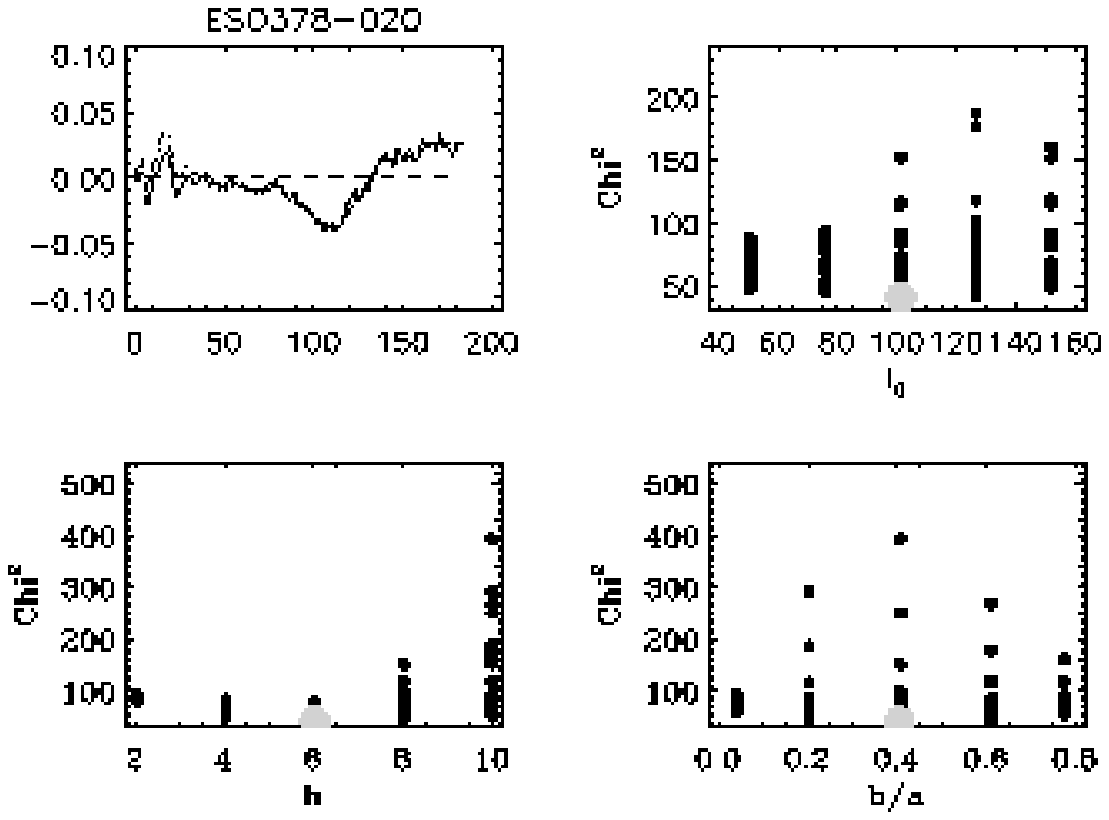}
\caption{Scorza and Bender decomposition of ESO352-057 on the left and  on the right. }
\end{center}
\end{figure*}

\begin{figure*}
\begin{center}
\includegraphics[width=3in]{ESO507-027p_lowR.ps}
\includegraphics[width=3in]{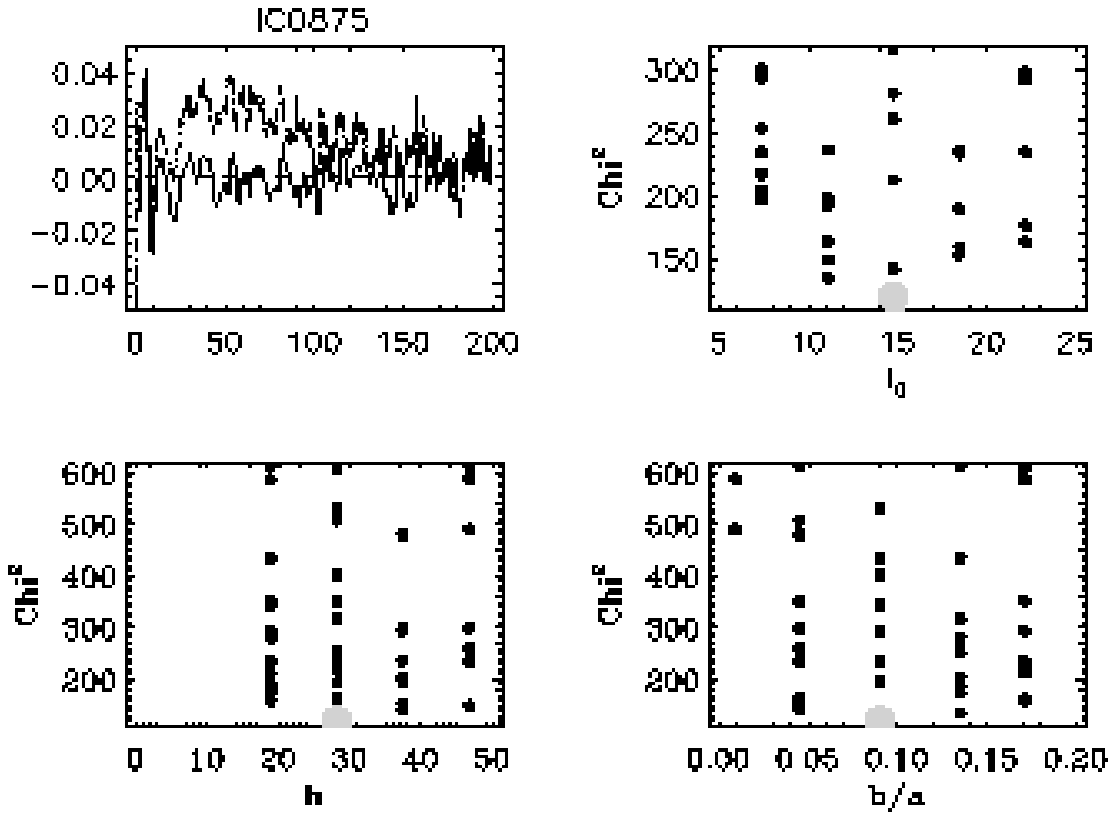}
\caption{Scorza and Bender decomposition of ESO507-027 on the left and IC0875 on the right.}
\end{center}
\end{figure*}

\begin{figure*}
\begin{center}
\includegraphics[width=3in]{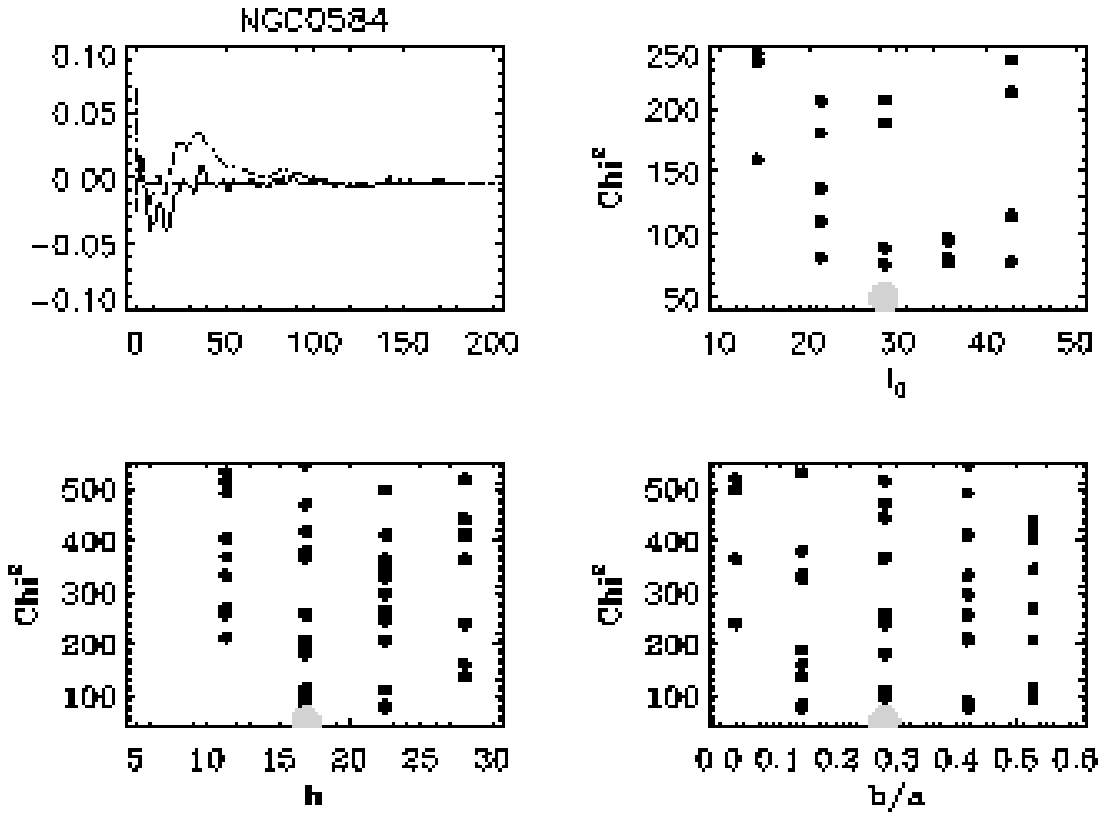}
\includegraphics[width=3in]{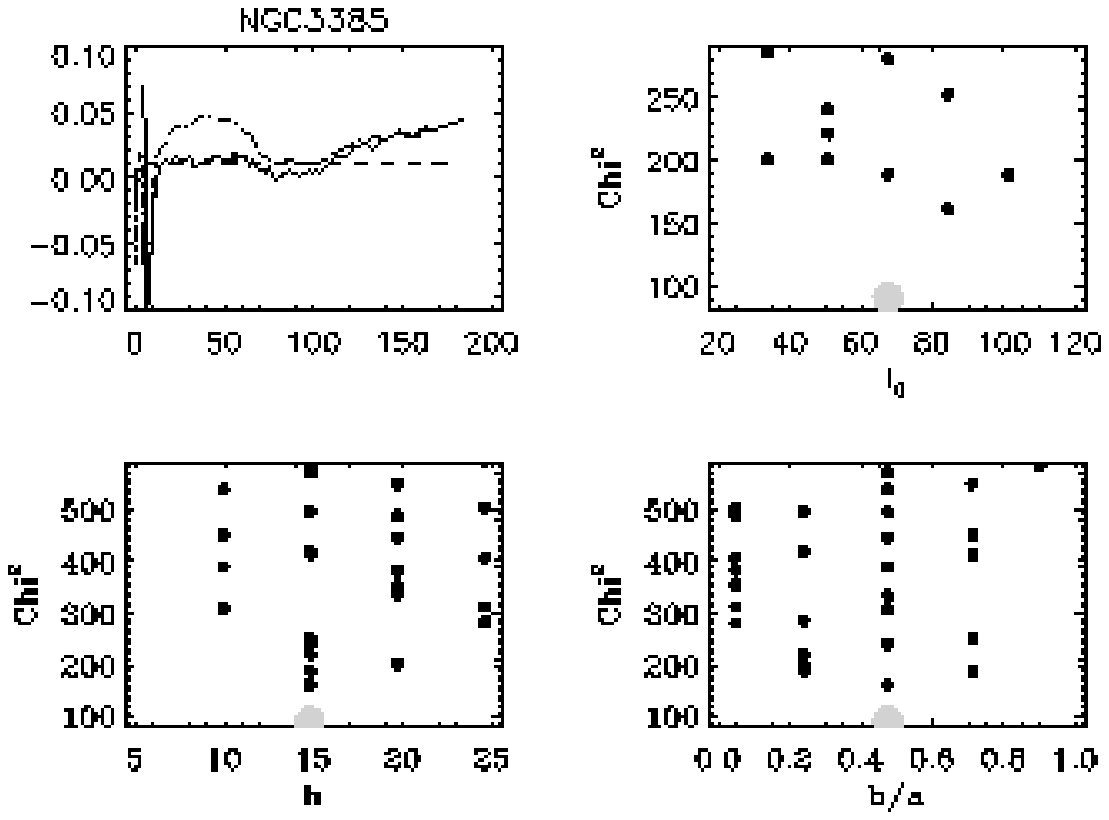}
\caption{Scorza and Bender decomposition of NGC0584 on the left and NGC3385 on the right.}
\end{center}
\end{figure*}

\begin{figure*}
\begin{center}
\includegraphics[width=3in]{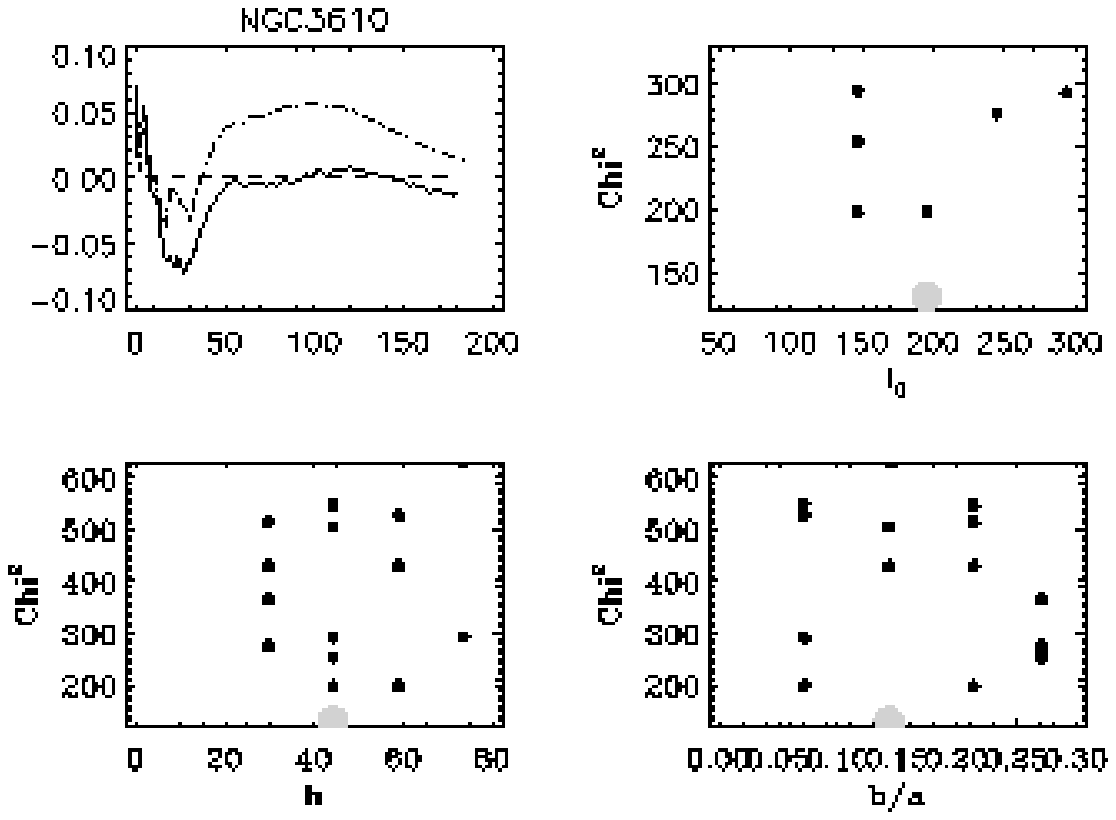}
\includegraphics[width=3in]{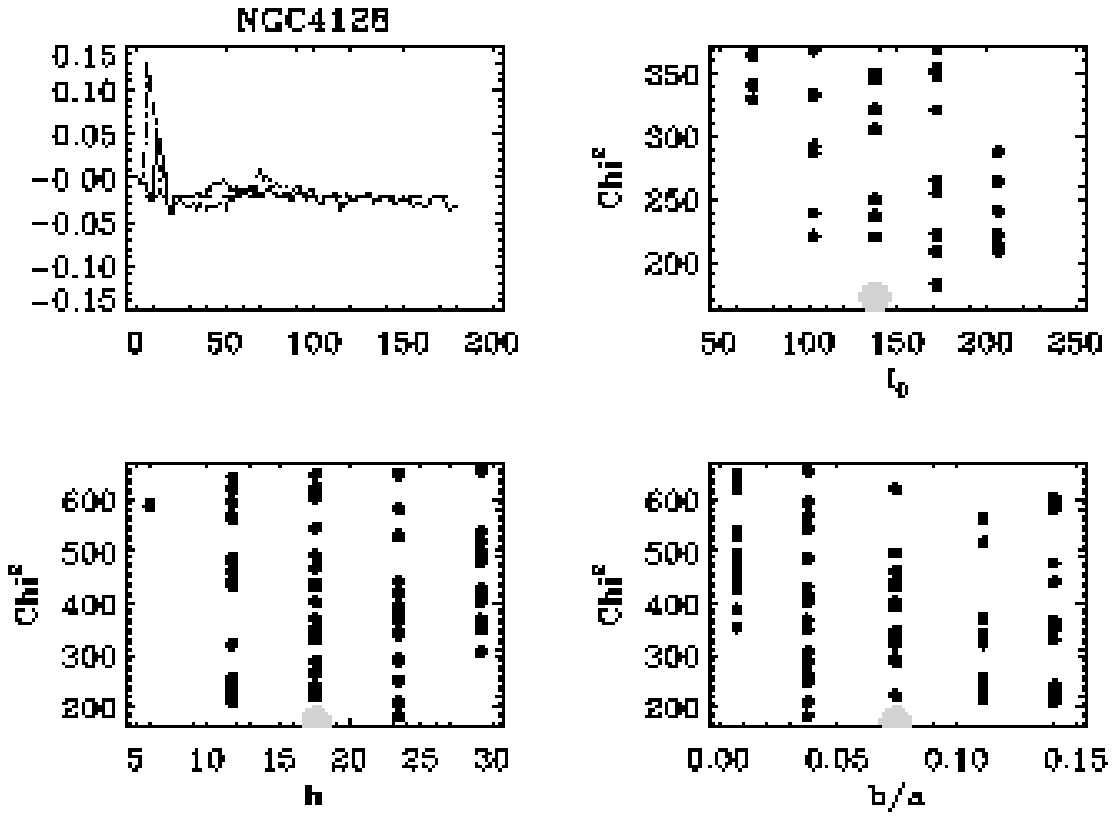}
\caption{Scorza and Bender decomposition of NGC3610 on the left and NGC4128 on the right.}
\end{center}
\end{figure*}

\begin{figure*}
\begin{center}
\includegraphics[width=3in]{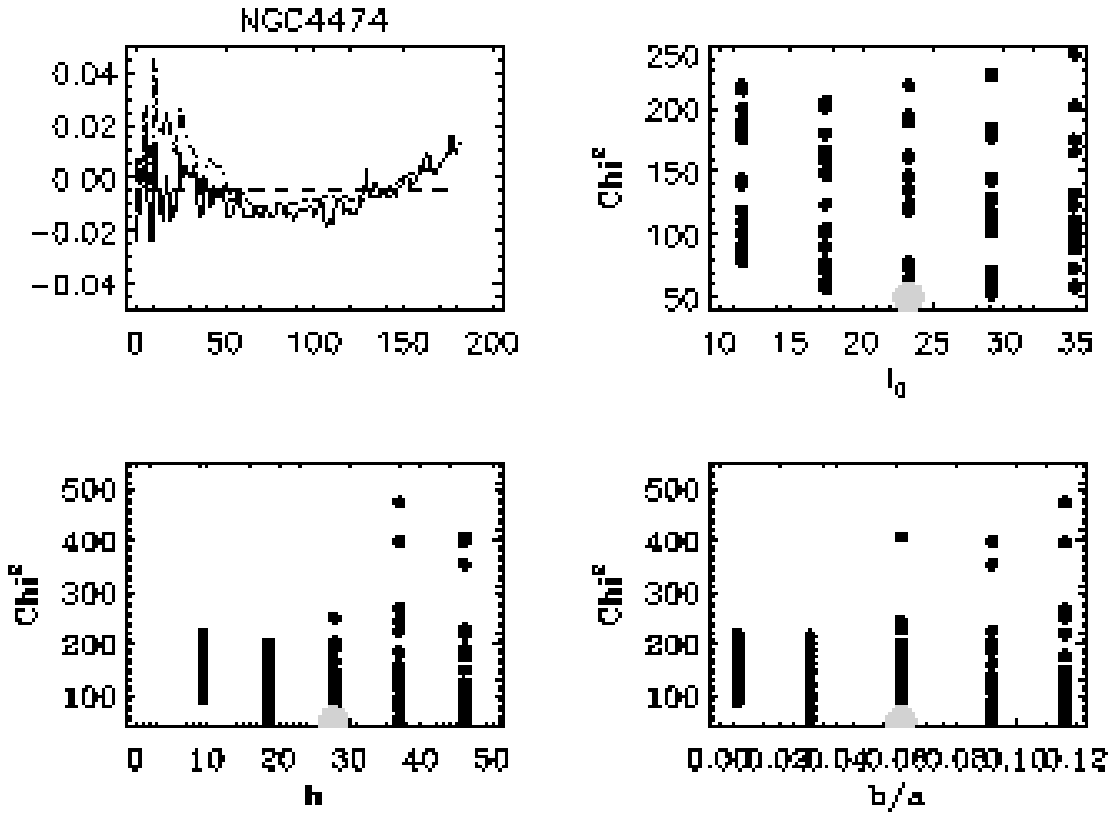}
\includegraphics[width=3in]{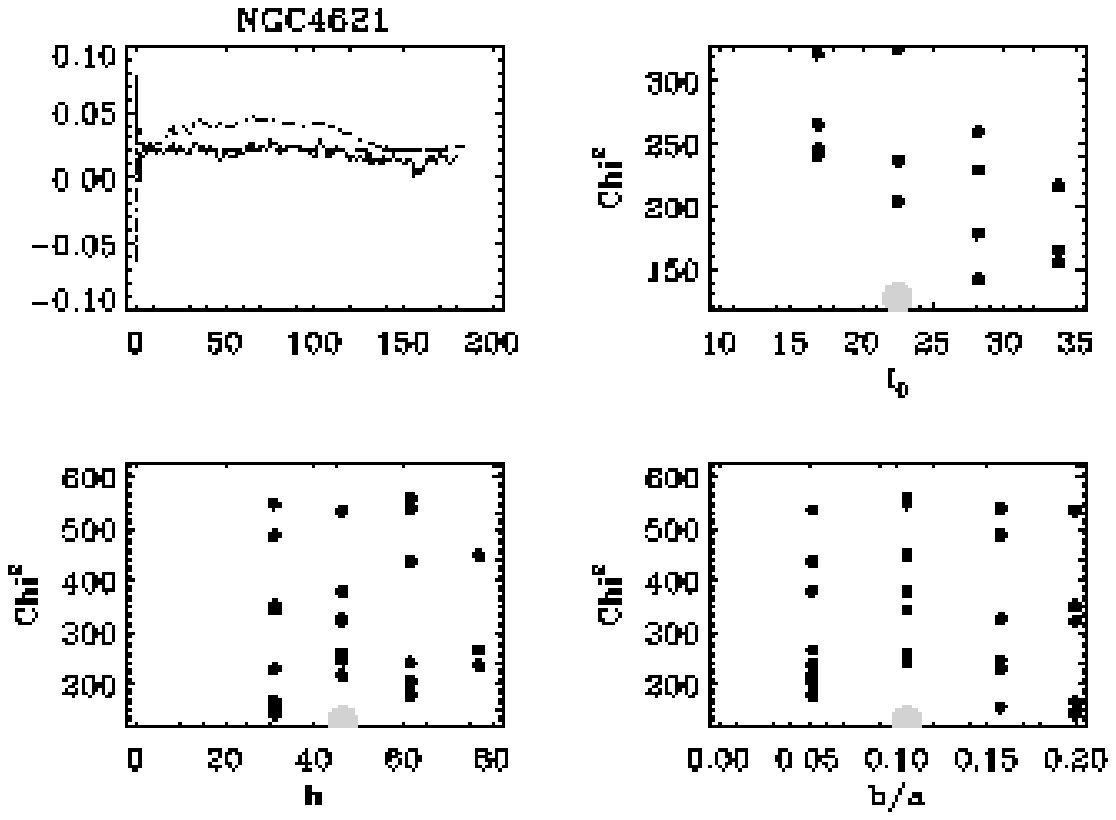}
\caption{Scorza and Bender decomposition of NGC4474 on the left and NGC4621 on the right.}
\end{center}
\end{figure*}

\begin{figure*}
\begin{center}
\includegraphics[width=3in]{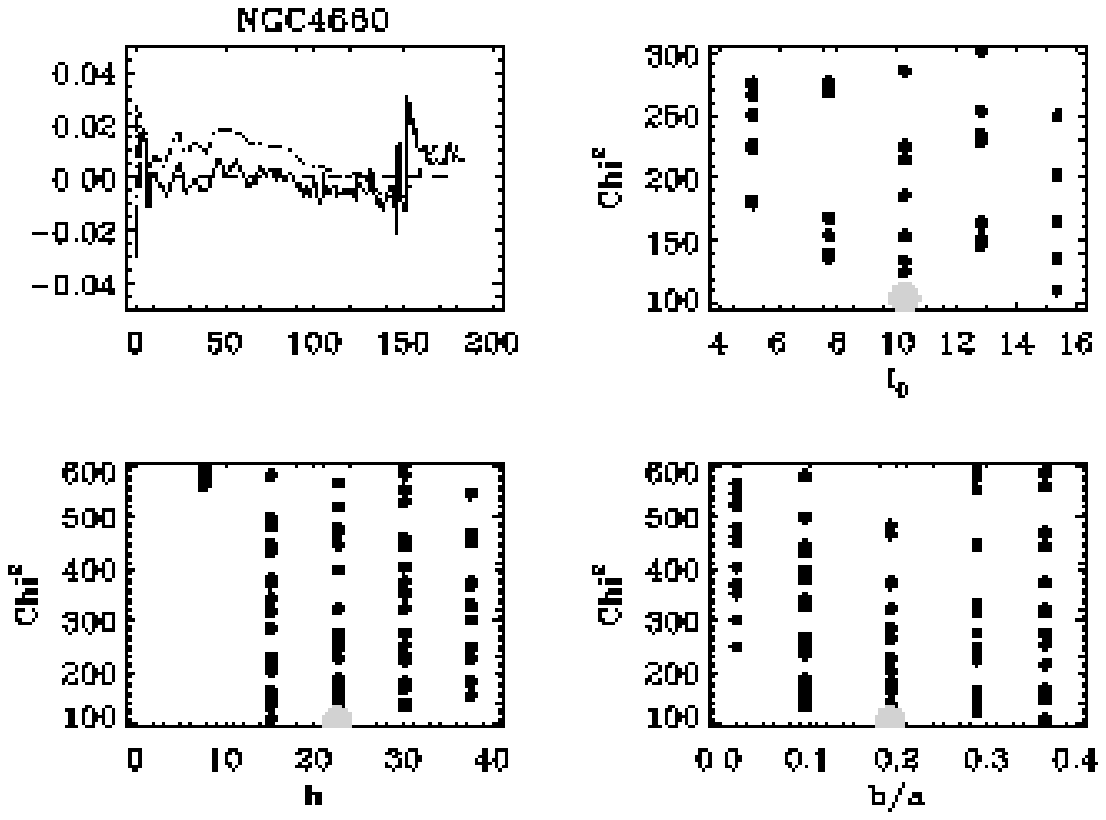}
\includegraphics[width=3in]{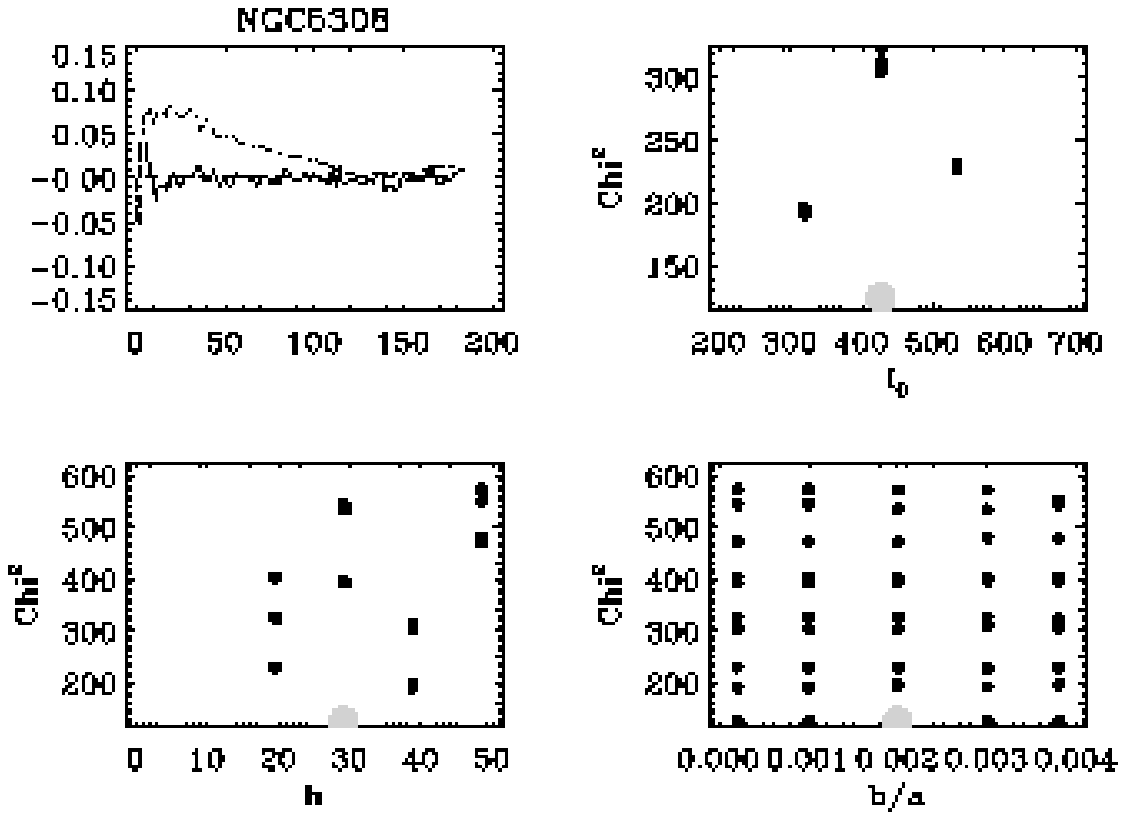}
\caption{Scorza and Bender decomposition of NGC4660 on the left and NGC5308 on the right.}
\end{center}
\end{figure*}

\label{lastpage}

\end{document}